\documentclass[12pt]{article}
\usepackage{amsmath,amsfonts,amssymb,booktabs,color,epsfig,graphicx,hyperref,url, amsthm}

\usepackage[margin=1 in]{geometry}
\usepackage{bm}
\usepackage{bbm}
\usepackage{float}
\usepackage[caption=false]{subfig}
\usepackage{float}
\usepackage{multicol}
\usepackage{multirow}
\usepackage{lscape}
\usepackage[utf8]{inputenc}
\usepackage{setspace}
\usepackage{placeins}
\usepackage{mathrsfs}
\usepackage{natbib}
\usepackage{enumitem}

\usepackage[algosection,linesnumbered,lined,boxed,commentsnumbered]{algorithm2e}
\usepackage[table,xcdraw]{xcolor}
\usepackage[caption=false]{subfig}

\theoremstyle{plain}
\newtheorem{theorem}{Theorem}

\newtheorem{corollary}[theorem]{Corollary}

\theoremstyle{remark}
\newtheorem{remark}{Remark}

\newcommand{\fpr}[1]{\left(#1\right)}
\newcommand{\tpr}[1]{\left[#1\right]}
\newcommand{\spr}[1]{\left\{#1\right\}}

\newcommand{\norm}[1]{\left\lVert#1\right\rVert}
\newcommand{\V}[1]{{\bm{\mathbf{\MakeLowercase{#1}}}}} 
\newcommand{\M}[1]{{\bm{\mathbf{\MakeUppercase{#1}}}}} 
\newcommand{\tr}{\operatorname{tr}} 
\newcommand{\abs}[1]{\left\lvert#1\right\rvert}

\def\hGam{h_{\gamma}}

\def\hxi{h_{\xi}}
\def\heta{h_{\eta}}
\def\ep{\textrm{E}}

\def\bG{\mathbf{G}}

\def\bW{\mathbf{W}}
\def\bH{\mathbf{H}}

\def\bX{\mathbf{X}}
\def\bY{\mathbf{Y}}

\def\bZ{\mathbf{Z}}

\def\bU{\mathbf{U}}

\def\mcS{\mathcal{S}}

\def\bY{\mathbf{Y}}

\def\sprime{s^{\prime}}
\def\rprime{r^{\prime}}
\def\jprime{j^{\prime}}
\def\lprime{l^{\prime}}
\def\vprime{v^{\prime}}
\def\tprime{t^{\prime}}
\def\soprime{s_{1}^{\prime}}

\def\bU{\mathbf{U}}

\def\heta{\widehat{\eta}}

\def\be{\mathbf{e}}
\def\bb{\mathbf{b}}
\def\ba{\mathbf{a}}

\def\bdelta{\boldsymbol{\delta}}

\def\bbeta{\boldsymbol{\beta}}

\def\bSigma{\boldsymbol{\Sigma}}

\def\bhatW{\widehat{\mathbf{W}}}

\def\bhbeta{\widehat{\bbeta}}

\def\bhatY{\widehat{\mathbf{Y}}}

\DeclareMathOperator*{\argmin}{arg\,min}

\def\Wkij{W_{k,ij}}
\def\Ykij{Y_{k,ij}}

\def\mfR{\mathfrak{R}}
\def\mfV{\mathfrak{V}}

\def\mftR{\widetilde{\mathfrak{R}}}

\def\tepsilon{\widetilde{\epsilon}}

\makeatletter
\renewcommand{\algocf@captiontext}[2]{#1\algocf@typo. \AlCapFnt{}#2} 
\def\@algocf@capt@plain{top}
\renewcommand{\algocf@makecaption}[2]{%
  \addtolength{\hsize}{\algomargin}%
  \sbox\@tempboxa{\algocf@captiontext{#1}{#2}}%
  \ifdim\wd\@tempboxa >\hsize
    \hskip .5\algomargin%
    \parbox[t]{\hsize}{\algocf@captiontext{#1}{#2}}
  \else%
    \global\@minipagefalse%
    \hbox to\hsize{\box\@tempboxa}
  \fi%
  \addtolength{\hsize}{-\algomargin}%
}
\makeatother

\usepackage[figuresright]{rotating}

\begin{document}

\title{PROFIT: Projection-based test in longitudinal functional data}

 \author
 {Salil Koner\thanks{ 
 Department of Biostatistics and Bioinformatics, Duke University, Durham NC USA, Email: salil.koner@duke.edu.} \;
 So Young Park\thanks{
 Affiliation: Elli Lilly and Company,  OH USA, Email: parksoyoung3859@gmail.com}\;
 Ana-Maria Staicu\thanks{
 Affiliation: Department of Statistics, North Carolina State University, Raleigh NC, USA, Email: astaicu@ncsu.edu}}
 
 \maketitle







\begin{abstract}
In many modern applications, a dependent functional response is observed for each subject over repeated time, leading to longitudinal functional data. In this paper, we propose a novel statistical procedure to test whether the mean function varies over time. Our approach relies on reducing the dimension of the response using data-driven orthogonal projections, and employs likelihood-based hypothesis testing. We investigate the methodology theoretically and discuss a computationally efficient implementation. The proposed test maintains the Type-1 error rate, and shows excellent power to detect departures from the null hypothesis in finite sample simulation studies. We apply our method to the longitudinal diffusion tensor imaging study of multiple sclerosis (MS) patients to formally assess whether the brain's healthy tissue, as summarized by the fractional anisotropy (FA) profile, degrades over time during the study period. 
\vspace{20 pt}
\end{abstract}

Longitudinal functional data analysis; Functional principal component analysis; Uniform convergence; Likelihood ratio test; Fractional Anisotropy; Multiple sclerosis.

\section{Introduction}
\label{sec:intro}
Modern longitudinal studies from a variety of fields include curves, images, or more generally, object-like variables measured repeatedly over time, for each individual of many. For example in circadian rhythm studies, the daily physical activity is recorded every minute of the day, for multiple days (not necessarily every day), for hundreds or thousands of adults \cite{xiao2015quantifying}; in longitudinal neuroimaging clinical studies often one-dimensional profiles summarizing the brain tissue health status at a hospital visit are collected on patients observed at multiple visits \citep{yuan2014fmem}. In our diffusion tensor imaging (DTI) study of multiple sclerosis (MS), brain images are taken at each hospital visit of a patient, for patients viewed at many visits during the study duration. An important scientific question in this framework is to formally assess whether the mean curve or image remains constant over time. For the motivating DTI study, the brain's health tissue, as imaged by DTI, is summarized by fractional anisotropy (FA) along corpus callosum (CCA) - the largest fiber bundle in the central nervous system, which is known to be affected by MS. Investigating whether the mean FA along the CCA changes over the studied time period has the potential to shed light about the MS progression during this time frame.

Longitudinal functional data (LFD) is a second-generation functional data structure where a response profile (function) is observed over a compact interval, repeatedly, at many times for each subject, inducing dependence among the functions corresponding to the same subject. Owing to this hierarchical structure, there has been an increasing interest in modelling and analysis of multilevel functional data, mostly focused on multi-way functional principal component analysis (FPCA); see \citet{scheffler2020hybrid}, \citet{lynch2018test}, \citet{chen2017modelling}, \citet{hasenstab2017multi}, \citet{park2015longitudinal}, \cite{greven2011longitudinal}, \cite{morris2006wavelet} to name a few. In the context of DTI study, assume that we observe $n$ iid copies of $ \{t_{j}, Y_{j}(\cdot) \}_{j = 1}^{ M }$, where $M$ is the number of repeated visits,  $t_{1}<\ldots < t_M$ are the times of the visits in the set $\mathcal{T}$, and for each $j$, $Y_j(\cdot) $ is one-dimensional `trajectory' corresponding to time $t_{j}$ and observed at fine grid of points $s_{j1}< \ldots< s_{jR_j}$ in the set $\mathcal{S}$. We assume that $Y_{j}(s)$, $s\in\{s_{j1}, \ldots, s_{jR_j}\}$ are observations of a smooth square integrable stochastic process $X_j(\cdot) \in \mathcal{L}^2(\mathcal{S})$ measured at discrete time points $s_{j1}, \ldots, s_{jR_j}$, possibly contaminated with measurement error, and furthermore $X_j(\cdot)$ corresponds to time $t_{j}$. Denote by $\mu(\cdot, t_j) = E[X_j(\cdot)]$, the mean of the latent profile at time $t_j$, for $t_j\in\mathcal{T}$.  

\begin{figure}
    \centering
    \subfloat{\includegraphics[scale=0.23]{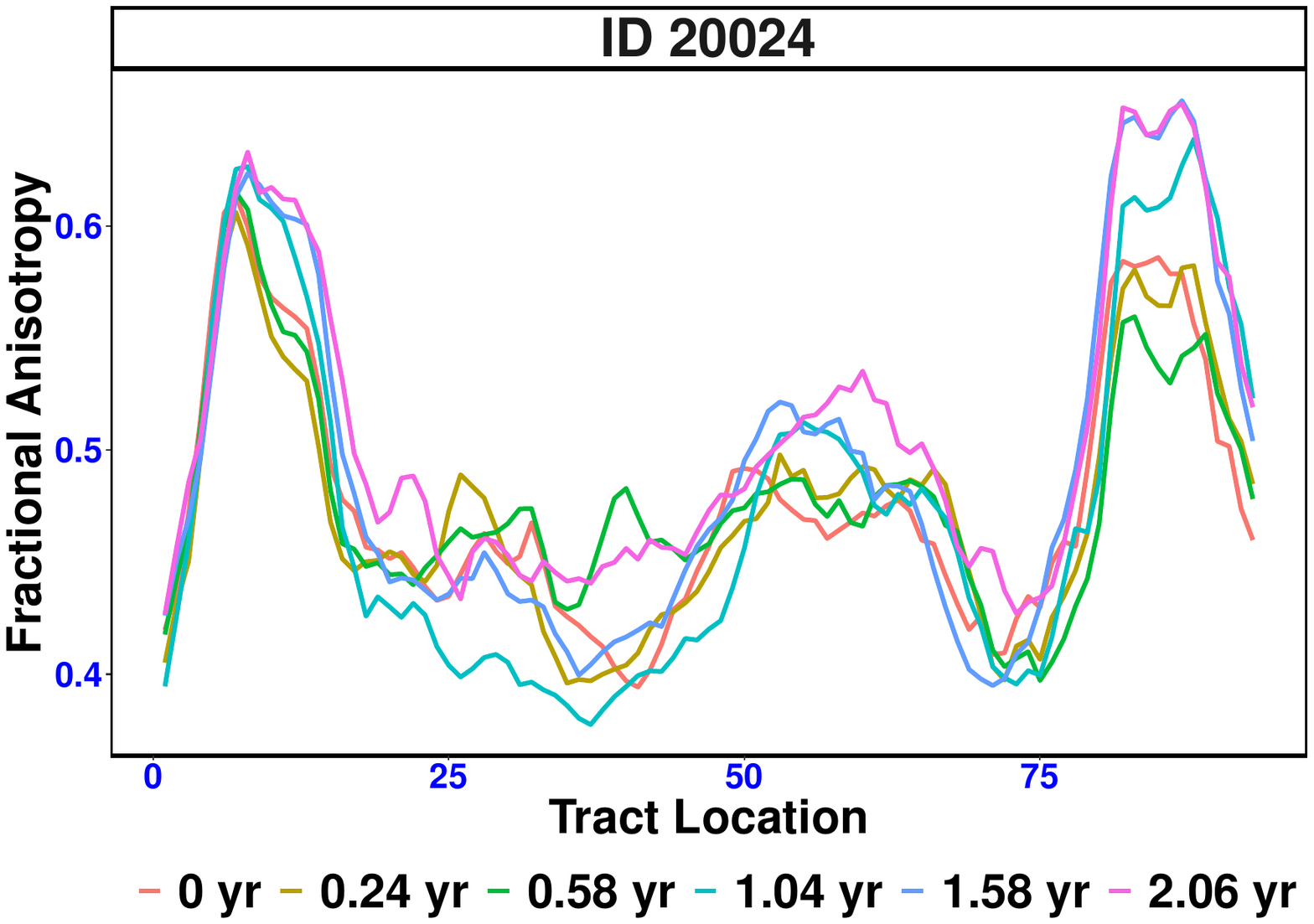}}\qquad
        \subfloat{\includegraphics[scale=0.23]{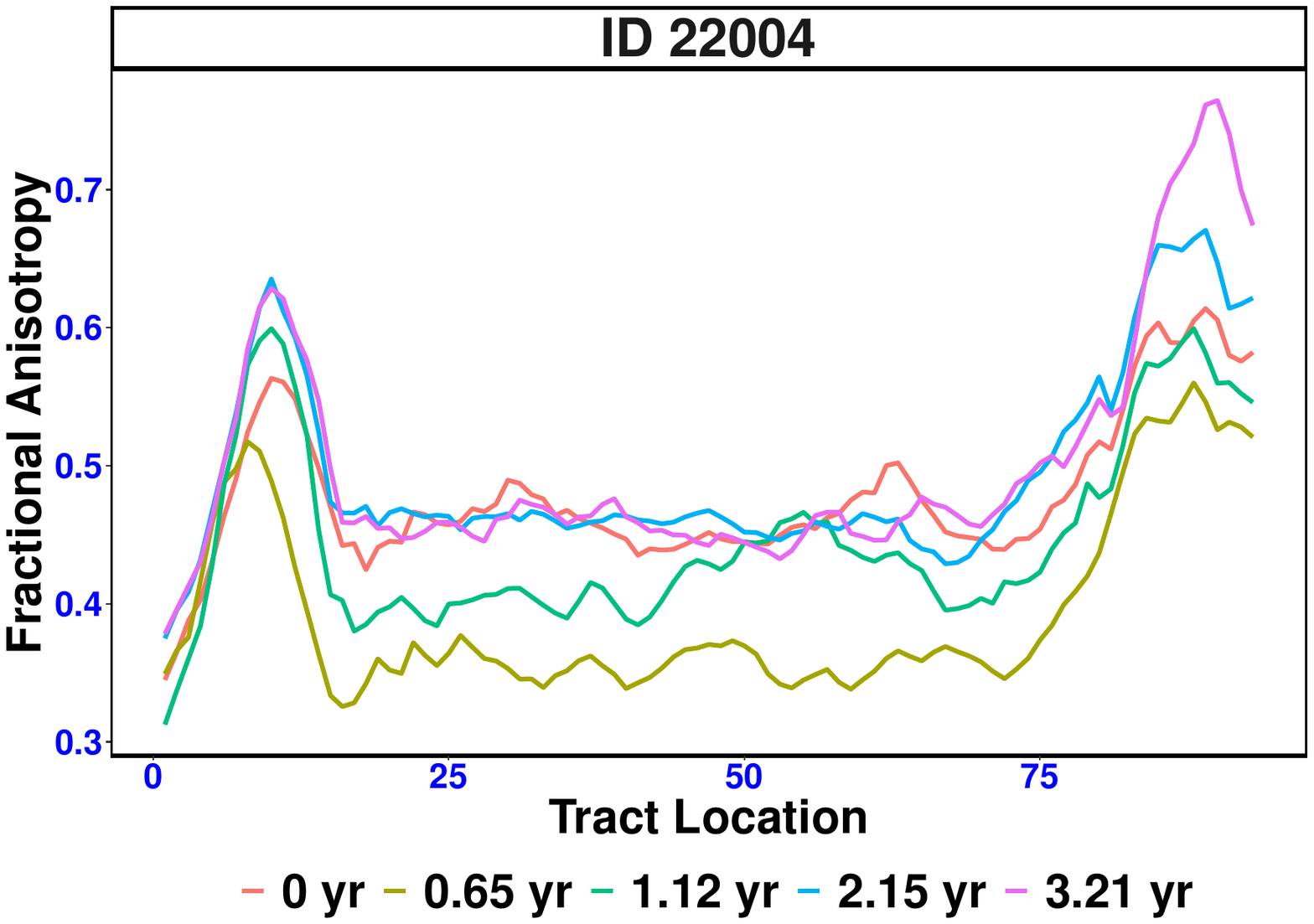}}
    \caption{Fractional anisotropy (FA) profile along the corpus callosum (CCA) of two MS patients at the different hospital visits over the duration of the study. The patients are observed at six visits (left panel) and five visits (right panel) respectively. The visit times (in years) are marked at the bottom for each patient with $0$ indicating baseline visit.}
    \label{fig:sampledsubj}
\end{figure}

Figure~\ref{fig:sampledsubj} shows one dimensional FA `trajectories' (profiles) for two randomly selected MS patients from the DTI study. We are interested to formally investigate whether the mean response changes over time. In the case of the DTI study, any change in the mean profile of the white matter tract during such a short time is equivalent to disease progression. We want to test the null hypothesis that the mean function $\mu(\cdot,t)$ is  time invariant, or
\begin{eqnarray}
H_0: \mu(\cdot, t) = \mu(\cdot,t'), \qquad \text{ for all }t\neq t'\in\mathcal{T}
\label{eqn:H0intro}
\end{eqnarray}  
versus the alternative
\begin{eqnarray*}
H_1: \mu(\cdot, t) \neq \mu(\cdot, t'), \qquad \text{ for some }t\neq t'\in\mathcal{T}.
\end{eqnarray*}
Hypothesis testing of the above kind has been discussed in non-parametric regression in the context of `test for omitted variables'; see \citet{chen1999consistent} \citet{fan2000consistent}, \citet{hall2007nonparametric}, \citet{delgado2014does} and the reference therein. Most of the methodologies inherently assume that the observations are independent and thus cannot be imported to our situation, as they fail to account for the existing dependence in the data; not accounting for data dependence yields to an inflated size, and hence is an invalid testing procedure. 

Over the past twenty years, there has been an explosion of research on modeling and inference for independent functional data. For correlated functional data, the interest has been in modeling the complex dependence of the data and the associations with covariates \citep{baladandayuthapani2008bayesian, di2009multilevel, scheipl2015functional, staicu2020longitudinal}. 
Testing procedures for independent functional data focus on assessing the significance of a smooth functional effect \citep{shen2004f, zhang2011statistical}, equality of the mean function for two groups \citep{cuesta2010simple, horvath2015introduction, zhang2019new}, equality of covariance functions  \citep{paparoditis2016bootstrap, guo2019new},  or equality in distribution \citep{pomann2016two, wynne2020kernel}. These tests are either constructed by taking point-wise supremum of the classical ANOVA based F-test or by considering an $L^2$-norm based Global Pointwise F (GPF) statistic that has a mixture of chi-square null distribution. For correlated functional data, to the best of authors knowledge, testing for the time-varying mean function has been only considered by \citet{park2018simple}, where the authors proposed an $L^2$-norm based distance between the estimated mean response under the null and the alternative hypotheses, and used bootstrap to approximate the test's null distribution. Therefore, the testing procedure is computationally intensive, making it unfeasible to study its size and power properties in large sample sizes. Developing a testing procedure for formally assessing the time behavior of the mean function in this complex dependence setting that is computationally feasible and has good testing properties is an important and pressing gap in the literature.

In this paper we propose a novel statistical procedure for testing whether the mean function varies over time in a longitudinally recorded functional data. The methodology accounts for the data dependence both over the time of the repeated visits, $t$, and the grid points at which the data profiles are observed, $s$. The idea is to represent the bivariate mean function $\mu(s, t)$ using a set of orthogonal basis functions along $s$ and coefficient functions that vary with $t$. Due to the basis functions' orthogonality, when the data profiles are projected along each direction in part, the $t$-varying coefficients corresponding to $\mu(s,t)$ are equal to the mean of the data-projections. This allows us to approximate the original null hypothesis by a set of simpler null hypothesis about the $t$-varying coefficients and leverage the existing inferential methods available in longitudinal data
literature to test each of these simpler hypotheses. 

We propose to use data-driven orthogonal basis functions obtained from the spectral decomposition of the so called `marginal covariance' function. Constructing the marginal covariance for longitudinal functional data \citep{chen2012modeling} has been discussed previously by \cite{park2015longitudinal} and \cite{chen2017modelling}. As a primary contribution of the paper, we develop a projection-based test for the time invariance of the mean function and derive its asymptotic distribution under the null hypothesis as well as under local alternative hypotheses. The asymptotic null distribution has a form similar to the restricted likelihood ratio statistic established by \cite{crainiceanu2004likelihood}. The proposed test maintains the Type-1 error rate, has excellent power in finite samples, and is computationally efficient. 

One of the crucial steps in deriving the null distribution of the test is establishing an uniform convergence rate of the estimated eigenfunctions of the marginal covariance. It is well-known that this rate directly depends on the smoothing method applied for the estimation of the covariance function. We derive an asymptotic uniform convergence rate of the marginal covariance function and of the resulting eigenfunctions for longitudinal functional data, under a general weighting scheme and sampling plan. \cite{chen2017modelling} derived a Hilbert-Schmidt norm convergence of the covariance function when the response data is densely observed in both arguments. When one sampling scheme is sparse - as in the framework studied here - we observe that a parametric uniform rate for the marginal covariance estimation cannot be achieved. In this situation, the convergence rate of the marginal covariance function is dominated by the rate of the mean function in longitudinally observed functional data.

The article is organized as follows. In Section~\ref{sec:big picture} we formulate the problem mathematically and introduce the model framework. The testing procedure along with the theoretical study of the asymptotic distributions under the null and local alternatives are described in Section~\ref{sec: methodology}. Simulation studies are presented in Section~\ref{sec: SimStudy} to demonstrate the finite sample performance of the test. Section~\ref{sec: DTI} contains the application of our test on DTI study. Assumptions related to the main theorems of the paper are in Appendix. Detailed proofs of the theorems as well as additional simulation results are provided in the Supplementary material. \textcolor{black}{R code to compute the power function of the test is publicly available at \url{https://github.com/SalilKoner/PROFIT}}. 

\section{Problem formulation and model framework} \label{sec:big picture}

\subsection{Alternate formulation of the original hypothesis}\label{sec: alternateform}

{\color{black} Assume the mean function $\mu(s,t)$ has mixed smoothness as follows:  $\mu(s, \cdot) \in C^{1}(\mathcal{T})$ for all $s \in \mathcal{S}$ and $\mu(\cdot, t)\in L^2[\mathcal{S}]$ for all $t\in\mathcal{T}$. Here, we use $L^2(\mathcal{S})$ to denote the usual space of square-integrable functions over the bounded and closed domain $\mathcal{S}$, on which we define the usual inner product $<f,g>=\int_{\mathcal{S}} f(s)g(s) ds$. Also $C^{1}(\mathcal{T})$ stands for the space of functions with continuous first derivatives over the bounded and closed domain $\mathcal{T}$. One can show that for an appropriately chosen orthogonal basis $\spr{\phi_k(s) : s \in \mathcal{S}}_{k \geq 1}$ where $\int_\mathcal{S} \phi_k(s) \phi_{k'}(s)ds=1(k=k')$ for $k, k'\geq 1$, the mean function $\mu(s,t)$ 
can be represented uniquely as $ \mu(s,t) = \sum_{k=1}^{\infty} \eta_k(t)\phi_k(s)$ for all $s \in \mathcal{S} $, where $\eta_k(t) = \int_\mathcal{S} \mu(s,t) \phi_k(s)ds$ is the coefficient function corresponding to $\phi_k(s)$ for $k\geq 1$.} Hereafter we use $\int f(s)ds$ to represent integration over $\mathcal{S}$ of some function $f$. It follows that the function $\mu(s,t)$ satisfies the null hypothesis (\ref{eqn:H0intro}) if and only if the corresponding coefficient functions with respect to the basis $\spr{\phi_k(s) : s \in \mathcal{S}}_{k \geq 1}$, are constant. In other words, the null hypothesis (\ref{eqn:H0intro}) can be equivalently written as
\begin{equation}\label{eqn:alternateH0}
   \eta_k(t) = \eta_k(t') \qquad \forall t\neq t', \text{ with } t,t'\in \mathcal{T },
\end{equation}
and for all $k\geq 1$. By the same logic, a function $\mu(s,t)$ for which $H_1$ is valid has the respective basis coefficients, $\eta_k(t)$, varying with $t$ for some $k \geq 1$. 

This representation allows us to simplify the testing problem: the initial formulation concerns testing that a bivariate function is a univariate function, while the representation (\ref{eqn:alternateH0}) implies testing a series of simpler hypotheses that a univariate function is constant. {\color{black} While the above representation of $\mu(s,t)$ holds for a specific basis system, the null hypothesis (\ref{eqn:H0intro}) implies that for any orthogonal basis system $\{\phi_k(\cdot)\}_k$ in $L^2[\mathcal{S}]$ the equivalent null hypotheses (\ref{eqn:alternateH0}) are true for all $k\geq 1$.} Naturally, there are several challenges with this latter approach. The first issue is the type of orthonormal basis used: pre-specified or data-driven. The second issue is the number of hypotheses to be tested. Regardless of the answer to these questions, once an orthogonal system is available, the simpler hypotheses are to be carried using the `projections' of the original data profiles on the respective directions $\phi_k(s)$'s. We discus this point next and consider the two remaining challenges related to the basis in Section \ref{sec: methodology}.

\subsection{Model framework}\label{subsec: datagenmodel}

Let the $i$th datum be $\left[t_{ij}, Y_{ij}(s): s\in \{ s_{ij1}, \ldots, s_{ijR_ {ij} } \} \right]_{j = 1}^{ m_i }$, for $i=1, \ldots, n$ where
$Y_{ij}(\cdot)$ is a one-dimensional curve observed at the $j$th time visit, $t_{ij}$, for $j=1, \ldots, m_i$.
In real world applications, the curves \(Y_{ij}(\cdot)\) are observed on a fine grid \(\{s_{ij1}, \ldots, s_{ijR_{ij}}\}\) for large $R_{ij}$; equivalently, for every $i,j$, the  set \(\{s_{ijr}: r = 1, \ldots, R_{ij}\}\) is dense in \(\mathcal{S}\). It is assumed that $m_i$ is small for every $i$, but $\{t_{ij}: i=1, \ldots, n; j=1, \ldots ,m_i\}$ is a dense set in \(\mathcal{T}\). Here \(\mathcal{S}\) and \(\mathcal{T}\) are closed and bounded intervals. Without loss of generality assume that \(s_{ijr} = s_{r}\) for all $r$ forms an equally spaced grid in \(\mathcal{S}\) and use the index \(s\) instead of \(s_{r}\). To distinguish between the two types of sampling designs, we refer to the functional design to be dense and call `\(s\)' the functional argument, and to the longitudinal design to be sparse and call `\(t\)' the longitudinal argument. 
We denote $\mu(s, t_{ij})=E[Y_{ij}(s)]$, where $\mu(\cdot, \cdot)$ is an unknown smooth intercept function defined on \(\mathcal{S} \times \mathcal{T}\), and write
\begin{equation}\label{eqn: assumedmodel}
    Y_{ij}(s) = \mu(s, t_{ij}) + \epsilon_{i}(s, t_{ij}),
\end{equation}
where \(\epsilon_i(\cdot,t_{ij})\) is a zero-mean random deviation, with $\varepsilon_i$ independent and identically distributed (iid) over $i$. The residual process is intended to capture the variability of the data over the functional argument $s$, over the repeated time $t_{ij}$, as well as due to measurement error. The proposed testing procedure makes weak assumptions in this regard. 

Let $\phi_k(s)$'s be continuous orthogonal functions, as described in Section \ref{sec: alternateform}, and denote by $Y_{k, ij} := \int Y_{ij}(s) \phi_k(s) ds$ the projected profile response onto this direction; in practice such projections can be calculated with high accuracy using numerical integration, since the profiles are observed over fine grids. The model (\ref{eqn: assumedmodel}) implies the decomposition of $Y_{k, ij}$ 
\begin{equation}\label{eqn: derivemodel}
    Y_{k,ij} = \eta_k(t_{ij}) +{\epsilon}_{ik}(t_{ij}),
\end{equation}
where $\eta_k(t)  =\int \mu(s,t) \phi_k(s) ds$, with $\eta_k(t) =E[Y_{k, ij}]$ and $\epsilon_{ik}(t_{ij}) = \int\epsilon_{i}(s, t_{ij})\phi_k(s) ds$ is zero-mean residual that is dependent over $j$.

\subsection{Testing framework under the projected model}\label{subsec: testingprocedure}

For each basis function $\phi_k(s)$, we approximate the corresponding projections of the observed profiles $Y_{ij}(\cdot)$ onto this directions using Riemann approximation as
$    Y_{k,ij} = \frac{1}{R}\sum_{r=1}^R Y_{ij}(s_r)\phi_k(s_r).$
Testing the null hypothesis (\ref{eqn:H0intro}) is equivalent to testing 
\begin{equation}\label{eqn: H04}
    H_{0k}^\prime : \eta_k(t) = C_k, \quad \text{for some constant }C_k \quad \textrm{vs} \quad H_{1k}^\prime: {\color{black}\eta_k(t) \neq \eta_k(t')}\quad \text{ for some }t\neq t', 
\end{equation}
for all $k\geq 1$. We choose a sufficiently large truncation $K$ and reduce  (\ref{eqn:H0intro}) to multiple testing of $K$ null hypotheses; to control the familywise error rate of the global testing procedure we adjust the significance level of each test using a Bonferroni correction. The test of each simplified hypothesis is carried in the implied model (\ref{eqn: derivemodel}) for $Y_{k,ij}$. Testing such null hypotheses in the context of dependent data has previously been studied in the literature; we consider the likelihood ratio test whose null distribution was first studied by \cite{crainiceanu2004likelihood} for independent data and later extended to dependent data \citep{wiencierz2011restricted, staicu2014likelihood}.  

To begin, we write the mean function $\eta_k(t)$ using the mixed effects model representation \cite[][Chapter 3]{ruppert2003semiparametric}. {\color{black} Specifically, we represent the unknown function $\eta_k(t)$ via the truncated linear basis and knots $\kappa_1, \dots, \kappa_Q$, and write 
$    \eta_k(t) = \beta_{0k} + \beta_{1k}t + \sum_{q=1}^Q b_{kq}(t-\kappa_q)_{+}$ for $k=1,\dots,K$, where the notation $x_{+} := x\mathbb{I}(x > 0)$ denotes the positive part of $x$. Representation of $\eta_k(t)$ using higher order of truncated power basis is possible, but that will restrict $\eta_k(t)$ to the space of functions with continuous higher order derivatives, and will unnecessarily increase the complexity of the model. Since $\eta_k(t) \in C^1(\mathcal{T})$, it can be approximated accurately by truncated linear basis, provided $Q$ is large enough to capture the complexity of the function \citep{xiao2019asymptotic}. Denote by $\V{\beta}_k = \fpr{\beta_{0k}, \beta_{1k}}^\top $ the vector of the coefficients corresponding to the polynomial basis and by $\V{b}_k = \fpr{b_{k1}, \dots, b_{kQ}}^\top$ the vector of spline coefficients. As it is common in the literature we treat $\V{\beta}_k$ as fixed but unknown parameters and the coefficients $\V{b}_k$ as random. Using the mixed model representation we can write $\eta_{k}(t_{ij}) = \bX^{\top}_{ij}\bbeta_{k} + \bZ^{\top}_{ij} \bb_{k}$, where $\bX^{\top}_{ij} = (1, \,t_{ij})$, $\bZ^{\top}_{ij} = ( (t_{ij} - \kappa_{1})_{+}, \ldots, (t_{ij} - \kappa_{Q})_{+})$ and  $b_{kq}$'s, are assumed to be iid with mean zero and variance $\sigma^2_{b,k}$ for $k=1, \ldots, K$.

Let $\M{X}_{i}$ be the ${m_i \times 2}$ fixed design matrix obtained by row-stacking $\bX^{\top}_{ij}$ and similarly let $\M{Z}_i$ be the ${m_i \times Q}$ random design matrix obtained by row-stacking $\M{Z}^{\top}_{ij}$. Denote by $N= \sum_{i=1}^n m_i$ the total number of time points for all the subjects, by $\M{X}= \tpr{\M{X}_1^\top, \dots, \M{X}_n^\top}^\top$ the $N \times 2$ matrix of $\M{X}_{i}$'s, and by $\M{Z} = \tpr{\M{Z}_1^\top, \dots, \M{Z}_n^\top}^\top$ the $N \times Q$ matrix of $\M{Z}_i$'s. Furthermore, let $\bY_k = \tpr{\M{Y}_{k,1}^\top, \dots, \M{Y}_{k,n}^\top}^\top$ with $\M{Y}_{k,i} = \fpr{Y_{k,i1}, \dots, Y_{k,im_i}}^\top$ be the vector of the projected responses for $i$th subject, $Y_{ij}(\cdot)$ for $j=1, \ldots, m_i$, onto the direction $\phi_k(\cdot)$. The projected model (\ref{eqn: derivemodel}) can be written in a mixed model representation as}
\begin{equation}\label{eqn: mixedmodel}
    \M{Y}_{k} = \M{X}\V{\beta}_k + \M{Z}\V{b}_k + \V{E}_{k},
\end{equation}
where the residual vector $\V{e}_k = \tpr{\V{e}_{1k}^\top, \dots, \V{e}_{nk}^\top}^\top$ is constructed in the same way, with $\V{e}_{ik} = \fpr{e_{i1k}, \dots, e_{im_ik}}^\top$ and $e_{ijk} = \epsilon_{ik}(t_{ij})$ as in model~(\ref{eqn: derivemodel}). In this model framework, the null hypothesis for testing (\ref{eqn: H04}) can be alternatively represented as
{\color{black}\begin{equation*}
    H_{0k} : \beta_{1k}=  0, \; \sigma^2_{b,k} = 0  \qquad \text{versus}\qquad  H_{1k}: \beta_{1k} \neq 0,  \text{ or } \sigma^2_{b,k} \neq 0.
\end{equation*}}
Hypothesis testing problem of this kind has been already investigated in the literature in the context of longitudinal and functional data under simple and complex correlation structure. For testing one variance component in linear mixed model with trivial residual dependence \cite{crainiceanu2004likelihood} derived both the likelihood ratio test (LRT) and restricted likelihood based test (RLRT) statistic along with their asymptotic and finite sample null distributions; see also \cite{crainiceanu2005exact}, \cite{greven2013likelihood}. 

The correlation structure of the error component $\V{e}_{ik}$ in the implied linear mixed model (\ref{eqn: mixedmodel}) is non-trivial. \cite{wiencierz2011restricted} developed both the LRT and RLRT testing procedure when the error dependence structure is known up to some parameters. We consider \cite{staicu2014likelihood} who assume a completely unknown dependence structure and use methods from functional data analysis to accommodate such dependence in a LRT-based framework. Specifically, let $\bSigma_{Y,ik}$ be the $m_i \times m_i$ covariance matrix of the error $\V{e}_{ik}$, with the $(j, j')$th element equal to $\textrm{Cov}(e_{ijk}, e_{ij^\prime k}) = \gamma_k(t_{ij}, t_{ij^\prime}) + \sigma_{e,k}^2$, where $\gamma_k(\cdot, \cdot)$ is an unknown continuous covariance function and $\sigma_{e,k}^2> 0$. Due to the independence of the subjects subject, $\textrm{Cov}(\V{e}_k) = \bSigma_{Y,k} = \textrm{diag}(\bSigma_{Y,k1}, \dots, \bSigma_{Y,kn})$. Using a Gaussian working assumption for $\V{e}_k$ and $\V{b}_k$, twice of the log-likelihood function in given by
$ 2\log L(\V{\beta}_k, \bSigma_k, \sigma^2_{b,k}) = -\log( |\bSigma_{Y,k} + \sigma^2_{b,k}\bZ\bZ^{\top}|  ) - ( \bY_{k} - \bX \bbeta_{k} )^{\top}(\bSigma_{Y,k} + \sigma^2_{b,k}\bZ\bZ^{\top})^{-1} (\bY_{k} - \bX \bbeta_{k} )$.

The covariance matrix of the residual $\bSigma_{Y, k}$ is typically unknown and \cite{staicu2014likelihood} extended the LRT by replacing the true covariance with an appropriate estimator. We estimate $\bSigma_{Y, k}$ by using bi-variate smoothing technique, e.g. kernel smoothing \citep{yao2005functional} or penalized splines smoothing \citep{xiao2013fast}. Let $\widehat{\bSigma}_{Y,k}$ be an estimator of the covariance matrix of $Y_{k,ij}$'s. We pre-whiten the response $Y_{k,ij}$ by the inverse square root of $\widehat{\bSigma}_{Y,k}$; repeat the procedure for the design matrices $\M{X}$ and $\M{Z}$ and let
$
\M{Y}^*_k = \widehat{\bSigma}_{Y,k}^{-1/2}\bY_k$, $\bX^*_k = \widehat{\bSigma}_{Y,k}^{-1/2}\bX$, and $\bZ^*_k = \widehat{\bSigma}_{Y,k}^{-1/2}\bZ$.
The pseudo-likelihood function is equal to
$$
2\log L_{\bY^*_k}(\V{\beta}_k, \sigma^2_{b,k}) = -\log( |\M{I}_N + \sigma^2_{b,k}\bZ^*_k\bZ^{*^{\top}}_k|  ) - ( \bY^*_{k} - \bX^*_k \bbeta_{k} )^{\top}(\M{I}_N + \sigma^2_{b,k}\bZ^*_k\bZ^{*^{\top}}_k)^{-1} (\bY^*_{k} - \bX^*_k \bbeta_{k} ).
$$
To test $H_{0k}$, we use the pseudo likelihood ratio test (pLRT) discussed in \cite{staicu2014likelihood}. In the next section, we propose a criterion to select the orthogonal basis that allows us to study the null distribution of the resulting pLRT under the multiple testing setting.

\section{Projection-based Functional Invariance Test (PROFIT)} \label{sec: methodology}

\subsection{Choice of the orthogonal basis}\label{subsec: basisfunc}

The orthogonal basis can be selected using a preset of basis functions, such as Fourier basis, or orthogonal B-spline basis, or wavelets, and so on, or data-driven basis. Using a preset basis makes the selection of the truncation $K$ very difficult. The level $K$ needs to be sufficiently large to capture the complexity of the bivariate mean $\mu(s,t)$ and in the absence of any information about the data, a preset basis would typically require a very large $K$. To bypass this issue, we choose the latter and consider the eigenbasis of an appropriate covariance. 

Assume that the residual process $\epsilon_i(\cdot, \cdot)$ in (\ref{eqn: assumedmodel}) has a covariance function represented as
\begin{equation}\label{eqn:covstructure}
    \textrm{Cov}\{ \epsilon_i(s,t), \epsilon_i(s',t') \}=c(s,s',t,t')+\Gamma(s,s') 1(t=t') + \sigma_e^2 1(s=s') 1(t=t'),
\end{equation}
where $c$ is a continuous covariance function over $\mathcal{S}^2 \times \mathcal{T}^2$,  $\Gamma$ is a continuous covariance function over $\mathcal{S}^2$ and $\sigma^2_e>0$. For example, a residual of the form described in \cite{park2015longitudinal} would have the covariance (\ref{eqn:covstructure}): $\epsilon_i(s,t_{ij}) = U_i(s, t_{ij}) + \epsilon^{sm}_{ij}(s) +\epsilon^{wn}_{ij}(s)$, where $U_i$ is iid continuous bivariate process with covariance function $c(\cdot, \cdot, \cdot, \cdot)$, $\epsilon^{sm}_{ij}$ is iid smooth process with covariance $\Gamma(\cdot, \cdot)$ and  $\epsilon^{wn}_{ij}$ is iid white noise process with covariance $\textrm{cov} \{\epsilon^{wn}_{ij}(s), \epsilon^{wn}_{ij}(s')\} = \sigma_e^2  1(s=s')$. 

We use this covariance to obtain the orthogonal basis used in the proposed testing procedure. \textcolor{black}{ Under the assumption that the residual trajectories and the $\mu(\cdot, t)$ are in the same functional space for all $t\in\mathcal{T}$, the eigenbasis of this marginal covariance is the required basis system for which the representation of $\mu(s,t)$ described in Section \ref{sec: alternateform} is valid.} Define by $f(t)$ the marginal density of the time points $t_{ij}$'s and consider the so-called `marginal covariance' \citep{park2015longitudinal} defined by
\begin{equation*}
    \Xi(s,\sprime) = \int_{\mathcal{T}} c(s,\sprime,t, t)f(t)dt + \Gamma(s, \sprime);
\end{equation*}
this quantity is a proper covariance function. Furthermore, $\Xi(s,\sprime) $ is a continuous covariance and thus, using Mercer's theorem, it has a spectral decomposition in terms of pairs of eigenvalues and eigenfunctions: $\Xi(s,\sprime) = \sum_{k=1}^\infty \lambda_k\phi_k(s)\phi_k(\sprime)$, 
where $\lambda_1\geq \lambda_2 \geq \ldots \geq 0$ and $\int\phi_k(s) \phi_{k'}(s)ds =1(k=k')$. Let $\{\phi_k(\cdot)\}_k$ be the orthogonal eigenbasis and define the finite truncation $K$ using the percentage of variation (PVE). By specifying a threshold for the PVE, the truncation $K$ is chosen such that $\sum_{k=1}^K\lambda_k/\tr(\Xi)\geq$ PVE but  $\sum_{k=1}^{K-1}\lambda_k/\tr(\Xi)< $PVE  where $\tr(\Xi)=\int  \Xi(s,s) ds$ is the trace of the covariance kernel $\Xi$. 

We discuss next the implications of using a data-driven basis within the testing procedure framework: we first present the estimation of the orthogonal basis, discuss the theoretical properties of the basis functions' estimator, and develop the asymptotic null distribution of the testing procedure based on projections of the data profiles onto the data-driven basis functions. Ultimately, we introduce the overall testing procedure for testing that the bivariate mean is time invariant, which we call `PROjection-based Functional Invariance Testing procedure', in short PROFIT, and justify its validity.

\subsection{Data-driven orthogonal basis}\label{subsec: challengeseigenfun}

Using data-driven basis functions creates challenges, primarily because the original data profile is projected on directions that are estimated based on the whole data set. As a result, the independence of the projections, which holds when pre-specified directions are used, is no longer valid. Nonetheless, we describe estimation of the marginal covariance (see also \cite{park2015longitudinal}, \cite{chen2017modelling}) and discuss the theoretical properties of the resulted estimator. Without loss of generality we assume that $\mathcal{T} =[0,1]$.  

We first obtain an estimator for the mean function $\mu(s,t)$ by pooling all data and performing a bivariate smoothing. Nonparametric smoothing can be done either by using local smoothing approaches such as kernel smoothing \citep{chen2017modelling}, or global smoothing methods \citep{wood2017generalized, xiao2013fast}. Denote the smooth bivariate mean estimator of $\mu(s,t)$ by $\widehat{\mu}(s,t)$. In the second step, we work with the de-meaned response, $\widetilde{Y}_{ij}(s_r) = Y_{ij}(s_r) - \widehat{\mu}(s_r, t_{ij})$, which we use to define the `raw marginal covariance'  $\widetilde{\Xi}(s_r, s_{\rprime})$,
\begin{equation}\label{eqn: rawcov}
    \widetilde{\Xi}(s_r, s_{\rprime}) = 
    \sum_{i=1}^n v_i \sum_{j=1}^{m_i}  \widetilde{Y}_{ij}(s_r) \widetilde{Y}_{ij}(s_{r'}),
\end{equation}
where the weights $v_i$ may vary across subjects, such that $\sum_{i=1}^n m_iv_i = 1$. The marginal covariance estimator, $\widehat{\Xi}(s,\sprime)$, is obtained by smoothing the marginal raw covariance by using bivariate kernel smoothing \citep{yao2005functional}. Let $K(\cdot)$ be a symmetric kernel in $[-1,1]$ and $\hxi > 0$ be a bandwidth parameter;  
the marginal covariance estimator is $\widehat{\Xi}(s, \sprime) = \widehat{a}_0$,
\begin{align} \label{eqn:bivkernel_covestimation}
     \fpr{\widehat{a}_0, \widehat{a}_1, \widehat{a}_2} &= \argmin_{a_{0}, a_{1}, a_{2}}\; 
     \sum_{1 \leq r \neq \rprime \leq R}   \left[\widetilde{\Xi}(s_r, s_{\rprime}) - a_0 - a_1 (s_r - s) - a_2 (s_{\rprime} - s^\prime)\right]^2  \nonumber \\
     & \qquad \qquad \qquad \qquad \qquad \times K\left(\frac{s_r-s}{{\hxi}} \right)K\left(\frac{s_{\rprime}-\sprime}{{\hxi}}\right).
\end{align}
\textcolor{black}{The bandwidth parameter is  selected by leave-one-subject-out cross validation (CV) or generalized CV (GCV) \citep{hall2006properties, yao2005functional}}. Finally, the data-driven basis, $\{ \widehat{\phi}_k(s)\}_k$, is obtained as the eignbasis of the estimated covariance $\widehat{\Xi}(\cdot,\cdot)$ and the truncation $K$ is chosen as discussed in the previous section.

For each $k\geq 1$, define the projections of the data in the direction $\widehat{\phi}_k(s)$ as $W_{k,ij} = R^{-1}\sum_{r=1}^R Y_{ij}(s_r)\widehat{\phi}_k(s_r)$; we refer to $W_{k,ij}$ by `quasi-projections', because they are transformations of the data using data-driven directions. The pseudo-likelihood function of Section \ref{subsec: testingprocedure}, obtained by substituting $\Ykij$ with the quasi projections $W_{k,ij}$, leads to a pseudo-likelihood ratio testing procedure; deriving the asymptotic null distribution of the resulting test is significantly more challenging compared to \cite{staicu2014likelihood}. This is because, unlike $\Ykij$, the quasi projections $\Wkij$ are no longer independent across $i$, as the eigen functions $\spr{\widehat{\phi}_k(s) : k=1,\dots, K}$ are estimated from the full data $Y_{ij}(s_r) : i=1,\dots, n\;\;$, $j=1,\dots m_i, r = 1,\dots, R$. Instead, as we show in Theorem~\ref{theorem: unifcov}, the recovery of the data-driven directions $\widehat \phi_k(\cdot)$ can be achieved with high accuracy, implying minimal changes in the asymptotic null distribution of interest; see Theorem~\ref{theorem: pLRTnulldist} in Section~\ref{subsec: profit}. 

The difference between the projected responses and their quasi counterparts equals 
\begin{equation*} 
    \Wkij - \Ykij = R^{-1}\sum_{r=1}^R Y_{ij}(s_r)\spr{\widehat{\phi}_k(s_r)-\phi_k(s_r)} \qquad k=1,2,\dots,K;
\end{equation*}
thus, in order to uniformly bound the difference $\lvert \Ykij - \Wkij \rvert$, we need to establish a uniform rate of convergence for the estimated eigenfunctions. \cite{park2015longitudinal}, who considered a similar residual covariance structure, made a first attempt to study the theoretical properties of the eigenfunction of the marginal covariance estimator. However, they only derived the consistency of the eigenfunctions estimators. Later, \cite{chen2017modelling} showed the Hilbert-Schmidt norm convergence rate of these quantities. No published work has hitherto established the strong uniform convergence rate of the eigenfunctions derived from the marginal covariance estimator. The next theorem discusses the uniform convergence rate of the marginal covariance estimator for longitudinal functional data, when the final marginal covariance is estimated through bivariate linear kernel smoothing. 

\begin{theorem} \label{theorem: unifcov}
Consider data $[ t_{ij}, Y_{ij}(s), s\in\{s_1, \ldots, s_R\}]_{j=1}^{ m_i}$ , $i=1, \ldots, n$ and assume the model (\ref{eqn: assumedmodel}) is correct. Assume that the mean $\mu(s,t)$ can be estimated by $\widehat\mu(s,t)$, where
\begin{equation*}
    \underset{s \in \mathcal{S}, t
\in \mathcal{T}}{\sup} \abs{\widehat{\mu}(s,t)-\mu(s,t)} = \mathcal{O}(\alpha_n) \quad a.s \qquad \text{for some sequence } \alpha_n = o(1). 
\end{equation*}

Let $K(\cdot)$ be a symmetric kernel in $[-1,1]$ and $\hxi > 0$ be a bandwidth parameter and estimate $\Xi(s,\sprime)$ as in (\ref{eqn:bivkernel_covestimation}), where $\widetilde{\Xi}(s_r, s_{\rprime})$ is the raw covariance estimator described in~(\ref{eqn: rawcov}). Suppose that $\sup_i m_i \leq B$ for some $B < \infty$ and the Assumptions \ref{assump: A1}-\ref{assump: A6} and \ref{assump: B1}-\ref{assump: B3} that are listed in the Appendix A 
hold true. Furthermore, if $v_i = 1/(\sum_{i=1}^n m_i)$ or $v_i = 1/(nm_i)$ and either of (i) $R/(n/\log n)^{1/4} \to C$, where $C > 0$ and $\hxi \approx \fpr{\log n/n}^{1/4}$ or (ii) $R/(n/\log n)^{1/4} \to \infty$, $\hxi = o\fpr{\fpr{\log n/n}^{1/4}}$ such that $\hxi R \to \infty$, holds, it follows that
\begin{equation*}
    \underset{s,\sprime \in \mathcal{S}}{\sup} \abs{\widehat{\Xi}(s,\sprime) - \Xi(s,\sprime)} = \mathcal{O}\fpr{\sqrt{\frac{\log(n)}{n}} + \alpha_n} \qquad a.s.
\end{equation*}
\end{theorem}

In Theorem \ref{theorem: unifcov}, the case \textit{(i)} corresponds to dense functional data, while case \textit{(ii)} is for ultra-dense functional data \citep{zhang2016sparse}. Since we assume that the trajectory $Y_{ij}(\cdot)$ is observed at a dense grid $\spr{s_1, \dots, s_r}$, we do not present the rate for the case when the trajectories are observed on a sparse grid regime. However, the rate for the sparse case can be easily derived from the general result proven in Supplementary material. The assumptions of the Theorem \ref{theorem: unifcov} are fairly standard in the literature of nonparameteric statistics; see Appendix A for more discussion on the conditions of the theorem. Bivariate mean estimation at the rate required by the theorem is discussed in \cite{chen2012modeling}, when bivariate kernel smoothing is used to obtain $\widehat \mu(s,t)$. In a hypothetical situation, when the mean function is known, and thus does not require estimation ($\alpha_n=0$), we can obtain an almost parametric rate for the marginal covariance function. This can be attributed to the fact that the estimation of the smoothed marginal covariance does not require smoothing over the sparsely observed longitudinal argument $t$; the smoothing is only done for the functional argument $s$. Thus, marginalizing over the longitudinal argument $t$ does not result in a loss of the convergence rate. In real world applications, the mean function $\mu(s,t)$ is not known and has to be estimated from the data; we need to smooth over both the sparse and the dense arguments to obtain a smoothed estimator $\widehat{\mu}(s,t)$. As a result, the uniform rate for the bivariate mean function, $\alpha_n$, is slower than the parametric rate \citep{chen2012modeling}, which ultimately implies that the uniform rate of the estimator $\widehat{\Xi}(s,\sprime)$ is dominated by that of the mean function estimator. In the next section we formulate the likelihood ratio based statistic to test the global null hypothesis in~(\ref{eqn:H0intro}) by using the quasi projections $\Wkij$.

An important implication of Theorem \ref{theorem: unifcov} is that the eigenfunctions of the marginal covariance $\Xi(s,\sprime)$ are estimated at the same rate by the eigenfunctions of the covariance estimator $\widehat{\Xi}(s,\sprime)$ \citep{li2010uniform}. In particular, $    \underset{s \in \mathcal{S}}{\sup} \abs{\widehat{\phi}_k(s) - \phi_k(s)} =o_p(1)$ for all $k$.

\subsection{Pseudo-likelihood ratio test using the quasi projections}\label{subsec: profit}

For each $k =1,\dots, K$, construct $\bW_k = \tpr{\bW_{k,1}^\top, \dots, \bW_{k,n}^\top }$ where $\bW_{k,i} = \fpr{W_{k, i1}, \dots, W_{k, im_i}}^\top$ is the vector of quasi projections. Based on the accuracy in the estimation of the eigenfunctions $\phi_k(s)$, the quasi projections $W_{k, ij}$ and the actual data projections $Y_{k, ij}$ are very close, and as a result we can model $W_{k, ij}$ using (\ref{eqn: derivemodel}), and furthermore we can assume model (\ref{eqn: mixedmodel}) for $\bW_k$. Let $\widehat{\bSigma}_{W,k}$ be the covariance estimator based on $W_{k,ij}$'s, calculated similarly to $\widehat{\bSigma}_{Y,k}$, by replacing $Y_{k,ij}$'s with  $W_{k,ij}$'s. In the Supplementary material we show that $\widehat{\bSigma}_{W,k}$ is a consistent estimator of ${\bSigma}_{Y,k}$. Compute the corresponding scaled versions of the data and design matrices using the inverse square root of $\widehat{\bSigma}_{W,k}$,
$    \widehat{\bW}_k = \widehat{\bSigma}_{W,k}^{-1/2}\bW_k$,  $\widehat{\bX}_k = \widehat{\bSigma}_{W,k}^{-1/2}\bX$ $\widehat{\bZ}_k = \widehat{\bSigma}_{W,k}^{-1/2}\bZ$, for $k = 1, \dots, K, 
$ and write the pseudo-likelihood function $L_{\widehat{\bW}_k}(\V{\beta}_k, \sigma^2_{b,k})$ as in the Section \ref{subsec: testingprocedure}. 
The pseudo-LRT statistic is constructed as
\begin{equation}\label{eqn: pLRThat}
    \widehat{pLRT}_{N,k} = \sup_{H_{0k} \cup H_{1k}}  2\log L_{\bhatW_{k}}( \bbeta_{k}, \sigma^2_{b,k}) - \sup_{H_{0k}} 2\log L_{\bhatW_{k}}( \bbeta_{k}, \sigma^2_{b,k}) \;\;\;\quad  k=1,2,\dots, K.
\end{equation}
The following theorem states that asymptotically, the null distribution of $\widehat{pLRT}_{N,k}$ remains the same as if the true eigenfunctions $\phi_k(s)$, and thus $Y_{k, ij}$ were used. 

\begin{theorem}\label{theorem: pLRTnulldist}
Consider the data $[ t_{ij}, Y_{ij}(s), s\in\{s_1, \ldots, s_R\}]_{j=1}^{ m_i}$ for $i=1, \ldots, n$ and suppose the model (\ref{eqn: assumedmodel}) is true. Assume that $\widehat{\phi}_k(s)$'s are consistent estimators of the eigenfunctions $\phi_k(s)$ in sup norm, i.e. $\sup_{s \in \mathcal{S}}\;\lvert\widehat{\phi}_k(s)-\phi_k(s)\rvert \to 0$ in probability as $n\rightarrow \infty$ and furthermore, assume that for every $k=1,\dots,K$ the conditions \ref{assump: C1}-\ref{assump: C3} hold for the elements of the projected model in (\ref{eqn: mixedmodel}). Under the null hypothesis (\ref{eqn:H0intro}), for each $k=1, \ldots, K$
\begin{equation}\label{eqn: Nulldistr}
    \widehat{pLRT}_{N,k} \overset{d}{\longrightarrow} \sup_{\lambda \geq 0} \spr{\sum_{q=1}^{Q} \dfrac{\lambda\zeta_{kq}\vartheta_{q}}{(1+\lambda \zeta_{kq})}  - \sum_{q=1}^{Q} \log (1+\lambda \xi_{ kq})} + \mathcal{X} \qquad \textrm{as} \; n \to \infty, R \to \infty;
\end{equation}
where $\vartheta_{q} \overset{iid}{\sim} \chi^2_1$, $\mathcal{X} \sim \textcolor{black}{\chi^2_1}$, and independently distributed with $\vartheta_{q}$ for all $q=1,\dots,Q$. The quantities $\spr{\zeta_{kq}}_{q=1}^Q$ and $\spr{\xi_{ kq}}_{q=1}^Q$ are specified in condition~\ref{assump: C3} of Appendix A.
\end{theorem}

The pseudo-likelihood ratio statistic is different from \cite{staicu2014likelihood} as it is calculated using the quasi projections ${W}_{k, ij}$. Furthermore, the derivation of the null asymptotic distribution accounts for the dependence of ${W}_{k, ij}$ across $i$. The main reason we obtain the same asymptotic null distribution of $\widehat{pLRT}_{N,k}$ as in \cite{staicu2014likelihood} is the uniform convergence of the eigenfunctions $\widehat\phi_k(s)$. A proof is provided in Section~\ref{sec: supp_proof_pLRT} of the Supplement. The asymptotic null distribution is not standard, but it can be easily simulated from it. In particular, one can empirically compute the $100(1-\alpha)\%$ percentile, $pLRT_{\infty, k \,;\, \alpha}$ by fast simulation from the null distribution using the R package \verb|RLRsim| \citep{RLRsim}. See \cite{crainiceanu2004likelihood} for details.

{\color{black}
\begin{remark}
    More generally, suppose we are interested in testing whether $(\partial^p/\partial t^p) \mu(s,t) = 0$, for some $p > 1$. Assuming that the mean function has a continuous partial derivatives with respect to $t$ up to order $p$, such hypothesis implies $(\partial^p/\partial t^p) \eta_k(t) = 0$, for all $k$, and $t \in \mathcal{T}$. For fixed $k$, to test this null hypothesis it makes sense to represent $\eta_k(t)$ in terms of the truncated power basis of order $p$, with knots $\kappa_1, \dots, \kappa_Q$, and write it as 
$    \eta_k(t) = \beta_{0k} + \beta_{1k}t + \dots + \beta_{p\,k}t^{p} + \sum_{q=1}^Q b_{kq}(t-\kappa_q)_{+}^{p}$ for $k=1,\dots,K$ \citep{crainiceanu2004likelihood}. In this representation, the null hypothesis $(\partial^p/\partial t^p) \eta_k(t) \equiv 0$ translates to
testing that $\beta_{p\,k} = 0, \sigma^2_{b} = 0$ versus one of them is not equal to zero. We remark that our proposed pseudo-likelihood ratio test statistic can also be used to test this null hypothesis, and the null distribution of the test has exactly the same form as in Theorem~\ref{theorem: pLRTnulldist} with appropriate modifications in the elements of design matrices $\bX$, and $\bZ$. 
\end{remark}
}

The projection based functional invariance test, or PROFIT, formally assess the null hypothesis (\ref{eqn:H0intro}), $H_0: \mu(\cdot, t)\equiv \mu(\cdot, t')$ for all $t\neq t'$ by simultaneous testing of (\ref{eqn: H04}) for $k=1, \ldots, K$ and using a Bonferroni's multiple test correction to control for the familywise error rate. For any $\alpha \in (0,1)$ and $K \geq 1$, PROFIT is defined by the rejection region 
\begin{equation*}
    \mathcal{R}^K_{\textrm{PROFIT}} =\spr{ \widehat{pLRT}_{N,k} > pLRT_{\infty, k \,;\, \alpha/K}, \text{for some } k=1,\dots,K }.
\end{equation*}
The next corollary states that this rejection region has a probability controlled by the nominal level $\alpha$, when the null hypothesis is true. 
\begin{corollary}
Assume the setup and the conditions of the Theorem~\ref{theorem: pLRTnulldist}, and furthermore, assume that the null hypothesis described in~(\ref{eqn:H0intro}) is true. Then, for any $K$  
\begin{equation*}
\mathrm{Pr}\fpr{ \mathcal{R}^K_{\textrm{PROFIT}}   } \leq \alpha. 
\end{equation*}
\end{corollary}
{\color{black}\begin{remark} \label{remark: sensitivityK}
    The choice of $K$ does not affect the size of the PROFIT; nonetheless it may affect its power. The impact of $K$ on the power is dependent on whether the mean function completely lies in the space of the residual trajectories, viewed as functions of $s$. For example if $\mu(\cdot,t)$ can be fully represented by $K_0$ eigenfunctions spanning the space of residuals, for all $t$, then choosing a value of $K > K_0$ will lead to a decrease in the power of the test. On the other hand, when $\mu(\cdot,t)$ is not accurately represented by the $K_0$ eigenfunctions of the residual trajectories, and that $\mu(\cdot,t) - \sum_{k=1}^{K_0} \phi_k(\cdot)\eta_k(t)$ is a non-constant function of $t$, and choosing a larger truncation $K > K_0$ based on a higher value of PVE (e.g. $99\%$) may lead to an increase in the test's power. This is justified by two reasons i) the additional eigenfunction/s can detect significant temporal variation of the mean function in the space orthogonal to the one spanned by the first $K_0$ eigenfunctions; ii) the additional eigenfunction/s project the data onto directions with smaller variance (compared to the first $K_0$'s) and a small temporal variation $\mu(s,t)$ can be detected with higher power along such directions. This feature is validated numerically later in Section~\ref{sec: sensitivityanalysis}. 
\end{remark}
In practice we propose to select $K$ using a reasonable level of PVE; in our numerical investigations we select $K$ using a PVE of $90\%$ and observe that PROFIT has competitive empirical size and power performance. The algorithm~\ref{alg: testing}
summarizes the steps of PROFIT.
}

{\small
 \begin{algorithm}[]
	\SetAlgoLined
	\KwIn{Data: $[ t_{ij}, Y_{ij}(s), s\in\{s_1, \ldots, s_R\}]_{j=1}^{ m_i}$ for $i=1, \ldots, n$, significance level $\alpha$, and a pre-specified PVE}
	    Construct a smooth estimator of the bivariate mean function as $\widehat{\mu}(s,t)$\;
	     Compute the demeaned response $\widetilde{Y}_{ij}(s_r) = Y_{ij}(s_r) - \widehat{\mu}(s_r, t_{ij})$\;
	     Obtain a smooth estimator of $\widehat{\Xi}(s,\sprime)$ using the demeaned responses\;
	     Get $\spr{\widehat{\phi}_k(s)}_{k=1}^K$ from the spectral decomposition of $\widehat{\Xi}(s,\sprime)$ with $K$ chosen by the specified PVE\;
		\For{$k \in \{1, \ldots, K\}$}{
		 Construct the projected data $\{ (t_{ij}, W_{k,ij} ): i = 1, \ldots, n \text{ and } j = 1, \ldots, m_{i} \}$ by calculating $W_{k,ij} = R^{-1}\sum_{r = 1}^{R} Y_{ij}(s_r)\widehat{\phi}_k(s_r)$\;
			 Compute the test-statistic $\widehat{pLRT}_{N,k}$ in~(\ref{eqn: pLRThat})\;
		 Calculate empirical p-value, $p_{k}$ using the asymptotic null distribution in~(\ref{eqn: Nulldistr})\;
		}
		 Reject $H_{0}$ in~(\ref{eqn:H0intro}) if $\min\{p_{k} : k = 1, \ldots, K\} < \alpha/K$ for a chosen significance level $\alpha$\;
		 \caption{PROjection-based Functional Invariance Testing (PROFIT) }
		  \label{alg: testing}
	\end{algorithm}
}
\subsection{Asymptotic power under local alternatives}
In this section, we assess the asymptotic power of PROFIT. We consider a sequence of local alternatives, 
\begin{align}
    H_0: \mu(s,t) = \mu_0(s) \qquad \text{vs} \qquad H_{1n}: \mu(s,t) = \mu_0(s) + n^{-a/2} \Delta(s,t), \label{eqn: localalter}
\end{align}
where $a > 0$, $\mu_0(s)$ is some function of $s$ solely, and $\Delta(s,t)$ is a fixed (known) bivariate function of both $s$ and $t$ with same smoothness properties to $\mu(s,t)$. To guarantee that $\Delta(s,t) \neq \Delta(s,t^\prime)$ for some $s$ and some $t \neq t^\prime$, we assume that the projection of $\Delta(s,t)$ onto $\phi_k(s)$ is a non-constant function of $t$, for some $k=1,\dots,K$. We represent the projection $\Delta_k(t) = \int \Delta(s,t)\phi_k(s)ds$ flexibly to describe a wide class of functions by expanding it in terms of truncated polynomial of order $p$ as,
{\color{black}
\begin{align*}
    \Delta_k(t) = d_{0k} + d_{1k}t + \sum_{q=1}^Q {b}_{kq}(t-\kappa_q)_{+},
\end{align*}
}
\textcolor{black}{where $\{d_{0k}, d_{1k}\}$ are the fixed coefficients and $\{{b}_{kq}\}_{q=1}^Q$ are the spline coefficients and $\kappa_1 < \dots < \kappa_Q$ are fixed knots, which are large enough to ensure the desired flexibility. The spline coefficients are assumed to be random and ${b}_{kq} \overset{iid}{\sim} (0, \sigma^2_{0,k})$. Under this mixed effect representation, $\Delta_k(t)$ is a non-constant function of $t$ if either $d_{1k} \neq 0$ and/or $\sigma^2_{0,k} > 0$ and for some $k=1, \dots, K$. Therefore, under the mean representation described in Section \ref{subsec: testingprocedure} testing~(\ref{eqn: localalter}) is equivalent to the following sequence of local alternatives,}
{\color{black}
\begin{equation}\label{eqn: H1kn}
\begin{aligned}
    H_{0k}: \beta_{1k}= 0, \; \sigma^2_{b,k} = 0 \quad \text{vs} \quad
     H_{1kn}: \beta_{1k} = n^{-a/2}d_{k}, \quad \sigma^2_{b,k} = n^{-a}\sigma^2_{0,k},
    \end{aligned}
\end{equation}}
\textcolor{black}{where at least one of $d_{k}$, and $\sigma^2_{0,k}$ is not zero.} The next theorem states the asymptotic power of the pseudo-likelihood ratio test to detect a local alternative of the form~(\ref{eqn: H1kn}).
\begin{theorem}\label{theorem: pLRTAltdist}
Consider the setup and assume that all the conditions specified in the statement of Theorem~\ref{theorem: pLRTnulldist} holds. Under the local alternative hypothesis (\ref{eqn: H1kn}), for each $k=1, \ldots, K$,
{\small\begin{equation*}\label{eqn: Altdistr}
   \widehat{pLRT}_{N,k}  \;\;\;\overset{d}{\longrightarrow}\;\; \begin{cases}
    M_k({d}_{k}, \sigma^2_{0,k}) \;\; &\text{if} \;\; a = \varrho \\
    \infty \;\; &\text{if} \;\; 0 \leq a < \varrho,
    \end{cases}
    \qquad \textrm{as} \; n \to \infty, R \to \infty;
\end{equation*}}
where 
{\color{black}\small\begin{align}
    \nonumber M_k({d}_k, \sigma^2_{0,k}) \;\;\; \overset{d}{=} \;\;\;\sup_{\lambda \geq 0} \spr{\sum_{q=1}^{Q} \dfrac{\lambda\zeta_{kq}(1+\sigma^2_{0,k}\zeta_{kq})}{(1+\lambda \zeta_{kq})}\vartheta_{kq}  - \sum_{q=1}^{Q} \log (1+\lambda \xi_{ kq})} 
   + \left(\mathscr{Z}_{k}\sqrt{1+\sigma^2_{0,k}\varphi_{k}} + \sqrt{\theta_{k}}d_k\right)^2, 
\end{align}}
{\color{black}with $\{\vartheta_{kq}\}_{q=1}^{Q} \overset{iid}{\sim} \chi^2_1\;$ for all $k$, and independently distributed with $\mathscr{Z}_{k} \overset{iid}{\sim} N(0,1)$ for all $k$ . The constant $\varrho$, and the quantities $\spr{\zeta_{kq}}_{q=1}^Q$, $\spr{\xi_{ kq}}_{q=1}^Q$, $\varphi_{k}$ and $\theta_{k}$ are functions of the design matrices of the underlying LMM~(\ref{eqn: mixedmodel}), specified in condition~\ref{assump: C3} and \ref{assump: C4} of Appendix A.}
\end{theorem}
Theorem~\ref{theorem: pLRTAltdist} is inspired by \cite{crainiceanu2005exact} and extends the work of \cite{staicu2014likelihood} by deriving the asymptotic power to detect departure from the null when the true fixed effects and the variance parameters converges to zero at a certain rate. 
\begin{corollary} \label{corollary: powerPROFIT}
Assume the setup and the conditions of the Theorem~\ref{theorem: pLRTAltdist}. Then, for any fixed $K$ and specified significance level $\alpha \in (0,1)$, under the local alternative hypotheses in~(\ref{eqn: localalter})
{\small\begin{align*}
&\lim_{n\to \infty} \mathrm{Pr}\,\fpr{ \mathcal{R}^K_{\textrm{PROFIT}} \mid H_{1n}  } \quad  \begin{cases}
\geq \displaystyle{\max_{k =1,\dots,K} \mathrm{Pr}\,\fpr{M_k(\V{d}_k, \sigma^2_{0,k}) > pLRT_{\infty,k\;;\; \alpha/K}}}  \quad &\text{if } \;\; a = \varrho \\
= 1 \quad & \text{if } \;\; 0 \leq a < \varrho, 
\end{cases}
\end{align*}}
where $pLRT_{\infty,k\;;\;\alpha}$ is the $100(1-\alpha)\%$ percentile of the asymptotic null distribution of $\widehat{pLRT}_{N,k}$ specified in~(\ref{eqn: Nulldistr}).
\end{corollary}


Corollary~\ref{corollary: powerPROFIT} provides an explicit formula for the asymptotic power curve of PROFIT to detect alternative when at least one of the projection of $\Delta(s,t)$ onto the first $K$ eigenfunctions is non zero. When $\Delta(s,t) \neq 0$, and $K$ is sufficiently large, it is unlikely that all of its projection on the first $K$ directions are equal to zero. If the true mean function departs from $\mu_0(s)$ by a factor that goes to zero at a rate of $n^{-a/2}$, the power of the test is an increasing function of the magnitude of the fixed effect $\mathbf{d}_k$ and $\sigma^2_{0,k}$, $=1,\dots,K$ in the truncated power basis representation of $\Delta_k(t)$. If the true mean function converges to $\mu_0(s)$ at a rate slower than $n^{-a/2}$, then PROFIT rejects $H_0$ with probability $1$ asymptotically. 

\subsection{Alternative approach to multiple testing}
To address the issue that controlling the level of global hypothesis by Bonferroni's corrections might lead to sub-optimal theoretical power behavior for large value of $K$, a feasible solution would be to construct a global test-statistic that sums the pseudo-likelihood ratio test-statistic over all $k=1, \dots, K$, as
\begin{equation}\label{eqn: sumstatistics}
    T^*_{\text{profit}} := \sum_{k=1}^K \widehat{pLRT}_{N,k}.
\end{equation}
 Under the null hypothesis, we expect that $T^*_{\text{profit}}$ converges to
\begin{equation*}
   T^*_{\text{profit}} \overset{d}{\longrightarrow} \sum_{k=1}^K \;\;\sup_{\lambda_k \geq 0} \spr{\sum_{q=1}^{Q} \dfrac{\lambda_k\zeta_{kq}\vartheta_{kq}}{(1+\lambda_k \zeta_{kq})}  - \sum_{q=1}^{Q} \log (1+\lambda_k \xi_{ kq})} + \mathcal{X}_k \qquad \textrm{as} \; n \to \infty, R \to \infty;
\end{equation*}
where $\mathcal{X}_k \sim \chi^2_1$, for all $k$, and all the other terms in the above equation are specified in the statement of Theorem~\ref{theorem: pLRTnulldist}. Although, we are not able to provide a theoretical proof of the statement, extensive simulation studies (Table \ref{tab: proposed only sumstat} and \ref{tab: proposed only NonGauss sumstat}) confirm that the above null distribution is correctly specified. Approximating the asymptotic null distribution of $T^*_{\text{profit}}$ is straightforward because generating sample from the null distribution of $\widehat{pLRT}_{N,k}$ for each $k=1, \dots, K$ can be done efficiently via R package \verb|RLRsim| \citep{RLRsim}. Using $T^*_{\text{profit}}$ as the single test-statistic for the test $H_0$ alleviates the problem of multiple testing and the sub-optimal power of test for large value of $K$. Additionally, increasing the value of $K$ will inevitably lead to increase in the power of the test since the value of test-statistic gets larger as $K$ increases. Therefore, if the number of estimated eigenfunctions $K = K(n)$ grows with $n$, then the power of test based on $T^*_{\text{profit}}$ grows to $1$.

\section{Simulation study}\label{sec: SimStudy}
\subsection{Data generation}\label{subsec: SSDatagen}
We study the empirical size and the power of the PROFIT in finite sample sizes $n$, for testing the null hypothesis (\ref{eqn:H0intro}) and when $n$ varies from $100$ to $400$. We generate data as in (\ref{eqn: assumedmodel}) with the mean function equal to $ \mu(s,t) = \cos(\pi s/2) + 5 \delta (t/4 - s)^3$
where $\delta$ quantifies the departure from the null hypothesis $H_0$. If $\delta=0$, $\mu(s,t)$ does not vary over $t$; thus $\delta=0$ represents $H_0$ and generating data from this setting allows us to study the size of the test. When $\abs{\delta}>0$, the mean $\mu(s,t)$ varies with $t$; this represents the alternative hypothesis, and thus generating data from this setting allows us to study the power behavior. 

For each subject $i$, the number of profiles, $m_i$, is generated randomly either from $\{8, 9, \ldots, 12\}$ (high sparsity level) or from $\{15, \ldots, 20\}$ (low sparsity level). For each $m_i$, the time points $t_{ij}$ are uniformly sampled from $\mathcal{T} =[0,1]$. The profiles $Y_{ij}(\cdot)$ are observed over a dense grid $R=101$ points equally spaced over $\mathcal{S} =[0,1]$. For each profile $Y_{ij}(s)$, the residual term is generated as
$
    \epsilon_i(s,t_{ij}) = \epsilon_{i1}(t_{ij}) \phi_{1}(s)+ \epsilon_{i2}(t_{ij})\phi_{2}(s) + r_{ij1}\phi_1(s) + r_{ij2}\phi_2(s) + \epsilon_{ij}^{wn}(s),
$
where $\phi_1(s) = \sqrt{2}\sin\fpr{2\pi s}$, $\phi_2(s) = \sqrt{2}\cos\fpr{2\pi s}$, $\epsilon_{ik}(t)$ are independently generated from
$\epsilon_{ik}(t) = \zeta_{i,k 1}\sqrt{2}sin(2\pi t) + \zeta_{i,k 2} \sqrt{2}cos(2\pi t), \;\; k = 1, 2, \;\;t \in [0,1]$, where $\zeta_{i,k\ell} \overset{\rm{iid}} \sim N(0, \sigma^2_{\zeta,k\ell})$ with $\{\sigma^2_{\zeta,11}, \sigma^2_{\zeta,12}\} = \{4,2\}$ and $\{\sigma^2_{\zeta,21}, \sigma^2_{\zeta,22} \} = \{3,1\}$.
The random coefficients and random noise are generated both from Gaussian and non-Gaussian models, to assess the performance of the method when Gaussian assumptions are not met. The Gaussian case is here: $r_{ij1} \overset{\rm{iid}}\sim N(0, 2)$, $r_{ij2}\overset{\rm{iid}}\sim N(0, 4/3)$ and $\epsilon_{ij}^{wn}(s_r)\overset{\rm{iid}}\sim N(0, 10)$; the description of the non-Gaussian case is presented in Section \ref{sec : numericalresult}.

\subsection{Computation details}\label{subsec: SSComp}

We carry out the steps of the Algorithm \ref{alg: testing} by first obtaining a smooth version of the bivariate mean function using the sandwich smoother \citep{xiao2013fast} implemented in \verb|fbps| function of \verb|refund| \citep{refund} package in R. After demeaning the response, we obtain the orthogonal basis functions $\widehat{\phi}_k(s) : k=1,\dots, K$ by applying the bivariate kernel smoother to the `raw covariances' using \verb|fpca.face()| function in the \verb|refund| package, for its computational efficiency over \verb|FPCA()| function in R package \verb|fdapace| \citep{fdapace}. The number of eigenbasis $K$ chosen based on PVE equal to $90\%$. The smoothing parameters in both the mean and the covariance estimation are chosen using generalized cross-validation (GCV). After projecting the response, we model the mean function $\eta_k(t)$ using truncated linear basis by placing the knots $\kappa_1, \dots, \kappa_Q$ at a equally spaced quantile levels of the observed visit times $\spr{\spr{t_{ij}}_{j=1}^{m_i}:i }$ with a number of knots $Q =\max \{ 20, \min ( 0.25 \times \text{ number of unique } t_{ij} , 40 ) \}$  \citep{ruppert2003semiparametric}. \textcolor{black}{Other choices of $Q$ have been considered, and the results are robust to the choice. More discussion and illustration of the results are presented in Section~\ref{sec: supp_senstivity_Q} of the Supplementary material.} For each $k$, the covariance function of the projected response, $\gamma_k(t,t^\prime)$, is estimated by representing the random component $\epsilon_{ik}(t)$ as a truncated KL expansion, i.e. ${\epsilon}_{ik}(t) = \sum_{\ell=1}^{L_{k}}c_{ik\ell}\psi_{k\ell}(t)$, where $\psi_{k\ell}(t)$ be the estimated eigenfunctions and the $c_{ik\ell}$ are estimated score coefficients. The eigenfunctions/eigenvalues corresponding to the projected data are estimated using \verb|fpca.sc()| function in \verb|refund| package. The number of eigenfunctions $L_k : k=1,\dots,K$ is chosen with PVE of $90\%$. After denoising the quasi projections $W_{k,ij}$ with the inverse square root of estimated covariance matrix $\widehat{\bSigma}_{W,k}$,  we fit the LMM in~(\ref{eqn: mixedmodel}) using the \verb|lme()| function in \verb|nlme| \citep{nlme} package and compute the test-statistic $\widehat{pLRT}_{N,k}$ along with the p-values via \verb|exactLRT()| function in \verb|RLRsim| package. The p-value with PROFIT is calculated as the minimum of these $K$ p-values. \textcolor{black}{The R code to replicate the simulation results is publicly available at \url{https://github.com/SalilKoner/PROFIT}.}  

\subsection{Competitive methods}\label{subsec: SSCompMethods}

We compare the performance of the PROFIT with two alternative approaches. The first method uses the multiple testing framework that we propose combined with the $L_2$ norm-based testing procedure discussed in \cite{zhang2007statistical}. Specifically, for each $k$, we use the projected data to test the hypotheses (\ref{eqn: H04}) and calculate the test statistic as $    T_{ZC,k} = \int_{0}^1 \fpr{\widehat{\eta}_k(t) - \widehat{C}_k}^2 dt$, $k=1,2,\dots, K$
where $\widehat{\eta}_k(t)$ and $\widehat{C}_k$ be the estimated mean functions for the projected data $\{W_{k,ij}:i, j\}$ under the alternative and the null hypotheses, respectively. The `smoothing first, then estimation' approach cannot be used because the sampling design of time points $\spr{t_{ij} : j=1, \dots, m_i}$ for each subject is sparse. Instead, we obtain a smooth estimate of $\widehat{\eta}_{k}(t)$ by fitting $10$ cubic basis functions to the projected data $W_{k,ij}$'s, and $\widehat{C}_{k}$ is the sample average of the projected responses, $\{W_{k,ij}: \forall \; i \text{ and } j \}$, pooling information from all subjects. GCV is used to select the smoothing parameters. The asymptotic null distribution of the test statistic is approximated in two ways: i) mixture of scaled chi-squared distribution ; $\sum_{r=1}^{L_k} \widehat{\nu}_{kr}  A_r$ where $A_1, \dots, A_{L_k}$ follows iid $\chi^2_1$ and $\widehat{\nu}_{kr}$ are eigenvalues of the estimated covariance $\widehat{\gamma}_k(t,\tprime)$ with $L_k$ chosen via a PVE of $90\%$; and ii) bootstrap of the subjects (see Section 3 of \cite{zhang2007statistical}). We use $B=1000$ bootstrap samples to estimate the null distribution of the test statistic. Bonferroni's correction is used to control the family-wise error rate in this multiple testing framework. We call these two methods as \verb|ZC-MC| (\textit{Zhen and Chen, mixture of Chi-square}) and \verb|ZC-BT| (\textit{Zhen and Chen, bootstrap}), respectively. 

The second method is the $L_2$ norm-based test statistic proposed in \cite{park2018simple},  $    T_{boot} = \int_{0}^1 \int_{0}^1 \fpr{\widehat{\mu}_A(s,t) - \widehat{\mu}_0(s)}^2 ds dt$, where $\widehat{\mu}_0(s)$ and $\widehat{\mu}_A(s,t)$ are the estimated mean functions under the null and the alternate hypotheses, respectively.
Both fits are estimated under a working independence assumption; the null distribution of the test $T_{boot}$ is estimated using $B$ bootstrap samples carried using the algorithm (3) of \cite{park2018simple}. Due to the massive computational time required by this method (see Table \ref{tab: computation time}, based on $B=300$ bootstrap samples, $10$ cubic B-splines to estimate ${\mu}_0(s)$ and tensor product of $10$ cubic B-splines in $s$ and $5$ cubic B-splines in $t$ to obtain $\widehat{\mu}_A(s,t)$) we do not include this method in our comparative study of size and power. 

\subsection{Assessing performance of the test} \label{sec : numericalresult}


\begin{table}
	\centering
	\caption{The empirical Type-1 error rates of PROFIT based on 10,000 simulations. Standard errors are presented in parentheses.}  
{
		\begin{tabular}{cc cccc}
			\hline	\hline
			\multicolumn{6}{c}{$m_i \sim \{8, \ldots, 12\}$} \\	\hline
			& & $\alpha = 0.01$ & 	$\alpha = 0.05$ & 	$\alpha = 0.10$ & 	$\alpha = 0.15$ \\ 
			\hline	\hline
	$n = $	100	 &&  0.013  (0.001) & 0.057  (0.002) & 0.110  (0.003) & 0.157  (0.004) \\  
	$n = $	150	 &&  0.010  (0.001) & 0.053  (0.002) & 0.105  (0.003) & 0.152  (0.004) \\ 
	$n = $	200	 && 0.010  (0.001) & 0.046  (0.002) & 0.095  (0.003) & 0.145  (0.004) \\  
	$n = $	300	 &&0.009  (0.001) & 0.047  (0.002) & 0.096  (0.003) & 0.142  (0.003) \\  
    $n = $	400	 && 0.009  (0.001) & 0.048  (0.002) & 0.095  (0.003) & 0.142  (0.003) \\ 
			\hline	\hline
			\multicolumn{6}{c}{$m_i \sim \{15, \ldots, 20\}$} \\	\hline
		&& $\alpha = 0.01$ & 	$\alpha = 0.05$ & 	$\alpha = 0.10$ & 	$\alpha = 0.15$ \\ 
			\hline	\hline
	$n = $100	  &  & 0.010  (0.001) & 0.052  (0.002) & 0.103  (0.003) & 0.152 (0.004) \\ 
	$n = $150	  &  & 0.008  (0.001) & 0.049  (0.002) & 0.101  (0.003) & 0.146 (0.004) \\  
	$n = $200	&     & 0.009  (0.001) & 0.047  (0.002) & 0.096  (0.003) & 0.140  (0.003)\\  
	$n = $300  &  &0.009  (0.001) & 0.048  (0.002) & 0.097  (0.003) & 0.143  (0.004) \\  
	$n = $400	& & 0.008  (0.001) & 0.046  (0.002) & 0.096  (0.003) & 0.145  (0.004)  \\
			\hline\hline
		\end{tabular}
	} \label{tab: proposed only}
\end{table}

\begin{table}
	\centering
	\caption{The empirical Type-1 error rates of PROFIT for non-Gaussian errors under low sparsity level, based on 5,000 simulations. Standard errors are presented in parentheses. }  
	\scalebox{1}{
		\begin{tabular}{cc cccc}
			\hline	\hline
			\multicolumn{6}{c}{Scenario (i) : $r_{ij,k} \sim$  mixture-normal} \\	\hline
			& & $\alpha = 0.01$ & 	$\alpha = 0.05$ & 	$\alpha = 0.10$ & 	$\alpha = 0.15$ \\ 
			\hline	\hline
	$n = $	100	 &&  0.008  (0.001) & 0.049  (0.003) & 0.103  (0.004) & 0.153  (0.005) \\  
	$n = $	150	 &&  0.008  (0.001) & 0.044  (0.003) & 0.091  (0.004) & 0.134  (0.005) \\ 
	$n = $	200	 && 0.007  (0.001) & 0.046  (0.003) & 0.095  (0.004) & 0.140  (0.005) \\  
	$n = $	300	 &&0.010  (0.001) & 0.044  (0.003) & 0.097  (0.004) & 0.149  (0.005) \\  
    $n = $	400	 && 0.008  (0.001) & 0.046  (0.003) & 0.096  (0.004) & 0.142  (0.005) \\
			\hline	\hline
			\multicolumn{6}{c}{Scenario (ii) : $\zeta_{i,k\ell} \sim$  mixture-normal} \\	\hline
		&& $\alpha = 0.01$ & 	$\alpha = 0.05$ & 	$\alpha = 0.10$ & 	$\alpha = 0.15$ \\ 
			\hline	\hline
	$n = $	100	 &&  0.009  (0.001) & 0.047  (0.003) & 0.102  (0.004) & 0.151  (0.005) \\  
	$n = $	150	 &&  0.008  (0.001) & 0.044  (0.003) & 0.092  (0.004) & 0.138  (0.005) \\ 
	$n = $	200	 && 0.012  (0.001) & 0.061  (0.003) & 0.109  (0.004) & 0.163  (0.005) \\  
	$n = $	300	 &&0.009  (0.001) & 0.049  (0.003) & 0.095  (0.004) & 0.141  (0.005) \\  
    $n = $	400	 && 0.008  (0.001) & 0.047  (0.003) & 0.099  (0.004) & 0.148  (0.005) \\ 
			\hline\hline
		\end{tabular}
	} \label{tab: proposed only NonGauss}
\end{table}

\textbf{Size}: \textcolor{black}{Quantile plots of ordered p-values obtained from the pseudo-likelihood ratio test versus the quantiles of Uniform$[0,1]$ distribution under the null hypothesis is presented in Section~\ref{sec: supp_validity_pvalues} of the Supplement, suggesting that the asymptotic null distribution of $\widehat{pLRT}_{N,k}$ in Theorem~\ref{theorem: pLRTnulldist} also has good finite sample properties.} Table~\ref{tab: proposed only} shows the empirical Type-1 error rate of the PROFIT for nominal levels $\alpha=0.01,0.05,0.1,0.15$ across different sample size $n$ and sparsity levels of the repeated measures $m_i$. Estimates and standard error (in parentheses) are obtained based on $10,000$ simulations. As the sample size gets larger ($n \geq 150$), the empirical size of the test are within twice standard error of the stipulated nominal level $\alpha$. However, we observe a slight inflation of the size of PROFIT when the sample size is $n=100$ and there are between 8 to 12 profiles per subject. As the number of curves per subject increases ($m_i$ is between 15 to 20), the test shows a correct empirical size even for $n=100$. This slight inflation is the result of poor quality estimation of the covariance of the projections, and as a result of the eigenfunctions, in this situation; Table~\ref{tab: proposed only all} in Section~\ref{sec: supp_additional_res_ZCMT_ZCBT} of the Supplement confirms that when the true eigenfunctions are used instead, PROFIT maintains the correct nominal size. We also study the size of the other competitors (see Table \ref{tab: size for all competitive methods} in Section~\ref{sec: supp_additional_res_ZCMT_ZCBT} of Supplement), and summarize the result here. \verb|ZC-MT| fails to maintain the size when there are few curves per subject  ($m_i$ is between 8 to 12) even for large sample sizes $n=400$; the performance does improve with increasing the number of curves per subject. In contrast, \verb|ZC-BT| shows a conservative behavior for all the sample sizes $n$ and number of curves per subject $m_i$ studied. 
\begin{figure}
    \centering
    \includegraphics[scale=0.39]{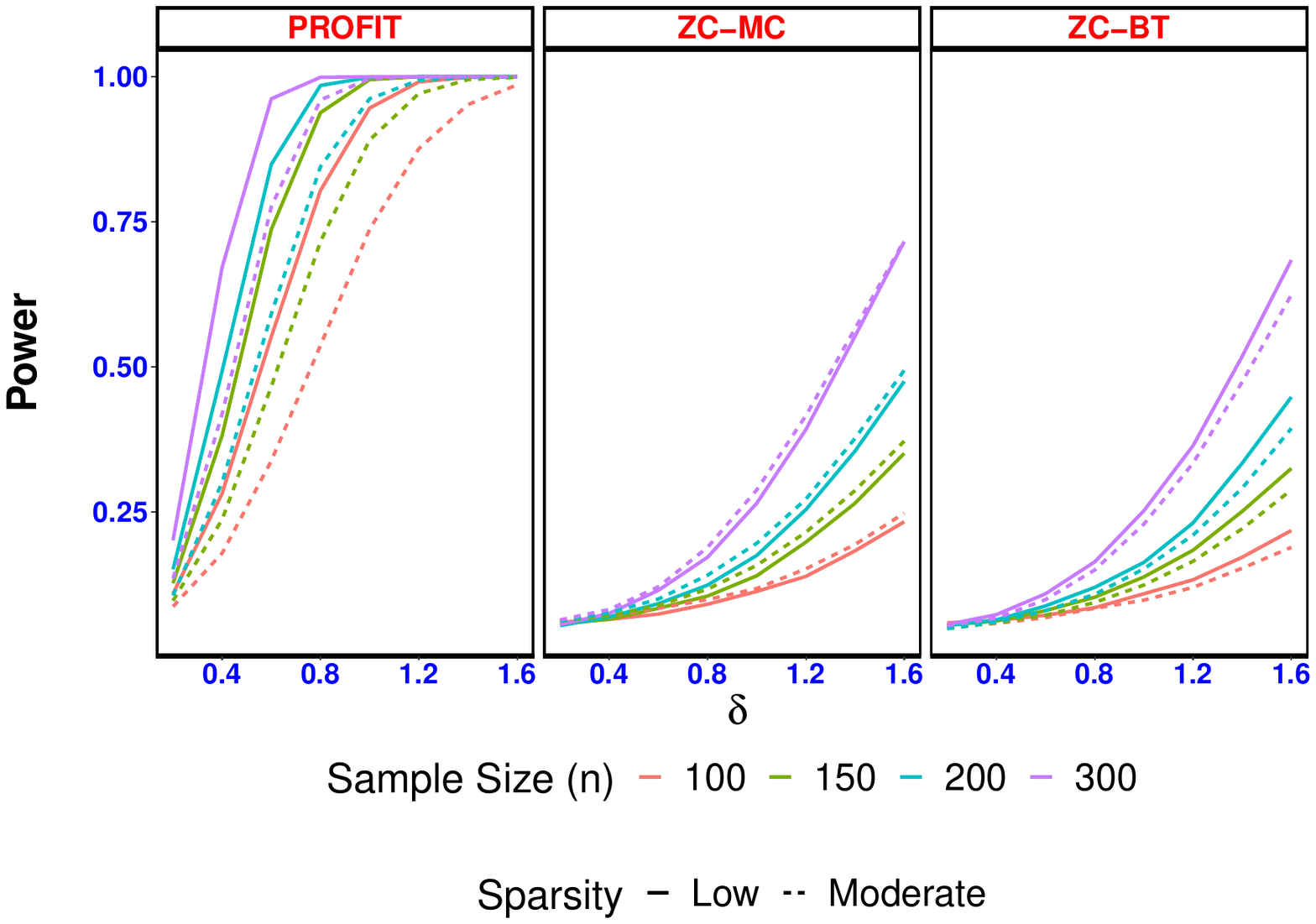}
    \caption{Empirical power curve for PROFIT, ZC-MC and ZC-BT as a function of $\delta$ across sample size $n$ varying from $100$ to $300$ when the sparsity levels of the visit times are \textit{low} (solid line) and \textit{moderate} (dashed line). The numbers are based on $5000$ simulations.}
    \label{fig:powercurve}
\end{figure}

Although the testing procedure assumes that the errors are Gaussian (condition \ref{assump: C1}), we further explore the size of the PROFIT for non-Gaussian errors. To this end, we consider two scenarios i) $r_{ij,k}$ are iid mixture of two normals, i.e. $N(\sigma_{e,k}/\sqrt{2},\sigma^2_{e,k}/2)$ with probability $1/2$ and $N(-\sigma_{e,k}/\sqrt{2},\sigma^2_{e,k}/2)$ with probability $1/2$, for $k=1,2$; ii) $\{\zeta_{i,k\ell} : k=1,2; \ell=1,2\}$ are generated from a mixture of $N(\sigma_{\zeta,k\ell}/\sqrt{2},\sigma^2_{\zeta,k\ell}/2)$ and $N(\sigma_{-\zeta,k\ell}/\sqrt{2},\sigma^2_{\zeta,k\ell}/2)$ with probability $0.5$ each. As reflected in Table~\ref{tab: proposed only NonGauss}, the test does an excellent job to maintain the empirical size as the sample size increases, numerically justifying its robustness when the assumption of Gaussian error may not be true. 

\textbf{Power}: Fix the level of significance $\alpha=0.05$. Figure~\ref{fig:powercurve} plots the empirical power of PROFIT as a function of $\delta$ for different sample size $n$ and sparsity levels of the repeated measurements. As expected, the power of the test increases with the sample size. The power increases with larger number of curves per subjects. Interestingly, the power of the test increases more with increasing the number of curves per subject than with increasing the number of subjects. This is due to the accuracy in estimating the covariance of the projected data, which is higher, when the number of curves per subject $m_i$ increases; the observation is likely specific to the setting considered and it may not generalize. The power curves of the competitive methods are shown in the middle and rightmost plots of Figure~\ref{fig:powercurve}; they show that PROFIT is much more powerful than the available competitors, irrespective of the number of curves per subject are observed. 

The computation times of the proposed method for $n= 200$ are tabulated in Table~\ref{tab: computation time} under different sparsity conditions. We can see that our method is scalable to the ZC-MC method, thanks to the fast simulation from the asymptotic null distribution by \verb|RLRsim| package \citep{RLRsim}, and is much faster compared to the analogous bootstrap based procedures. The bootstrap based approach \citep{park2018simple} takes enormous amount of time (about $150$ times slower than PROFIT), by comparison, as it requires fitting the bivariate smoother multiple times for each simulation. 
\begin{table}
	\centering
	\caption{Computation time (in seconds) for one simulation when $n = 200$}
	\scalebox{1}{
	\begin{tabular}{ccccc}
		\hline	\hline
		&	PROFIT & ZC-MC & ZC-BT & \citeauthor{park2018simple} \\ 
		\hline
	$ m_{i} \sim \{ 8, \ldots, 12\}$ &	 7.546 & 2.317 & 24.412 & 1126.880 \\ 
		$ m_{i} \sim \{ 15, \ldots, 20\}$&	8.010 & 2.376 & 30.081 & 1841.730 \\ 
		\hline	\hline
	\end{tabular}}\label{tab: computation time}
\end{table}

The empirical size of the test using the `summed' statistic $T^*_{\text{profit}}$ in~(\ref{eqn: sumstatistics}) is provided in the Table~\ref{tab: proposed only sumstat} and \ref{tab: proposed only NonGauss sumstat}. The numbers establish that the null distribution of is $T^*_{\text{profit}}$ correctly computed as the test maintains the Type-1 error for all the scenarios. The empirical power curve of test using $T^*_{\text{profit}}$ is also presented in left panel of the Figure~\ref{fig:powercurve_sum}, which demonstrates a similar behavior to the power curve of Bonferroni-corrected test (presented in the right panel).

{\color{black}\subsection{Sensitivity of the size and power to the choice of truncation, $K$} \label{sec: sensitivityanalysis}
Table~\ref{tab: supp_sens_PVE} displays the results of our test's size and power as we vary the truncation parameter $K$ based on different levels of PVE, ranging from $90\%$ to $99\%$. In our simulation study, a PVE set to $90\%$ leads to selecting $K$ equal to 2-3. Due to the way we designed the generative model, a PVE set to $95\%$ does not, in general increase $K$; however a PVE set to $99\%$ results in selecting $K$ equal to 3-4.

From Table~\ref{tab: supp_sens_PVE}, we draw two important conclusions. As expected, the empirical size of PROFIT remains consistent even as we increase the truncation parameter $K$.  Secondly, since the mean function $\mu(\cdot,t)$ in our simulation design cannot be accurately represented by the two leading eigenfunctions that span the space of the residual trajectories for every $t$, increasing PVE to $99\%$ leads to identification of additional eigenfunction/s that project the data along directions with smaller variance. Consequently, the projected response $W_{ijk}$ for $k=3$/$k=4$ exhibits significantly lower variance, enabling us to detect slight deviations from the null hypothesis with greater power. Therefore, increasing the PVE to $99\%$ enhances the power of the testing procedure considerably. This reinforces the point we made in Remark~\ref{remark: sensitivityK}. }

\begin{table}[]
\centering
    \caption{Effect of large truncation parameters on the size and the power of the test at $\alpha=0.05$ for the mean function $\mu(s,t) = \cos(\pi s/2) + 5 \delta (t/4 - s)^3$. This mean function do not completely lie in the space of residuals. Selecting a higher PVE of $99\%$ detects the temporal variation of the mean along the directions orthogonal to the leading eigenfunctions, detected by PVE of $90\%$. The number of knots is fixed at $Q=20$. The columns with $\delta=0$ corresponds to the case when the null hypothesis is true.}
    \label{tab: supp_sens_PVE}
\begin{tabular}{ccccccccccccc}
\hline \hline 
\multicolumn{12}{c}{$m_i \sim \{8,\dots, 12\}$}                                                      \\ \hline
\multicolumn{3}{c}{\multirow{2}{*}{\begin{tabular}[c]{@{}c@{}} PVE \end{tabular}}} & & \multicolumn{4}{c}{$n = 100$} & & \multicolumn{4}{c}{$n = 200$} \\ 
\cline{5-13} 
\multicolumn{3}{c}{}           & & $\delta=0$ & $0.05$ & $0.1$ & $0.2$  & & $\delta=0$ & $0.05$ & $0.1$ & $0.2$ \\ \hline
\multicolumn{3}{c}{$90\%$}  & &  0.05 &	0.05	& 0.06	& 0.08 &	& 0.05 &	0.06	& 0.07 & 0.11  \\
\multicolumn{3}{c}{$95\%$}  & & 0.05 &	0.05	& 0.06	& 0.08 &	& 0.05 &	0.06	& 0.07 & 0.11  \\
\multicolumn{3}{c}{$99\%$} & & 0.05 & 0.13 & 0.52 & 1.00 &&  0.05	& 0.29	& 0.87 & 1.00	 \\ \hline
\multicolumn{12}{c}{$m_i \sim \{15,\dots, 20\}$}                                                      \\ \hline
\multicolumn{3}{c}{\multirow{2}{*}{\begin{tabular}[c]{@{}c@{}} PVE \end{tabular}}} & & \multicolumn{4}{c}{$n = 100$} & & \multicolumn{4}{c}{$n = 200$} \\ 
\cline{5-13} 
\multicolumn{3}{c}{}           & & $\delta=0$ & $0.05$ & $0.1$ & $0.2$  & & $\delta=0$ & $0.05$ & $0.1$ & $0.2$ \\ \hline
\multicolumn{3}{c}{$90\%$}  & &  0.05 &	0.05	& 0.06	& 0.11 &	& 0.04 &	0.06	& 0.07 & 0.15  \\
\multicolumn{3}{c}{$95\%$}  & & 0.05 &	0.05	& 0.06	& 0.11 &	& 0.04 &	0.06	& 0.07 & 0.15  \\
\multicolumn{3}{c}{$99\%$} & & 0.05 & 0.23 & 0.82 & 1.00 &&  0.05	& 0.47	& 0.99 & 1.00	 \\ \hline
\end{tabular}
\end{table}

\section{Diffusion tensor imaging study} \label{sec: DTI}

Multiple sclerosis (MS) is an autoimmune disease that is associated with physical and cognitive disorder. It is characterized by the disruption of myelin sheaths and axonal loss, which impairs the integrity of the white matter (WM) tracts \citep{basser2011microstructural}. DTI is a powerful tool to visualize the microstructural organization of the WM, by describing the diffusion of the molecule of water in the white matter; acute demyelination leads to preferential direction of the water motion. Fractional Anisotropy (FA) measures the degree of anisotropy of the water diffusivity process and reflects myelinal and axonal disintegrity in the normally appearing WM of MS patients \citep{hasan2005diffusion}. This study involves 162 MS patients aged between 20 and 70 years, for whom FA profiles along the corpus callosum tract (FA CCA) are measured at each of their hospital visits for approximately 4.3 years, the length of the study. The visit times of patients vary between $1$ and $8$, with almost $90\%$ of the patients having less than $6$ visits. A segment of the full data is freely available in the R package \verb|refund| \citep{refund}.

FA is known to be scientifically correlated to the state of MS progression \citep{goldsmith2011penalized, ciccarelli2003study}. Studying the dynamics of the FA profiles over times can be helpful to detect the progression of MS, as in healthy individuals FA is not expected to change much during this short study. In this paper we formally investigate whether the mean FA along the CCA changes over time. This is very important, as it can inform better interventions to prevent the rapid progression of the disease \citep{coote2009getting}.

First, we align and scale the visit times, so that $t_{i1}=0$ and $t_{ij}\in [0,1]$ for all patients. For each $i$ and $j$ the FA CCA profiles $Y_{ij}(\cdot)$ are observed at $R=93$ locations equally spaced along the CCA tract. To adjust for the patients' age, we posit the model
\begin{equation*} \label{eqn: modelforDTI}
    Y_{ij}(s) = \mu(s, t_{ij}) + \textrm{Age}_i \; \alpha(s) + \epsilon_i(s, t_{ij}),
\end{equation*}
where $\mu(\cdot, t)$ is the mean FA along CCA at time $t$, $\alpha(\cdot)$ is a smooth effect of the age along the tract location that is assumed constant during the duration of the study, and $\epsilon_i(\cdot, \cdot)$ is the residual process. The objective of study can be equivalently written as $H_0: \mu(s, t) =\text{does not vary over $t$}$ versus the alternative $H_A: \mu(s, t) =\text{varies over $t$} $; we use PROFIT as described in the paper. 

\begin{figure}
    \centering
    \subfloat[Estimated mean function $\widehat{\mu}(s,t)$]{\includegraphics[width=7 cm, height = 5.8 cm]{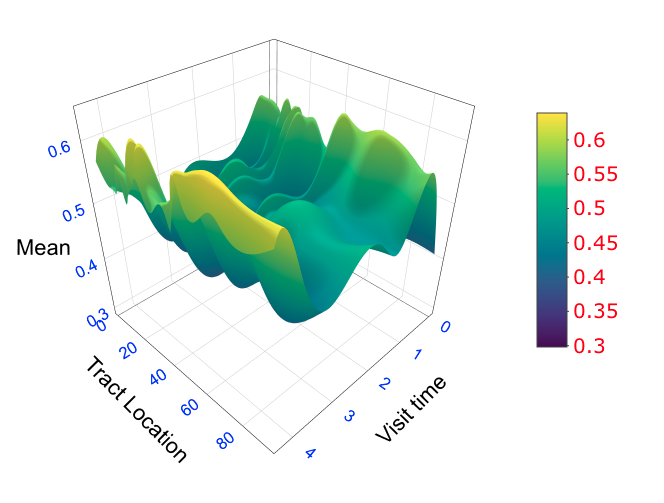}}\qquad
        \subfloat[Univariate crosssection of $\widehat{\mu}(\cdot,t)$]{\includegraphics[scale=0.23]{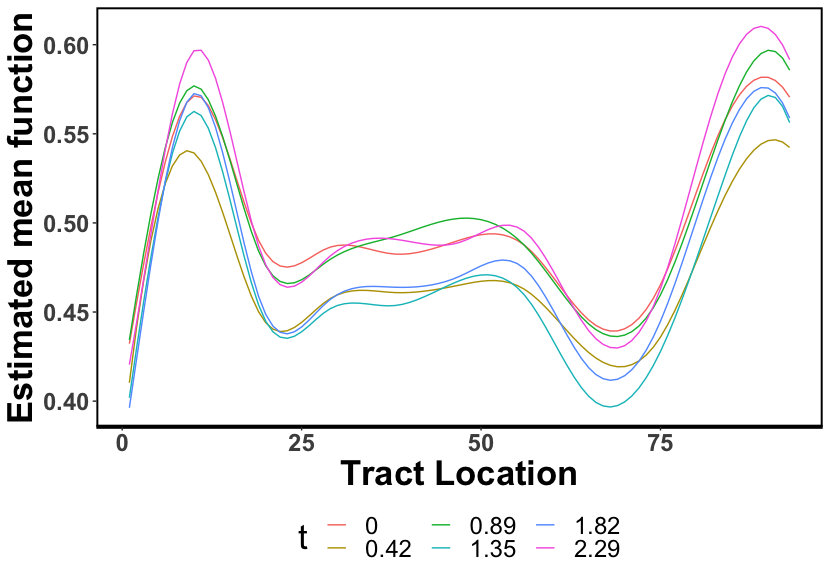}}
    \caption{(a) Estimated smooth bivariate mean $\widehat{\mu}(s,t)$ as a function of the tract location,  (in the y-axis) and the visit times in years (in x-axis). (b) The univariate restriction of estimated mean FA trajectory, $\widehat{\mu}(\cdot, t)$ for different $t$, starting from the baseline $t=0$, and $t=0.42, 0.89, 1.35, 1.82$ and $2.29$ years.}
    \label{fig:EstMean}
\end{figure}

We estimate the bivariate mean function $\widehat{\mu}(s, t)$ nonparametrically using tensor product of univariate B-splines via \verb|mgcv| package in R \citep{wood2017generalized}. The distribution of the hospital visits of the subjects is highly right-skewed, we placed the knots at $20$ points for $t$ based on the quantiles of the visit times $t_{ij}$. For the dense component $s$, the knots are placed at $10$ equidistant points. The smoothing parameter is selected via REML. The smooth version of the estimated bivariate mean function after adjusting for the baseline age of the patients is presented in Figure \ref{fig:EstMean}(a). The contour of estimated FA profile $\widehat{\mu}(\cdot, t)$ at different values of $t$, as presented in figure~\ref{fig:EstMean}(b) shows a preliminary evidence of departure from null hypothesis.  


\begin{figure}
    \centering
    \subfloat[Estimated correlation structure]{{\includegraphics[width=7 cm, height = 5.7 cm]{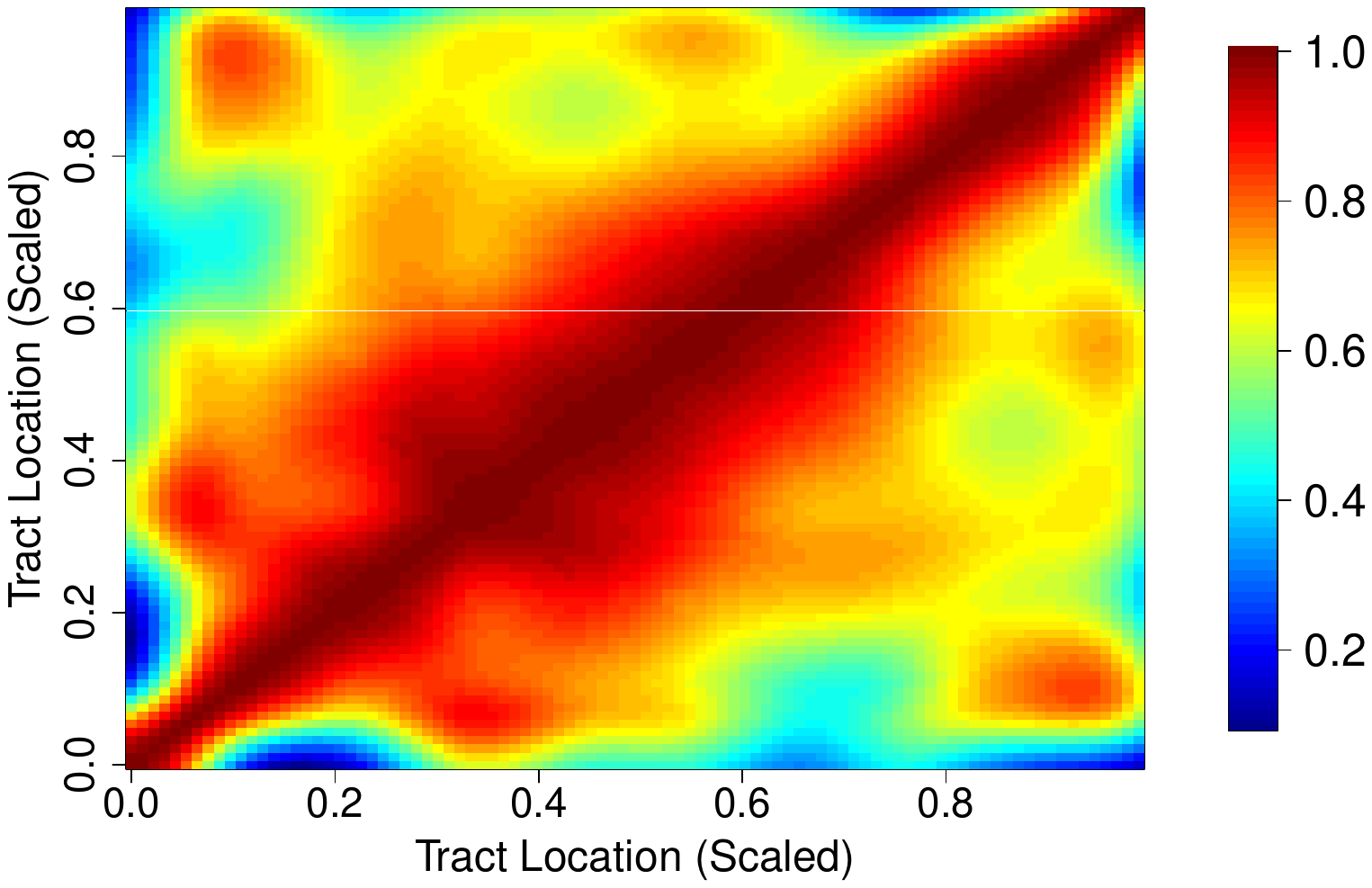} }} \qquad
    \subfloat[Estimated eigen functions]{{
    \includegraphics[scale=0.22]{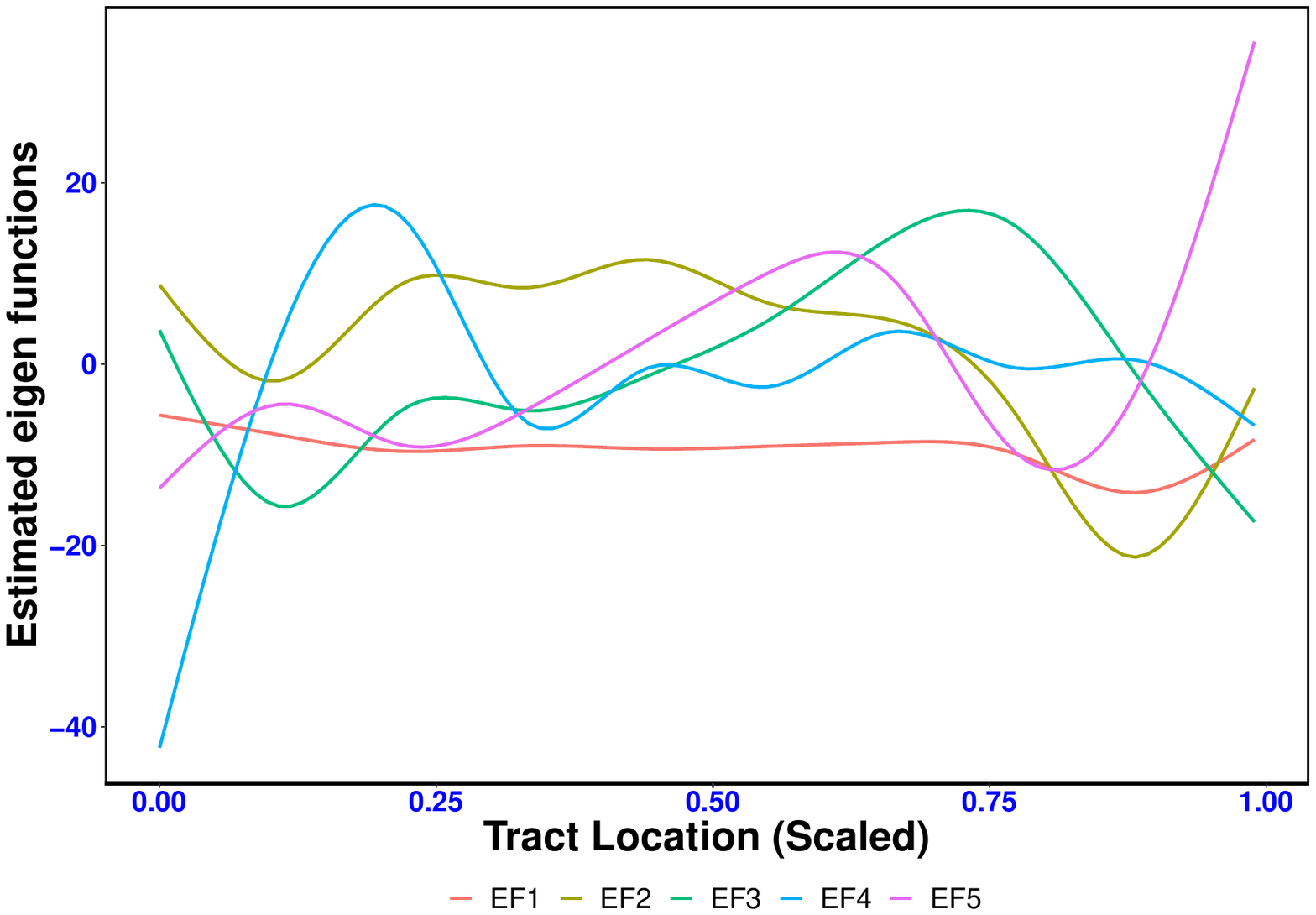} }}
    \caption{(a) Heat map of the estimated marginal correlation structure of FA trajectories as a function of scaled tract location $s,\sprime \in [0,1] \times [0,1]$. (b) Estimated eigenfunctions $\widehat{\phi}_k(s)$ from the spectral decomposition of marginalized covariance function $\widehat{\Xi}(s,s^\prime)$ as a function of $s \in [0,1]$. The leading $5$ eigenfunctions correspond to PVE $= 90\%$}
    \label{fig:EstEigen}
\end{figure}

Once the population level effects are estimated, we demean the response, $Y_{ij}(s) - \widehat \mu(s, t_{ij}) -Age_i\widehat \alpha(s)$ and estimate the marginal covariance function $\widehat{\Xi}(s,\sprime)$ via penalized spline smoother \citep{di2009multilevel} implemented in \verb|fpca.sc| function in \verb|refund| by taking $10$ B-spline basis in the direction of tract location $s$. Figure \ref{fig:EstEigen}(a) presents the heatmap of the estimated marginal correlation structure, indicating a strong correlation along the nearby points in the tract and decaying as the distance increases. The spectral decomposition of the estimated marginal covariance function yields $K=5$ eigenfunctions $\{\widehat{\phi}_k(\cdot) \}_{k=1}^K$ that explain $90\%$ of the data variation; Figure \ref{fig:EstEigen}(b).


For each $k$ in part, let ${W}_{k, ij} = \frac{1}{93}\sum_{r=1}^{93} Y_{ij}(s_r)\widehat{\phi}_k(s_r)$ and consider the implied approximated model 
$
    W_{k,ij} = \eta_k(t_{ij}) + \textrm{Age}_i\alpha_k + \epsilon_{ik}(t_{ij}),
$
where $\alpha_k = \int \alpha(s)\phi_k(s) ds$ be the effect of age in the projected model. PROFIT allows to approximate the testing procedure by simultaneous testing of $5$ simpler hypothesis of the form (\ref{eqn: H04}), $H_{0k}^\prime: \eta_k(t) =\text{constant}$. We use truncated linear splines to model the smooth mean $\eta_k(\cdot)$ and the test continues as explained in Section \ref{subsec: testingprocedure}. The p-values of all the five pseudo-likelihood ratio tests are $< 0.0001, 0.002, 0, 0.24$ and $< 0.0001$ respectively, suggesting a very strong evidence of mean FA profile for MS patients deteriorate over the duration of the study. Additionally it adds insights into the directions along meaningful changes occur. In particular, the testing results jointly with Figure \ref{fig:EstEigen} show evidence that there is a significant horizontal shift in FA as well as a significant contrast between the FA in the middle and at the ends of the WM tract. The results are in agreement with the ones obtained with the single `summed` statistic $T^*_{\text{profit}}$ in~(\ref{eqn: sumstatistics}), where the p-value is $< 0.0001$.  

{\color{black}\section{Conclusion and discussion}
In conclusion, this paper introduces a novel statistical approach for testing the variability of the mean function over time in longitudinally recorded functional data, addressing data dependence both across time and grid points. The method employs orthogonal basis functions derived from the spectral decomposition of the marginal covariance function, leading to a projection-based test with well-established asymptotic properties. The paper also tackles the challenge of establishing an asymptotic uniform convergence rate for the estimated eigenfunctions of the marginal covariance, shedding light on the influence of the smoothing method and sampling plan on this rate. Overall, this work contributes valuable tools to the analysis and understanding of longitudinal functional data, providing a robust and computationally efficient means of assessing temporal changes in the mean function.}

\textcolor{black}{Our proposed methodology relies on representing the mean function $\mu(s,t)$ using an appropriate orthogonal basis system. The choice of basis system or the choice of $K$ does not affect the size performance of the test, but they may affect the power performance. When the space of the residual trajectories, viewed as functions over $\mathcal{S}$, is the same space as $\mu(\cdot, t)$ for every $t$, then using the eigenfunctions of the marginal covariance ensures the optimal selection for the truncation $K$. Increasing PVE too much, thus possibly $K$, may result in a decrease of the power performance of the testing procedure. If the space described by $\mu(\cdot, t)$ for all $t$ includes that of the residuals, then our methodology allows us to recover the optimal basis for the representation of the mean, by choosing a larger PVE. Increasing PVE is likely to result in an increase in the power to detect a correct alternative hypothesis. Specifically, if there is a temporally varying part of the mean function that lies in the space orthogonal to say the leading $K$ eigenfunctions selected with a certain PVE, increasing PVE has the advantage of identifying the orthogonal eigenfunctions that detect the temporal variation of the mean, that was not previously discovered. However, if the space of the residual trajectories is much larger than the space of $\mu(\cdot, t)$ for all $t$, then, not surprisingly, increasing PVE may result in a decrease of power.  }

\textcolor{black}{In practice we recommend to start with a reasonable PVE (say PVE=90\%), identify the corresponding leading eigenfunctions, and then decide on whether or not to increase PVE based on the the $L_2$ difference between the estimated mean and its approximation on the space of the identified leading eigenfunctions. One should increase PVE to larger values, and thus increase $K$, only if the test does not show evidence to reject the null based on the leading eigenfunctions corresponding to the prespecified PVE and the time-varying approximation error $\int_{\mcS} \{\widehat{\mu}(s,t) - \sum_{k=1}^K\widehat{\eta}_k(t)\widehat{\phi}_k(s)\}^2\;ds$ is not constant, or equivalently the initial eigenfunctions are insufficient to accurately capture the estimated mean function variation over time. 
%
}

\section*{Acknowledgments}
The authors would like to acknowledge the support of NSF, grant number DMS 1454942.  

\section*{Disclosure statement}
The authors confirm that there are no relevant financial or non-financial competing interests to report.

\section*{Data availability statement}
The full data that support the findings of this study can not be made available to public due to legal restrictions. However, a part of the full data is available publicly in the R package \verb|refund| \citep{refund}.

\begin{appendix}

\section{Assumptions required for theoretical results}

We list the regularity conditions required for the proof of Theorem~\ref{theorem: unifcov} and \ref{theorem: pLRTnulldist}. Without loss of generality we assume that the timepoints $\spr{t_{ij}}_{j=1}^{m_i}$ and the observation points of the trajectories $s_1, \dots, s_r$ are random. For the rest of the paper, we denote them in capital letter and denote them as $\spr{T_{ij}}_{j=1}^{m_i}$ and $\spr{S_r}_{r=1}^R$ respectively to account for their randomness. 

\vskip 8 pt

\noindent\textbf{Kernel functions and true process}
\begin{enumerate}[label={(A\arabic*)}, ref={(A\arabic*)}]
\item \label{assump: A1} The kernel function $K(\cdot)$ is a $L$-Lipschitz continuous symmetric pdf on $[-1,1]$ with finite second moment $\sigma_K^2 = \int u^2K(u)du$.
\item \label{assump: A2} The observation points $\{S_{r}: r=1,\dots,R\}$ are iid copies of a random variable $S$ with support $\mathcal{S}$ and density $g(s)$ such that
\begin{equation*}
    0 < m_g \leq \underset{s \in \mathcal{S}}{\inf}\; g(s) \leq \underset{s \in \mathcal{S}}{\sup}\; g(s) \leq M_g,
\end{equation*}
and the second derivative of $g(\cdot)$ is bounded in $\mathcal{S}$.
\item \label{assump: A3} The time points $\{T_{ij}: i=1,\dots,n, j=1,\dots, m_i\}$ are iid copies of a random variable $T$ with support $\mathcal{T}$ and density $f(t)$ with
\begin{equation*}
    0 < m_f \leq \underset{t \in  \mathcal{T}}{\inf}\; f(t) \leq \underset{t \in  \mathcal{T}}{\sup}\; f(t) \leq M_f;
\end{equation*}
Moreover, the second derivative of $f(\cdot)$ is bounded in $\mathcal{T}$.
\item \label{assump: A4} The random variable $T$ and $S$ are independently distributed. The error process $\epsilon$ is independent of $S$ and $T$.
\item \label{assump: A5} The true mean function $\mu(s,t)$ is differentiable for each $(s,t) \in \mathcal{S} \times \mathcal{T}$.
\item \label{assump: A6} The marginal covariance function $\Xi(s, \sprime)$ is twice differentiable and all partial second derivatives $\Xi(s, \sprime)$ with respect to $s$ and $\sprime$ are uniformly bounded on $ \mathcal{S} \times  \mathcal{S}$.
\end{enumerate}
Assumptions \ref{assump: A1} is typical in kernel smoothing literature. Assumptions \ref{assump: A2}-\ref{assump: A6} are adaptation of the conditions assumed in \cite{yao2005functional}, and are some of the essential regularity conditions for mean and the marginal covariance function, to prove uniform convergence.

\noindent\textbf{Uniform convergence of covariance functions.}  


\begin{enumerate}[label=(B\arabic*), ref=(B\arabic*)]
    \item \label{assump: B1} $\sup_n (n \;\textrm{max}_i \;m_i v_i) \leq M^{\prime} < \infty$.
    \item \label{assump: B2} $\hxi \to 0$, $\spr{\log(n)\hxi^{-2}/R(R-1)}\sum_{i=1}^n m_i^2v_i^2 \to 0$,  $\spr{\log(n)/R\hxi}\sum_{i=1}^n m_i^2v_i^2 \to 0$, \\ $\log(n)\sum_{i=1}^n m_i^2v_i^2 \to 0$.
    \item[]  Let $\norm{Y} := \underset{s \in \mathcal{S}}{\sup} \abs{Y(s)}$ for univariate (random) function $Y$ and $\norm{Y} := \underset{(s,t) \in \mathcal{S} \times \mathcal{T}}{\sup} \abs{Y(s,t)}$ for a bivariate (random) function $Y$.
    \item \label{assump: B3}  There exists a $\tau > 2$ such that $\ep\norm{\epsilon}^{2\tau} < \infty$ and
 \begin{align*}
     n\tpr{\frac{\log(n)}{n}}^{2/\tau-1}\spr{R(R-1)}^{-1}\hGam^{2}\tpr{1 + (R-2)\hGam + (R-2)(R-3)\hGam^2}\sum_{i=1}^n m_i^2v_i^2 \to \infty.
 \end{align*}
\end{enumerate}
Assumption \ref{assump: B1}-\ref{assump: B3} are very common in the FDA literature and crucial for the uniform convergence of the marginal covariance function. Assumption \ref{assump: B1} is automatically satisfied if $v_i = (nm_i)^{-1}$. Moreover, when $v_i = (\sum_i m_i)^{-1}$ for all $i$, \ref{assump: B1} is satisfied when $\sup_i m_i$ is bounded above (sparse case). Similar versions of Assumption~\ref{assump: B2} are also assumed by \cite{zhang2016sparse}. The moment condition in \ref{assump: B3} is closely related to the continuity of the sample paths of bivariate error process $\epsilon(s,t)$ and assumed by \cite{li2010uniform} and \cite{xiao2020asymptotic}. For example, if the error follows a Gaussian process, this condition is equivalent to having continuous sample paths in any compact interval \citep{landau1970supremum}. 

\vskip 8 pt

\noindent\textbf{Asymptotic null and alternate distribution of pseudo-likelihood ratio statistic} 
\begin{enumerate}[label=(C\arabic*), ref=(C\arabic*)]
        \item \label{assump: C1} The random components $\V{b}_k$ and the error $\V{e}_k$ are jointly Gaussian.
    	\item \label{assump: C2} The minimum eigenvalue of $\bSigma_{Y, k}$ is bounded away from zero as $n$ diverges. Let the estimator $\widehat{\bSigma}_{W,k}$  of $\bSigma_{Y,k}$ satisfies $\ba^{\text{T}} \widehat{\bSigma}_{W,k}^{-1} \ba - \ba^{\text{T}} \bSigma^{-1}_{Y,k} \ba = o_p(1)$, and $\ba^{\text{T}} \widehat{\bSigma}_{W,k}^{-1} \be_{k} - \ba^{\text{T}} \bSigma^{-1}_{Y,k} \be_{k} = o_p(1)$, where $\ba$ is any non-random $N \times 1$ vector of unit norm. 
		\item \label{assump: C3} There exists positive constant, $\varrho$, such that  $N^{-\varrho}\bZ^{\top}\bZ$ and $N^{-\varrho}\bX^{\top}\bX$ converge to non-zero matrices. For every eigenvalue $\xi_{N, kq}$, $\zeta_{N, kq}$ of the matrices $N^{-\varrho} \bZ^{\top} \bSigma^{-1}_{Y,k}\bZ$, $N^{-\varrho} \{ \bZ^{\top}\bSigma^{-1}_{Y,k}\bZ - \bZ^{\top} \bSigma^{-1}_{Y,k}\bX (\bX^{\top}\bSigma^{-1}_{Y,k}\bX)^{-1}  \bX^{\top} \bSigma^{-1}_{Y,k}\bZ \}$ respectively,  $\xi_{N,kq} \rightarrow\xi_{ kq}$, $\zeta_{N,kq} \rightarrow \zeta_{kq}$ for some $\{\xi_{ kq}\}_{q=1}^Q$ and $\{\zeta_{kq}\}_{q=1}^Q$  that are not all zero.


		\item \label{assump: C4} {\color{black}Partition the design matrix as $\bX = [\V{1}_N \mid \bX_{(2)}]$.  Define, $\bH = \bSigma^{-1/2}_{Y,k}\{\bX (\bX^{\top}\bSigma^{-1}_{Y,k}\bX)^{-1}  \bX^{\top} -  \V{1} (\V{1}^{\top}\bSigma^{-1}_{Y,k}\V{1})^{-1}  \V{1}^{\top}  \}\bSigma^{-1/2}_{Y,k}$ to be the rank $1$ projection matrix. Let $\varphi_{N,k}$ be the non-zero eigenvalue of  $N^{-\varrho} \bZ^{\top}\bSigma^{-1/2}_{Y,k}\bH\bSigma^{-1/2}_{Y,k}\bZ$, and $\theta_{N,k} = N^{-\varrho} \bX_{(2)}^{\top}\bSigma^{-1/2}_{Y,k}\bH \bSigma^{-1/2}_{Y,k}\bX_{(2)}$. For every $k=1,\dots, K$, $\lim_{N \to \infty} \varphi_{N,k} =  \varphi_{k}$, and $\lim_{N \to \infty} \theta_{N,k} =  \theta_{k} \geq 0$, with at least one of $\varphi_{k}$ and $\theta_{k}$ being non-zero, for some $k$.}
  

\end{enumerate}
Assumptions \ref{assump: C1}-\ref{assump: C3}, inspired from \cite{staicu2014likelihood}, relates the elements to the projected model~(\ref{eqn: mixedmodel}) and necessary to derive the null distribution of PROFIT. It is important to note that the Gaussian assumption in \ref{assump: C1} is taken for the unobserved projected response $Y_{k,ij}$, not for the quasi projection the $W_{k,ij}$, which is a function of the entire data. Moreover, $e_{ijk} = \int \epsilon_i(s,t_{ij})\phi_k(s)ds$ will be Gaussian if the error $\epsilon(s, t_{ij})$ in~(\ref{eqn: assumedmodel}) is a Gaussian process with continuous sample paths for every $t_{ij}$ and the eigen function $\phi_k(s)$ is continuous in $\mathcal{S}$, which is true by Assumption~\ref{assump: A6} and application of the Mercer's theorem. Assumption~\ref{assump: C4} is specific to computation of the local asymptotic power of pseudo-likelihood ratio statistic, and similar to those specified in Assumption~\ref{assump: C3}.  

Assumption~\ref{assump: C2} is similar to the condition of \cite{staicu2014likelihood} but $\widehat{\bSigma}_{Y,k}$ is replaced by $\widehat{\bSigma}_{W,k}$. This is connected to consistency in the estimation of the true covariance matrix $\bSigma_{Y,k}$ based on the quasi projections $W_{k,ij}$. \cite{staicu2014likelihood} identified sufficient conditions for~\ref{assump: C2} to hold under different sampling design and estimation procedure of the covariance. For example, in the case of sparse functional data, they remarked that when the estimation of $\bSigma_{Y,k}$ is done through KL expansion (Proposition 3.2 and 3.3) of $\epsilon_{ik}(t_{ij})$ as  $\tpr{\widehat{\bSigma}_{W,ki}}_{j\jprime} = \sum_{\ell=1}^{L_{k}}\widehat{\nu}_{k\ell}\widehat{\psi}_{k\ell}(t_{ij})\widehat{\psi}_{k\ell}(t_{i\jprime})$, if $\spr{\psi_{k\ell}(t)}_{\ell \geq 1}$, eigen function of the covariance $\gamma_k(t,\tprime)$, are estimated uniformly at a certain rate then \ref{assump: C2} holds automatically. Following \cite{yao2005functional} or \cite{li2010uniform} a uniform convergence rate of order $n^{-\alpha}$ can be derived under usual regularity conditions if $\spr{\psi_{k\ell}(t)}_{\ell \geq 1}$ are estimated by the unobserved projected response $Y_{k,ij}$. However, following the implications of Theorem~\ref{theorem: unifcov}, a similar uniform convergence rate can also be achieved even when $\psi_{k\ell}(t)$ are estimated using the quasi projections $W_{k,ij}$, implying that condition~\ref{assump: C2} is reasonable even when $\bSigma_{Y,k}$ is estimated by $\widehat{\bSigma}_{W,k}$. We provide a mathematical justification for this in Section~\ref{sec: supp_IntJustSigma} of the Supplementary material.

\section{Representation of the mean function using orthogonal basis in $L^2[\mathcal{S}]$ }
\textcolor{black}{
Not any orthogonal basis in the space of square integrable functions $L^2[\mathcal{S}]$ would allow to represent $\mu(s,t)$ in this way. Specifically,  define $G(s,s') = \int_{\mathcal{T}} \{\mu(s,t) - \bar\mu(s)\} \{\mu(s',t) - \bar\mu(s')\} dt$, where $\bar\mu(s) = \int_{\mathcal{T}} \mu(s,t) dt$, which exists as $\mu(\cdot, \cdot)$ is continuous and $\mathcal{
T}$ is a closed interval. The quantity $G(\cdot, \cdot)$ is continuous, symmetric and positive semi-definite, on $\mathcal{S}^{\otimes2}$ and thus it can be decomposed using a spectral decomposition. Let  $\{\phi_k(\cdot)\}_k$ be the orthogonal basis identified from the spectral decomposition of $G(\cdot, \cdot)$: i.e. $G(s,s') =\sum_{k\geq 1} \lambda_k \phi_k(s) \phi_k(s') $ where $\{\phi_k(\cdot)\}$ is an orthonormal basis in $L^2[\mathcal{S}]$ arranged to correspond to the decreasing order of $\lambda_1\geq \lambda_2>\ldots\geq 0$. One can readily see that this basis allows us to represent
$$
\mu(s,t) = \sum_{k\geq 1} \phi_k(s)\eta_k(t)
$$
and the summation holds in $L^2$, as $\int_\mathcal{S} \int_\mathcal{T}  \left \{ \mu(s,t) - \sum_{k=1}^K \phi_k(s)\eta_k(t)\right\}^2 dsdt \rightarrow 0$ for $K\rightarrow \infty$. 
}

\end{appendix}

\begin{table}
	\centering
	\caption{The empirical Type-1 error rates of PROFIT using the `summed' statistic $T^*_{\text{profit}}$ based on 10,000 simulations. Standard errors are presented in parentheses.}  
	\scalebox{1}{
		\begin{tabular}{cc cccc}
			\hline	\hline
			\multicolumn{6}{c}{$m_i \sim \{8, \ldots, 12\}$} \\	\hline
			& & $\alpha = 0.01$ & 	$\alpha = 0.05$ & 	$\alpha = 0.10$ & 	$\alpha = 0.15$ \\ 
			\hline	\hline
	$n = $	100	 &&  0.012  (0.001) & 0.059  (0.002) & 0.114  (0.003) & 0.167  (0.004) \\  
	$n = $	150	 &&  0.011  (0.001) & 0.053  (0.002) & 0.105  (0.003) & 0.159  (0.004) \\ 
	$n = $	200	 && 0.011  (0.001) & 0.048  (0.002) & 0.099  (0.003) & 0.151  (0.004) \\  
	$n = $	300	 &&0.009  (0.001) & 0.047  (0.002) & 0.096  (0.003) & 0.146  (0.003) \\  
    $n = $	400	 && 0.010  (0.001) & 0.050  (0.002) & 0.098  (0.003) & 0.148  (0.003) \\ 
			\hline	\hline
			\multicolumn{6}{c}{$m_i \sim \{15, \ldots, 20\}$} \\	\hline
		&& $\alpha = 0.01$ & 	$\alpha = 0.05$ & 	$\alpha = 0.10$ & 	$\alpha = 0.15$ \\ 
			\hline	\hline
	$n = $100	  &  & 0.011  (0.001) & 0.053  (0.002) & 0.105  (0.003) & 0.156 (0.004) \\ 
	$n = $150	  &  & 0.009  (0.001) & 0.049  (0.002) & 0.102  (0.003) & 0.153 (0.004) \\  
	$n = $200	&     & 0.009  (0.001) & 0.048  (0.002) & 0.099  (0.003) & 0.146  (0.003)\\  
	$n = $300  &  &0.008  (0.001) & 0.047  (0.002) & 0.098  (0.003) & 0.148  (0.004) \\  
	$n = $400	& & 0.008  (0.001) & 0.046  (0.002) & 0.097  (0.003) & 0.149  (0.004)  \\
			\hline\hline
		\end{tabular}
	} \label{tab: proposed only sumstat}
\end{table}

\begin{table}
	\centering
	\caption{The empirical Type-1 error rates of PROFIT using the `summed' statistic $T^*_{\text{profit}}$ for non-Gaussian errors under low sparsity level, based on 10,000 simulations. Standard errors are presented in parentheses. }  
	\scalebox{1}{
		\begin{tabular}{cc cccc}
			\hline	\hline
			\multicolumn{6}{c}{Scenario (i) : $r_{ij,k} \sim$  mixture-normal} \\	\hline
			& & $\alpha = 0.01$ & 	$\alpha = 0.05$ & 	$\alpha = 0.10$ & 	$\alpha = 0.15$ \\ 
			\hline	\hline
	$n = $	100	 &&  0.009  (0.001) & 0.053  (0.003) & 0.104  (0.004) & 0.158  (0.005) \\  
	$n = $	150	 &&  0.010  (0.001) & 0.050  (0.003) & 0.102  (0.004) & 0.154  (0.005) \\ 
	$n = $	200	 && 0.010  (0.001) & 0.043 (0.003) & 0.093 (0.004) & 0.146 (0.005) \\  
	$n = $	300	 && 0.009  (0.001) & 0.048 (0.003) & 0.102 (0.004) & 0.153 (0.005) \\  
    $n = $	400	 && 0.011  (0.001) & 0.053 (0.003) & 0.106 (0.004) & 0.156 (0.005) \\
			\hline	\hline
			\multicolumn{6}{c}{Scenario (ii) : $\zeta_{i,k\ell} \sim$  mixture-normal} \\	\hline
		&& $\alpha = 0.01$ & 	$\alpha = 0.05$ & 	$\alpha = 0.10$ & 	$\alpha = 0.15$ \\ 
			\hline	\hline
	$n = $	100	 &&  0.006  (0.001) & 0.047  (0.003) & 0.094  (0.004) & 0.144  (0.005) \\  
	$n = $	150	 &&  0.011  (0.001) & 0.056  (0.003) & 0.109  (0.004) & 0.159  (0.005) \\ 
	$n = $	200	 && 0.010  (0.001) & 0.054 (0.003) & 0.100 (0.004) & 0.152 (0.005) \\  
	$n = $	300	 &&0.009  (0.001) & 0.048 (0.003) & 0.097 (0.004) & 0.146 (0.005) \\  
    $n = $	400	 && 0.011  (0.001) & 0.050  (0.003) & 0.103  (0.004) & 0.152  (0.005) \\ 
			\hline\hline
		\end{tabular}
	} \label{tab: proposed only NonGauss sumstat}
\end{table}

\begin{figure}
    \centering
    \includegraphics[scale=0.45]{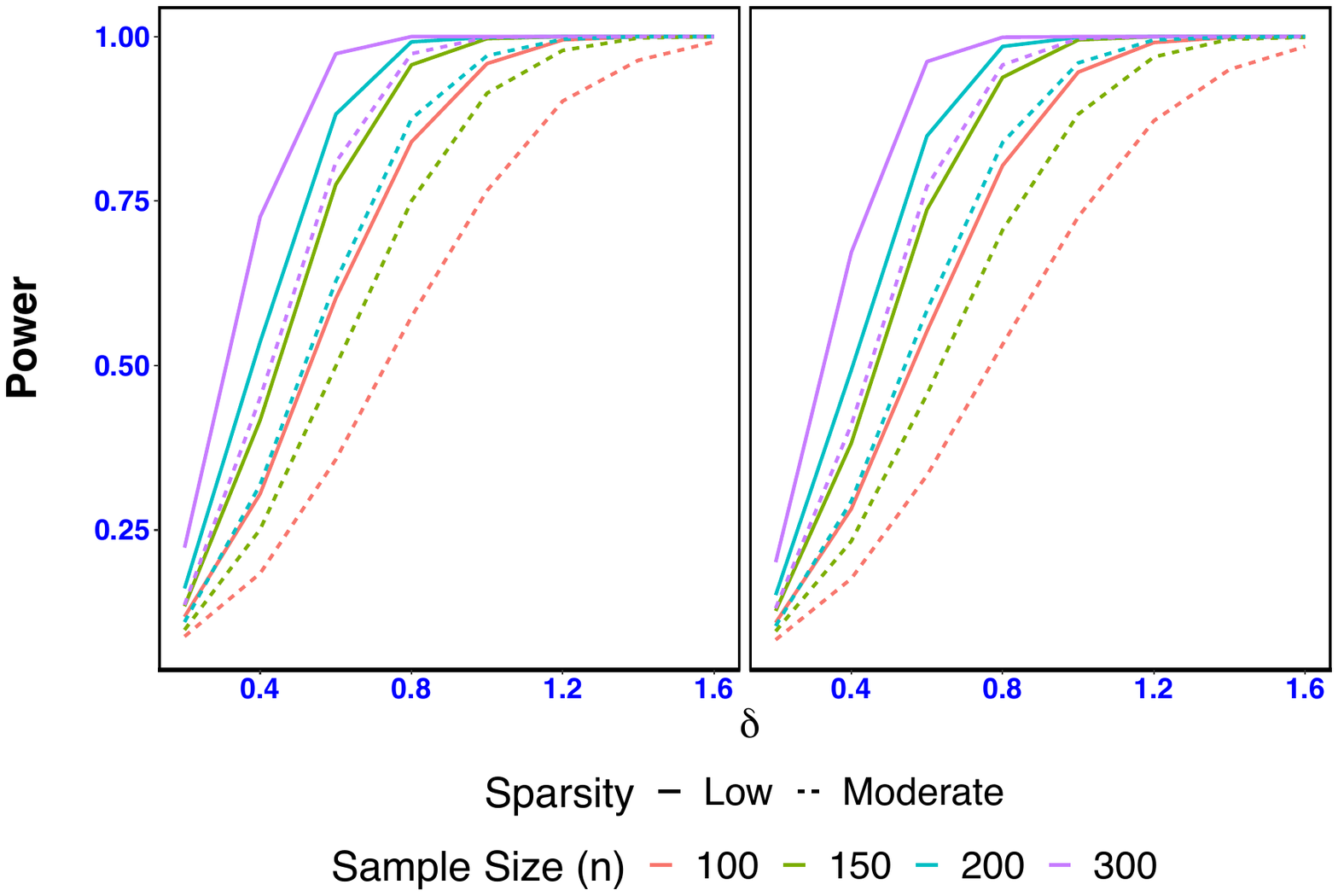}
    \caption{Empirical power curve for PROFIT using the `summed' test-statistic $T^*_{\text{profit}}$ (left panel) and the usual Bonferroni-correction rejection region (right panel) as function of $\delta$ across sample size $n$ varying from $100$ to $300$ when the sparsity levels of the visit times are \textit{low} (solid line) and \textit{moderate} (dashed line). The numbers are based on $5000$ simulations.}
    \label{fig:powercurve_sum}
\end{figure}

\bibliographystyle{agsm} 
\bibliography{ref}

\newpage

\begin{center}
\bf Supplementary material for `PROFIT: Projection-based test in longitudinal functional data'
\end{center}


\renewcommand{\thesection}{S\arabic{section}}
\renewcommand{\thefigure}{A\arabic{figure}}
\renewcommand{\thetable}{A\arabic{table}}







This supplementary material is divided into four parts. In Section~\ref{sec: supp_proof_unifconv}, we present the proof of Theorem~\ref{theorem: unifcov}, followed by the proof of Theorem~\ref{theorem: pLRTnulldist} in Section~\ref{sec: supp_proof_pLRT}. The proof of Theorem~\ref{theorem: pLRTAltdist} follows the similar steps of Theorem~\ref{theorem: pLRTnulldist} and Theorem 1 of \cite{crainiceanu2005exact}, hereby omitted. Section~\ref{sec: supp_IntJustSigma} provides a detailed justification of the feasibility of the Assumption~\ref{assump: C2} introduced in Appendix. Additional simulation result are presented in Section~\ref{sec: supp_additionalsimresults}.


\section{Proof of Theorem \ref{theorem: unifcov}} \label{sec: supp_proof_unifconv}
We will prove a general result that can be applied to the case when the functional trajectory is observed over sparse, dense or ultra-dense grid and under a general weighting scheme $v_i$. Let $w_i = v_i/R(R-1)$ be the weights applied to each observations in the smoother. Without loss of generality, assume that $\mathcal{S} = [0,1]$ and $\mathcal{T} =[0,1]$. Under the assumptions specified in Theorem~\ref{theorem: unifcov} we prove that
\begin{align*}
    \underset{s,\sprime \in [0,1]}{\sup} \abs{\widehat{\Xi}(s,\sprime) - \Xi(s,\sprime)} &= \mathcal{O}\left(\left\{\log(n)\left[\frac{R(R-1)}{\hxi^{2}} + \frac{R(R-1)(R-2)}{\hxi} \right. \right. \right. \\
    & \quad + \left. \left. \left.  R(R-1)(R-2)(R-3)\right]\sum_{i=1}^n m_i^2w_i^2\right\}^{1/2} + \hxi^2 + \alpha_n \right) \;\; a.s.
\end{align*}

The rate specified in Theorem~\ref{theorem: unifcov} follows from the above general rate if the order of the kernel bandwith parameter $\hxi$, number of repeated measures $m_i$, $R$ and the weights $v_i$ satisfies the conditions specified in the statement of theorem.

\begin{proof}
The proof of the theorem mimics the general steps for the uniform convergence of covariance functions established by \cite{zhang2016sparse}. At first we write the difference between $\widehat{\Xi}(s,\sprime)$ and $\Xi(s,\sprime)$ in a compact form, that involves the quadratic form 
Define, $K_h(\cdot) = (1/h)K(\cdot/h)$ and
$$
R_{pq}(s,\sprime) = \sum_{i=1}^n w_i \sum_{j=1}^{m_i} \sum_{1 \leq r \neq \rprime \leq R}  K_{\hxi}\fpr{S_r - s}K_{\hxi}\fpr{S_{\rprime} - \sprime} \fpr{\frac{S_r - s}{\hxi}}^p\fpr{\frac{S_{\rprime} - \sprime}{\hxi}}^q \widetilde{Y}_{ijr}\widetilde{Y}_{ij\rprime},
$$
and 
$$
V_{pq}(s, \sprime) = \frac{1}{R(R-1)}\sum_{1 \leq r \neq \rprime \leq R}  K_{\hxi}\fpr{S_r - s}K_{\hxi}\fpr{S_{\rprime} - \sprime} \fpr{\frac{S_r - s}{\hxi}}^p\fpr{\frac{S_{\rprime} - \sprime}{\hxi}}^q. 
$$
For brevity we might omit the dependence of $s$ and $\sprime$ in $R_{pq}(s,\sprime)$ and $V_{pq}(s,\sprime)$ and just simply write $R_{pq}$ and $V_{pq}$ respectively. By standard weighted least squares calculations (tedious but straight forward), we get
\begin{equation}\label{eqn:betahat}
    \widehat{a}_0 = \widehat{\Xi}(s, \sprime) =\frac{\fpr{V_{20}V_{02}-V_{11}^2}R_{00} - \fpr{V_{10}V_{02}-V_{01}V_{11}}R_{10} - \fpr{V_{01}V_{20}-V_{10}V_{11}}R_{01}}{\fpr{V_{20}V_{02}-V_{11}^2}V_{00} - \fpr{V_{10}V_{02}-V_{01}V_{11}}V_{10} - \fpr{V_{01}V_{20}-V_{10}V_{11}}V_{01}}. 
\end{equation}
Let's call the denominator of the expression in (\ref{eqn:betahat}) as $D_n$. After some simplification we write, 
\begin{align}
    \nonumber &\widehat{\Xi}(s, \sprime) - \Xi(s,\sprime) \\
    \nonumber &= D_n^{-1}\left\{ \fpr{V_{20}V_{02}-V_{11}^2}\tpr{R_{00} - \Xi(s,\sprime) V_{00} - \hxi \frac{\partial \Xi}{\partial s}(s,\sprime)V_{10} - \hxi \frac{\partial \Xi}{\partial \sprime}(s,\sprime)V_{01}} \right. \\
    \nonumber & \;\;\qquad\quad -  \fpr{V_{10}V_{02}-V_{01}V_{11}}\tpr{R_{10} - \Xi(s,\sprime) V_{10} - \hxi \frac{\partial \Xi}{\partial s}(s,\sprime)V_{20} - \hxi \frac{\partial \Xi}{\partial \sprime}(s,\sprime)V_{11}} \\
    & \;\;\qquad\quad \left. - \fpr{V_{01}V_{20}-V_{10}V_{11}}\tpr{R_{01} - \Xi(s,\sprime) V_{01} - \hxi \frac{\partial \Xi}{\partial s}(s,\sprime)V_{11} - \hxi \frac{\partial \Xi}{\partial \sprime}(s,\sprime)V_{02}} \right\} \label{eqn: Xiform}.
\end{align}
Before moving on to the lines of the proof in details, we will provide an outline of the proof. 
\begin{enumerate}
    \item[(A)] First, we will show the rate of convergence of each of the term in the square braces in equation (\ref{eqn: Xiform}). The rate of convergence for all the three terms in the square braces can be shown in exactly the same way, hence we will show it only for the first term in the square bracket. Note that, only the terms of the form $R_{pq}$ involves the response $Y_{ijr}$. To derive the convergence of the terms involving $R_{pq}$,
It is helpful to observe that by the assumed model in (\ref{eqn: assumedmodel}), 
{\small\begin{align*}
    &\widetilde{Y}_{ijr}\widetilde{Y}_{ij\rprime} \\ &= \fpr{Y_{ijr} - \mu(S_r, T_{ij}) + \mu(S_r, T_{ij}) - \widehat{\mu}(S_r, T_{ij})}\fpr{Y_{ij\rprime} - \mu(S_{\rprime}, T_{ij}) + \mu(S_{\rprime}, T_{ij}) - \widehat{\mu}(S_{\rprime}, T_{ij})} \\
    &=\epsilon_{ijr}\epsilon_{ij\rprime} +  \epsilon_{ijr}\fpr{\mu(S_{\rprime}, T_{ij}) - \widehat{\mu}(S_{\rprime}, T_{ij})}  +  \epsilon_{ij\rprime}\fpr{\mu(S_{r}, T_{ij}) - \widehat{\mu}(S_r, T_{ij})}  \\
    & \qquad + \fpr{\mu(S_{r}, T_{ij}) - \widehat{\mu}(S_r, T_{ij})}\fpr{\mu(S_{\rprime}, T_{ij}) - \widehat{\mu}(S_{\rprime}, T_{ij})}.
\end{align*}}
Using the above expansion and by direct mathematical calculation it  follows that 
{\begin{align}
    \nonumber &R_{00} - \Xi(s,\sprime)V_{00} - \hxi V_{10}\frac{\partial }{\partial s}\Xi(s,\sprime) - \hxi V_{01}\frac{\partial }{\partial \sprime}\Xi(s,\sprime) \\
    &= \Psi_1(s, \sprime) + \Psi_2(s, \sprime) + \Psi_3(s, \sprime) + \hxi^{-2} L(s, \sprime) + \hxi^{-2} H(s, \sprime) + A_1(s, \sprime) + A_2(s, \sprime), \label{eqn: breakdown}
\end{align}}
where each of the terms defined above are as follows
{\begin{align} 
    \Psi_1(s, \sprime) &= \sum_{i=1}^n w_i \sum_{j=1}^{m_i} \sum_{r \neq \rprime}  \epsilon_{ij\rprime}\fpr{\mu(S_{r}, T_{ij}) - \widehat{\mu}(S_r, T_{ij})}  K_{\hxi}\fpr{S_r - s}K_{\hxi}\fpr{S_{\rprime} - \sprime} \label{eqn: psi1}; \\
   \Psi_2(s, \sprime) &=   \sum_{i=1}^n w_i \sum_{j=1}^{m_i} \sum_{r \neq \rprime}  \epsilon_{ijr}\fpr{\mu(S_{\rprime}, T_{ij}) - \widehat{\mu}(S_{\rprime}, T_{ij})} K_{\hxi}\fpr{S_r - s}K_{\hxi}\fpr{S_{\rprime} - \sprime} \label{eqn: psi2}; \\
    \nonumber \Psi_3(s, \sprime) &= \sum_{i=1}^n w_i \sum_{j=1}^{m_i} \sum_{1 \leq r \neq \rprime \leq R} \fpr{\mu(S_{r}, T_{ij}) - \widehat{\mu}(S_r, T_{ij})}\fpr{\mu(S_{\rprime}, T_{ij}) - \widehat{\mu}(S_{\rprime}, T_{ij})} \\
    & \quad \quad \quad \quad \quad \quad \quad  \times K_{\hxi}\fpr{S_r - s}K_{\hxi}\fpr{S_{\rprime} - \sprime} \label{eqn: psi3}; \\
    L(s,\sprime) &=  \sum_{i=1}^n w_i \sum_{j=1}^{m_i} \sum_{r \neq \rprime} \spr{ \epsilon_{ijr} \epsilon_{ij\rprime} - c(S_r,S_{\rprime},T_{ij})-\Gamma(S_r,S_{\rprime})} K\fpr{\frac{S_r - s}{\hxi}}K\fpr{\frac{S_{\rprime} - \sprime}{\hxi}} \label{eqn: Ls}; \\
    H(s,\sprime) &=  \sum_{i=1}^n w_i \sum_{j=1}^{m_i} \sum_{r \neq \rprime} K\fpr{\frac{S_r - s}{\hxi}}K\fpr{\frac{S_{\rprime} - \sprime}{\hxi}} \spr{c(S_r,S_{\rprime},T_{ij}) - \int c(S_r,S_{\rprime},t)g(t)dt} \label{eqn: Hs}; \\
    \nonumber A_1(s,\sprime) &= \frac{1}{R(R-1)}\sum_{r \neq \rprime}  K_{\hxi}\fpr{S_r - s}K_{\hxi}\fpr{S_{\rprime} - \sprime}\left\{  \int \left[c(S_r, S_{\rprime}, t) - c(s,\sprime, t) \right. \right. \\
    & \qquad \qquad \left. \left. - (S_r - s)\frac{\partial }{\partial s}c(s,\sprime, t) - (S_{\rprime} - \sprime)\frac{\partial }{\partial \sprime}c(s,\sprime, t) \right] f(t)dt \right\} \label{eqn: A1s}; \\
    \nonumber A_2(s,\sprime) &= \frac{1}{R(R-1)}\sum_{r \neq \rprime}  \left[ \Gamma(S_r, S_\rprime) -(S_r-s)\frac{\partial }{\partial s}\Gamma(s,\sprime) - (S_{\rprime}-\sprime)\frac{\partial }{\partial \sprime}\Gamma(s,\sprime) \right] \\
    & \quad \qquad \qquad \qquad \qquad \times K_{\hxi}\fpr{S_r - s}K_{\hxi}\fpr{S_{\rprime} - \sprime}
    \label{eqn: A2s}.
\end{align}}

The convergence rate of the term in~(\ref{eqn: breakdown}) is obtained by the rates of each of the terms in equations (\ref{eqn: psi1})-(\ref{eqn: A2s}).

\item[(B)] In the second step, we show that each of the terms $V_{pq}, p = 0, 1, 2; q = 0, 1, 2$ are uniformly bounded
and the denominator $D_n$ is bounded away from zero as $n$ goes to $\infty$.
\end{enumerate}

Now, we will proceed to proving Step A. In each of the sub-steps below we will derive the rate of convergence of the terms in equations (\ref{eqn: psi1})-(\ref{eqn: A2s}). \\

\noindent\textbf{Step A1}: We will show that
\begin{equation}\label{eqn:rateLs}
    \underset{s,\sprime \in [0,1]}{\sup} \abs{L(s,\sprime)} = \mathcal{O}(b_n(\hxi)) \qquad a.s,
\end{equation}
where, 
$$b_n(\hxi) = \fpr{\log(n) \spr{R(R-1)\hxi^2 + R(R-1)(R-2)\hxi^3 +  R(R-1)(R-2)(R-3)\hxi^4} \sum_{i=1}^n w_i^2 m_i^2 }^{1/2}.$$
\begin{proof}
With an abuse of notation,
$\tepsilon_{ijr\rprime} := \epsilon_{ijr} \epsilon_{ij\rprime} - c(S_r,S_{\rprime},T_{ij})-\Gamma(S_r,S_{\rprime})$
and
$
B_n(\hxi) = \fpr{\log n}^{-1}nb_n(\hxi)
$.
To prove the rate, we will truncate $L(s,\sprime)$ in two parts as
$$
L(s,\sprime) = L_1(s, \sprime) + L_2(s,\sprime),
$$
where
$$
L_1(s,\sprime) = \sum_{i=1}^n w_i \sum_{j=1}^{m_i} \sum_{1 \leq r \neq \rprime \leq R} K\fpr{\frac{S_r - s}{\hxi}}K\fpr{\frac{S_{\rprime} - \sprime}{\hxi}} \tepsilon_{ijr\rprime}I\fpr{\abs{\tepsilon_{ijr\rprime}} \leq B_n(\hxi)},
$$
and
$$
L_2(s,\sprime) = \sum_{i=1}^n w_i \sum_{j=1}^{m_i} \sum_{1 \leq r \neq \rprime \leq R} K\fpr{\frac{S_r - s}{\hxi}}K\fpr{\frac{S_{\rprime} - \sprime}{\hxi}} \tepsilon_{ijr\rprime}I\fpr{\abs{\tepsilon_{ijr\rprime}} > B_n(\hxi)}.
$$
The asymptotic convergence rate for $L_1$ will be shown by an application of Bernstein concentration inequality for bounded random variables in conjunction with Borel-Cantelli lemma and for $L_2$ we will show that the summand can not grow very fast using Assumption~\ref{assump: B3}. We will first derive the rates for $L_2(s,\sprime)$. Note that,
$$
\abs{\tepsilon_{ijr\rprime}} \leq  \fpr{\underset{s, t \in [0,1]}{\sup}\abs{\epsilon_i(s,t)}}^2 +  \underset{s , t \in [0,1]}{\sup} c(s,\sprime,t) + \underset{s, \sprime \in [0,1]}{\sup}\Gamma(s,\sprime).
$$
Because $\tau > 2$, using the fact that the function $x \mapsto x^\tau$ is convex for $x > 0$, we get $(a+b+c)^\tau \leq 3^{\tau-1}(a^\tau + b^\tau + c^\tau) $ for $a, b, c > 0$. Using the inequality,
\begin{align*}
   \abs{\tepsilon_{ijr\rprime}}I\fpr{\abs{\tepsilon_{ijr\rprime}} > B_n(\hxi)} &\leq \fpr{B_n(\hxi)}^{1-\tau}\abs{\tepsilon_{ijr\rprime}}^{\tau} \\
   &\leq \fpr{B_n(\hxi)}^{1-\tau}3^{\tau-1}\spr{ \norm{\epsilon_i}^{2\tau} + \norm{c}^{\tau} + \norm{\Gamma}^{\tau}}.
\end{align*}
By Assumption~\ref{assump: A1}, $K(x) \leq M_K$ for all $x \in [0,1]$ for some constant $M_K > 0$. By~\ref{assump: A6}, $\norm{c} < \infty$ and $\norm{\Gamma}< \infty$. Then,
\begin{align*}
    \abs{L_2(s,\sprime)} &= \sum_{i=1}^n w_i \sum_{j=1}^{m_i} \sum_{1 \leq r \neq \rprime \leq R} K\fpr{\frac{S_r - s}{\hxi}}K\fpr{\frac{S_{\rprime} - \sprime}{\hxi}} \abs{\tepsilon_{ijr\rprime}}I\fpr{\abs{\tepsilon_{ijr\rprime}} > B_n(\hxi)} \\
    & \leq M_K^2  \fpr{B_n(\hxi)}^{1-\tau}3^{\tau-1}\fpr{\sup_n n \;\max_{i} w_i m_i R(R-1)} \tpr{\frac{1}{n}\sum_{i=1}^n\norm{\epsilon_i}^{2\tau} + \norm{c}^{\tau} + \norm{\Gamma}^{\tau}} \\
    &\leq M_K^2 M^{\prime} \fpr{B_n(\hxi)}^{1-\tau}3^{\tau-1} \tpr{\frac{1}{n}\sum_{i=1}^n\norm{\epsilon_i}^{2\tau} + \norm{c}^{\tau} + \norm{\Gamma}^{\tau}};
\end{align*}
By strong law of large numbers $\frac{1}{n}\sum_{i=1}^n\norm{\epsilon_i}^{2\tau} \to \ep\norm{\epsilon_i}^{2\tau} < \infty$. By Assumption~\ref{assump: B1}, \ref{assump: B3}, we can conclude 
\begin{equation}\label{eqn:rateL2}
    \underset{s, \sprime \in [0,1]}{\sup} \abs{L_2(s,\sprime)} = o(b_n) \qquad a.s.
\end{equation}
 Now we move onto proving the rate for $L_1(s,\sprime)$. Since we want to show the rate for the supremum over an uncountable set, $[0,1]$, we will partition the interval into a finite grid and control for each of them. By Assumption~\ref{assump: B2}, there exists a $\rho > 0$ such that $n^{\rho}\hxi b_n(\hxi) \to \infty$. Let $\chi(\rho)$ be an equidistant partition of $[0,1]$ with grid length $n^{-\rho}$. That is $\chi(\rho) := \spr{n^{-\rho}, 2n^{-\rho}, \dots, 1 - n^{-\rho}, 1}$. 
 Note that, for every $s,\sprime \in [0,1]$ there exists a $s_1, s_{1}^{\prime} \in \chi(\rho)$ such that $\abs{s-s_1} \leq n^{-\rho}$ and $\abs{\sprime-s_{1}^{\prime}} \leq n^{-\rho}$. By triangle inequality,
\begin{align*}
     \abs{L_1(s,\sprime) } &\leq \abs{L_1(s_1,\sprime_1) } + \abs{L_1(s_1,\sprime_1) - L_1(s,\sprime)} \\ \\
     &\leq \underset{s,\sprime \in \chi(\rho)}{\sup} \abs{L_1(s,\sprime)} + \underset{\abs{s-s_1},\abs{\sprime-\soprime} \leq n^{-\rho}; s, \sprime \in [0,1] }{\sup} \abs{L_1(s,\sprime) - L_1(s_1,\soprime)}
\end{align*}
 Taking supremum over $s$ and $\sprime$ on the left hand side of the above equation,
\begin{equation}
    \underset{s,\sprime \in [0,1]}{\sup} \abs{L_1(s,\sprime) } \leq \underset{s,\sprime \in \chi(\rho)}{\sup} \abs{L_1(s,\sprime)} +  \underset{\abs{s-s_1},\abs{\sprime-\soprime} \leq n^{-\rho}; s, \sprime \in [0,1] }{\sup} \abs{L_1(s,\sprime) - L_1(s_1,\soprime)}. \label{eqn: supL1}
\end{equation}
We will control each of the two terms above separately. The second term will be controlled using that the length between $s$ and $s_1$ does not exceed $n^{-\rho}$, the kernel is Lipschitz continuous and the fourth moment of $\widetilde{\epsilon}_{ijr\rprime}$ is bounded. We will apply Bernstein inequality to control the rate for the first term. First we will control the second term as it is easier to handle. Define,
\begin{equation*}
    D_1 := \underset{\abs{s-s_1},\abs{\sprime-\soprime} \leq n^{-\rho} }{\sup} \abs{L_1(s,\sprime) - L_1(s_1,\soprime)}.
\end{equation*}
Now,
\begin{align*}
    D_1 &\leq \underset{\abs{s-s_1},\abs{\sprime-\soprime} \leq n^{-\rho} }{\sup} \sum_{i=1}^n w_i \sum_{j=1}^{m_i} \sum_{1 \leq r \neq \rprime \leq R} \abs{K\fpr{\frac{S_r - s}{\hxi}} - K\fpr{\frac{S_r - s_1}{\hxi}}} K\fpr{\frac{S_{\rprime} - \sprime}{\hxi}} \abs{\tepsilon_{ijr\rprime}} \\
    & \;\; + 
    \underset{\abs{s-s_1},\abs{\sprime-\soprime} \leq n^{-\rho} }{\sup} \sum_{i=1}^n w_i \sum_{j=1}^{m_i} \sum_{1 \leq r \neq \rprime \leq R} \abs{K\fpr{\frac{S_{\rprime} - \sprime}{\hxi}} - K\fpr{\frac{S_{\rprime} - \soprime}{\hxi}}} K\fpr{\frac{S_{r} - s_1}{\hxi}} \abs{\tepsilon_{ijr\rprime}} \\
    &\leq \underset{\abs{s-s_1},\abs{\sprime-\soprime} \leq n^{-\rho} }{\sup} \sum_{i=1}^n w_i \sum_{j=1}^{m_i} \sum_{1 \leq r \neq \rprime \leq R} \frac{L M_K}{\hxi}(s-s_1)  \abs{\tepsilon_{ijr\rprime}}  \\
    &+ \;\; \underset{\abs{s-s_1},\abs{\sprime-\soprime} \leq n^{-\rho} }{\sup} \sum_{i=1}^n w_i \sum_{j=1}^{m_i} \sum_{1 \leq r \neq \rprime \leq R} \frac{L M_K}{\hxi}(\sprime-\soprime)  \abs{\tepsilon_{ijr\rprime}} \\
    &\leq \frac{2L M_K}{\hxi n^{\rho}}\sum_{i=1}^n w_i \sum_{j=1}^{m_i} \sum_{1 \leq r \neq \rprime \leq R} \abs{\tepsilon_{ijr\rprime}} \\
    &\leq \frac{2L M_K}{\hxi n^{\rho}}\fpr{\sum_{i=1}^n w_i \sum_{j=1}^{m_i} \sum_{1 \leq r \neq \rprime \leq R} \abs{\tepsilon_{ijr\rprime}}^{\tau}}^{1/\tau} \\
    &\leq \frac{2L M_K}{\hxi n^{\rho}} \fpr{\sum_{i=1}^n w_i \sum_{j=1}^{m_i} \sum_{1 \leq r \neq \rprime \leq R}   3^{\tau-1}\spr{\norm{\epsilon_i}^{2\tau} + \norm{c}^{\tau} + \norm{\Gamma}^{\tau}} }^{1/\tau},
\end{align*}
where we use Lipschitz continuity of the kernel in the second inequality, and the fourth line follows by Hölder's inequality. By Assumption~\ref{assump: B1},~\ref{assump: B3} and the strong law of large numbers
\begin{align*}
    &\sum_{i=1}^n w_i \sum_{j=1}^{m_i} \sum_{1 \leq r \neq \rprime \leq R}  \spr{\norm{\epsilon_i}^{2\tau} + \norm{c}^{\tau} + \norm{\Gamma}^{\tau}} \\
 &\leq \fpr{\sup_n n \;\max_{i} w_i m_i R(R-1)} \fpr{ \frac{1}{n}\sum_{i=1}^n\norm{\epsilon_i}^{2\tau} + \norm{c}^{\tau} + \norm{\Gamma}^{\tau}} \\
 &\longrightarrow M^\prime\fpr{ \ep\norm{\epsilon}^{2\tau} + \norm{c}^{\tau} + \norm{\Gamma}^{\tau}} < \infty.
\end{align*}
 Since, $n^{\rho}\hGam b_n(\hxi) \to \infty$, we can conclude that 
\begin{equation}\label{eqn:rateD1}
    D_1 = o(b_n(\hxi)) \qquad  a.s.
\end{equation}
Finally, we control the first term in equation~(\ref{eqn: supL1}). Observed that,
$
    L_1(s,\sprime) = \sum_{i=1}^n \Lambda_i 
$
where $\Lambda_i$ are independently distributed random variable as
\begin{equation*}
    \Lambda_i = w_i \sum_{j=1}^{m_i} \sum_{1 \leq r \neq \rprime \leq R} K\fpr{\frac{S_r - s}{\hGam}}K\fpr{\frac{S_{\rprime} - \sprime}{\hGam}} \tepsilon_{ijr\rprime}I\fpr{\abs{\tepsilon_{ijr\rprime}} \leq B_n(\hxi)}.
\end{equation*}
Then, $\abs{\Lambda_i} \leq m_i R(R-1)w_iM_K^2 B_n(\hxi) \leq 2M^{\prime}M_K^2B_n(\hxi)/n$. Furthermore we bound the variance of the $\Lambda_i$ as,
\begin{align*}
    & \ep(\Lambda_i^2) \\
    &\leq w_i^2  \sum_{j=1}^{m_i} \sum_{\jprime=1}^{m_i} \sum_{1 \leq r \neq \rprime \leq R}\sum_{1 \leq l \neq \lprime \leq R} \ep\left[ \abs{\tepsilon_{ijr\rprime}}\abs{\tepsilon_{i\jprime l\lprime}}K\fpr{\frac{S_r - s}{\hxi}} \right. \qquad \qquad \\
    & \left. \qquad \qquad \qquad \qquad \qquad \qquad \qquad \qquad \times K\fpr{\frac{S_{\rprime} - \sprime}{\hxi}} K\fpr{\frac{S_l - s}{\hxi}}K\fpr{\frac{S_{\lprime} - \sprime}{\hxi}} \right] \\
    &\leq  w_i^2 m_i^2 \left\{ R(R-1)\;\ep\tpr{K^2\fpr{\frac{S_1 - s}{\hxi}}K^2\fpr{\frac{S_{2} - \sprime}{\hxi}} \abs{\epsilon(S_1, T_1)\epsilon(S_2, T_1)\epsilon(S_1, T_2)\epsilon(S_2, T_2)}  } \right.\\
    & \qquad \qquad \;\;\; +  R(R-1)(R-2)\;\ep\left[ \abs{\epsilon(S_1, T_1)\epsilon( S_2,T_1)\epsilon(S_1,T_2)\epsilon(S_3,T_2)}K^2\fpr{\frac{S_1 - s}{\hxi}} \right. \\
    & \left. \qquad \qquad \qquad \qquad \qquad \qquad \qquad \qquad \times K\fpr{\frac{S_{2} - \sprime}{\hxi}} K\fpr{\frac{S_{3} - \sprime}{\hxi}} \right] \\
    & \qquad \qquad \;\; +   R(R-1)(R-2)(R-3)\ep\left[\abs{\epsilon(S_1,T_1)\epsilon(S_2, T_1)\epsilon(S_3, T_2)\epsilon(S_4, T_2)} K\fpr{\frac{S_1 - s}{\hxi}} \right. \\
    & \qquad \qquad \qquad \qquad \qquad \qquad \qquad \qquad \times \left. \left. K\fpr{\frac{S_{2} - \sprime}{\hxi}} K\fpr{\frac{S_3 - s}{\hxi}}K\fpr{\frac{S_{4} - \sprime}{\hxi}} \right] \right\} \\
    &\leq M_\epsilon m_i^2w_i^2\tpr{R(R-1)\hxi^2 + R(R-1)(R-2)\hxi^3 + R(R-1))(R-2)(R-3)\hxi^4},
\end{align*}
for some constant $0 < M_\epsilon < \infty$, which is a function of the fourth moment of $\lVert \epsilon \rVert$. For example by Assumption~\ref{assump: A3}, 
\begin{align*}
    &\textrm{E} \left\{ K^2\fpr{\frac{S_1 - s}{\hGam}}K\fpr{\frac{S_{2} - \sprime}{\hxi}} \abs{\epsilon(S_1, T_1)\epsilon(S_2, T_1)\epsilon(S_1, T_2)\epsilon(S_2, T_2)} \right\}\\
    &\leq \ep\tpr{\norm{\epsilon}^{4}}\hxi^2 \int \int K\fpr{v}K\fpr{\vprime}  g(v + s\hxi) g(\vprime + \sprime \hxi)dv d\vprime 
   = \mathcal{O}(\hxi^2).
\end{align*}
Thus, the variance
$
    \sum_{i=1}^n \textrm{Var}(\Lambda_i) \leq \sum_{i=1}^n \ep\Lambda_i^2 \leq M_\epsilon \fpr{b_n(\hxi)}^2/\log n
$.
By Boole's inequality for probability measure,
\begin{align*}
    &P\fpr{\underset{s,\sprime \in \chi(\rho)}{\sup} \abs{L_1(s,\sprime)} > Mb_n(\hxi)} \\
    &=P\fpr{ \abs{L_1(s,\sprime)  } > Mb_n(\hxi) \textrm{ for some } s,\sprime \in \chi(\rho)} \leq n^{2\rho}P\fpr{ \abs{L_1(s,\sprime)} > Mb_n(\hxi)};
\end{align*}
By Bernstein concentration inequality for bounded random variables $\Lambda_i$, 
\begin{align*}
    &P\fpr{ \abs{L_1(s,\sprime)} > Mb_n(\hxi)} 
    \leq  P\fpr{\abs{\sum_{i=1}^n \Lambda_i} > Mb_n(\hxi)} \\
    &\leq 2 \exp\fpr{- \frac{M^2\fpr{b_n(\hxi)}^2/2}{\sum_{i=1}^n \textrm{Var}(\Lambda_i) + 2M^{\prime}M_K^2B_n(\hxi) Mb_n(\hxi)/3n}} \\
    &\leq 2 \exp\fpr{- \frac{M^2\fpr{b_n(\hxi)}^2/2}{M_U\fpr{b_n(\hxi)}^2/\log n + 2M^{\prime}M_K^2M \fpr{b_n(\hxi)}^2/3\log n}} \\
    &\leq 2n^{-M^*},
\end{align*}
where, $M^* = M^2/(2M_U + 4MM^{\prime}M_K^2/3)$.
If we choose $M$ large enough so that $M^*-2\rho > 1$; then
$$
\sum_{n=1}^\infty P\fpr{\underset{s,\sprime \in \chi(\rho)}{\sup} \abs{L_1(s,\sprime)} > Mb_n(\hxi)} \leq \sum_{n=1}^\infty 2n^{2\rho-M^*} < \infty;
$$
By Borel-Cantelli lemma, 
\begin{equation}\label{eqn:rateL1}
    \underset{s,\sprime \in \chi(\rho)}{\sup} \abs{L_1(s,\sprime)} = \mathcal{O}(b_n(\hxi)) \qquad a.s.
\end{equation}
Equation (\ref{eqn:rateL2}), (\ref{eqn:rateD1}), (\ref{eqn:rateL1}) completes the proof of the statement in equation (\ref{eqn:rateLs}).
\end{proof}
\noindent\textbf{Step A2}: We will show that
\begin{equation}\label{eqn:rateHs}
    \underset{s,\sprime \in [0,1]}{\sup} \abs{H(s,\sprime)} = \mathcal{O}(b_n(\hxi)) \qquad a.s.
\end{equation}
\begin{proof}
The proof follows exactly in the same way as the proof in step 1 and hence it is omitted. The proof here is much easier because the random variables are bounded as
$
\abs{c(S_r,S_{\rprime},T_{ij}) - \int c(S_r,S_{\rprime},t)g(t)dt } \leq 2\norm{c} < \infty
$.
\end{proof}

\noindent\textbf{Step A3}: We will show that
\begin{equation}\label{eqn:rateA1s}
    \underset{s,\sprime \in [0,1]}{\sup} \abs{A_1(s,\sprime)} = \mathcal{O}(\hxi^2)\qquad a.s.
\end{equation}
\begin{proof}
Observe that, $K\fpr{(S_r-s)/\hxi} = 0$ if $\abs{S_r - s} > \hxi$ as the kernel $K$ has its support between $[-1,1]$. By Taylor series expansion up to second order, when $\abs{S_r - s} > \hxi$ and $\abs{S_{\rprime} - \sprime} > \hxi$,
\begin{align*}
   & \abs{\int \tpr{c(S_r, S_{\rprime}, t) - c(s,\sprime, t)
    - (S_r - s)\frac{\partial }{\partial s}c(s,\sprime, t) - (S_{\rprime} - \sprime)\frac{\partial }{\partial \sprime}c(s,\sprime, t)}f(t)dt} \\
    &\leq \int \abs{(S_r-s)^2 c_{ss}(s_*, \sprime_*,t) + (S_{\rprime}-\sprime)^2 c_{\sprime \sprime}(s_*, \sprime_*,t) + 2(S_r-s)(S_{\rprime}-\sprime) c_{s\sprime}(s_*, \sprime_*,t)}f(t)dt \\
    &\leq M\hxi^2,
\end{align*}
for some $M > 0$ as the density function and partial second derivatives are bounded by Assumption~\ref{assump: A3} and \ref{assump: A6}. The differentiability of $c$ follows by the fact that the marginal covariance $\Xi(s,\sprime)$ is differentiable and the range of integration $\mathcal{T}$ is compact. This implies,
\begin{align*}
    \underset{s,\sprime \in [0,1]}{\sup} \abs{A_1(s,\sprime)} \leq  M\hxi^2\underset{s,\sprime \in [0,1]}{\sup} V_{00}(s,\sprime).
\end{align*}
By equation~(\ref{eqn: unifratesV}), $\sup_{s,\sprime} V_{00}(s,\sprime) = \mathcal{O}(1)$. This completes the proof of the statement in equation (\ref{eqn:rateA1s}).
\end{proof}

\noindent\textbf{Step A4}: We will show that
\begin{equation}\label{eqn:rateA2s}
    \underset{s,\sprime \in [0,1]}{\sup} \abs{A_2(s,\sprime)} = \mathcal{O}(\hxi^2)\qquad a.s.
\end{equation}
\begin{proof}
The proof is exactly same as the proof of (\ref{eqn:rateA1s}) and hence is omitted.
\end{proof}

\noindent\textbf{Step A5}: We will show that
\begin{equation}\label{eqn:ratePsi1s}
    \underset{s,\sprime \in [0,1]}{\sup} \abs{\Psi_1(s,\sprime)} = \mathcal{O}(\alpha_n)\qquad a.s.
\end{equation}
\begin{proof}
It is straight forward to see that 
\begin{equation}\label{eqn: Psi1der}
     \abs{\Psi_1(s,\sprime)} \leq  \underset{s,t \in [0,1]}{\sup} \abs{\widehat{\mu}(s,t) - \mu(s,t)}  \sum_{i=1}^n w_i \sum_{j=1}^{m_i} \sum_{1 \leq r \neq \rprime \leq R} K_{\hxi}\fpr{S_r - s}K_{\hxi}\fpr{S_{\rprime} - \sprime}\abs{\epsilon_{ij\rprime}}.
\end{equation}
Following the same lines as in step 1 we can prove an asymptotic almost sure convergence rate for the second term in the right hand side of equation (\ref{eqn: Psi1der}) and that implies that 
$$
\sum_{i=1}^n w_i \sum_{j=1}^{m_i} \sum_{1 \leq r \neq \rprime \leq R} K_{\hxi}\fpr{S_r - s}K_{\hxi}\fpr{S_{\rprime} - \sprime} \abs{\epsilon_{ij\rprime}} = \mathcal{O}(1) \qquad a.s;
$$
The proof follows as 
$
\underset{s,t \in [0,1]}{\sup} \abs{\widehat{\mu}(s,t) - \mu(s,t)}  = \mathcal{O}(\alpha_n) \qquad a.s.
$.
\end{proof}
By similar arguments, one can show that 
\begin{equation}\label{eqn:ratePsi2s}
    \underset{s,\sprime \in [0,1]}{\sup} \abs{\Psi_2(s,\sprime)} = \mathcal{O}(\alpha_n)\qquad  \underset{s,\sprime \in [0,1]}{\sup} \abs{\Psi_3(s,\sprime)} = \mathcal{O}(\alpha_n^2)\qquad a.s.
\end{equation}
Equation (\ref{eqn:rateLs}), (\ref{eqn:rateHs}), (\ref{eqn:rateA1s}), (\ref{eqn:rateA2s}), (\ref{eqn:ratePsi1s}), (\ref{eqn:ratePsi2s}) in conjunction with equation (\ref{eqn: breakdown}) implies that 
\begin{equation}\label{eqn: finalrateR00}
    \underset{s,\sprime \in [0,1]}{\sup}\abs{R_{00} - \Xi(s,\sprime)V_{00} - hV_{10}\frac{\partial }{\partial s}\Xi(s,\sprime) - hV_{01}\frac{\partial }{\partial \sprime}\Xi(s,\sprime)} = \mathcal{O}(b_n(\hxi)/\hxi^2 + \hxi^2 + \alpha_n).
\end{equation}

\noindent\textbf{Step B}: Using Taylor series expansion and Assumption~\ref{assump: A1} and \ref{assump: A2}, we derive that for $s, \sprime \in [\hxi, 1-\hxi]$, $\ep{V_{00}} = g(s)g(\sprime) + \mathcal{O}(\hxi^2)$, $\ep{V_{10}} = \mathcal{O}(\hxi) = \ep{V_{01}}$, $\ep{V_{11}} = \mathcal{O}(\hxi^2)$, $\ep{V_{20}} = \sigma_K^2g(s)g(\sprime) + \mathcal{O}(\hxi) = \ep{V_{02}}$. Following the similar steps in A1, for $s, \sprime \in [\hxi, 1-\hxi]$, we have the following uniform almost sure rates,
\begin{equation}\label{eqn: unifratesV}
\begin{aligned}
        V_{00} &= g(s)g(\sprime) + \mathcal{O}(\hxi + b_n(\hxi)) \qquad  &V_{01} = \mathcal{O}(\hxi + b_n(\hxi)) \\
    V_{20} &= \sigma^2_K g(s)g(\sprime) + \mathcal{O}(\hxi + b_n(\hxi)) \qquad &V_{10} = \mathcal{O}(\hxi + b_n(\hxi)) \\
        V_{02} &= \sigma^2_K g(s)g(\sprime) + \mathcal{O}(\hxi + b_n(\hxi)) \qquad &V_{11} = \mathcal{O}(\hxi + b_n(\hxi)).
\end{aligned}
\end{equation}

Similarly, for $s,\sprime \in [0,\hxi]$ and $s,\sprime \in [1-\hxi,1]$, we can show that the similar uniform rates hold. We refer the reader to see the proof of Theorem $3.3$ of \cite{li2010uniform} for more details. This implies that the first term in the denominator $D_n$, $(V_{20}V_{02} - V_{11}^2)V_{00}$ is bounded away uniformly from $0$ for all $s, \sprime \in [0,1]$. In fact one can show that for every $s,\sprime \in [0,1]$, $(V_{20}V_{02} - V_{11}^2)V_{00}$ converges to $\sigma_K^2 \fpr{g(s)g(\sprime)}^3$ uniformly. The other two terms in the denominator converges to zero uniformly, by Assumption~\ref{assump: A2} the denominator is bounded away from $0$. Moreover, $V_{pq}, p=0,1,2, q=0,1,2$ are uniformly bounded on $[0,1] \times [0,1]$ almost surely. Equation~(\ref{eqn: finalrateR00}) and~(\ref{eqn: unifratesV}) proves that the first term in equation (\ref{eqn: Xiform}) is of the order $\mathcal{O}(b_n/\hxi^2 + \hxi^2 + \alpha_n)$. After similar calculation, one can show that the other terms in the equation (\ref{eqn: Xiform}) is of lower order, which completes the proof of the theorem.  
\end{proof}

\section{Proof of Theorem~\ref{theorem: pLRTnulldist}}\label{sec: supp_proof_pLRT}

\begin{proof}

Derivation of null distribution of $\widehat{pLRT}_{N,k}$ is non-trivial because 1) it is constructed upon $\widehat{W}_{ijk}$ which are not independently distributed across the subjects 2) the random variable $W_{ijk}$ is unobserved. However, denoting $ \delta_{k, ij} = \int Y_{ij}(s) \fpr{\widehat{\phi}_{k}(s)-\phi_k(s)} ds$, we can establish that
\begin{equation} \label{eq: w_what_rel}
    \bhatW_{k} =\widehat{\M{\Sigma}}_{W, k}^{-1/2}\bW_{k} =  \widehat{\M{\Sigma}}_{W, k}^{-1/2}\fpr{\M{Y}_{k} + \bdelta_k} = \widehat{\bY}_k + \widehat{\bdelta}_{k},
\end{equation}
where $\bdelta_k=\fpr{\bdelta_{k1}^\top, \dots, \bdelta_{kn}^\top}^\top$ and $\bdelta_{ki}^\top = \fpr{\delta_{k, i1},\dots,\delta_{k, im_i}}^\top$ as the stacked vector of $\delta_{k, ij}$.
Now, let's define the test-statistic of the pseudo-likelihood ratio test-statistic based on 
$\widehat{\bY}_{k}$ as
\begin{equation} \label{eq: OrigpLRT}
pLRT_{N,k}= \sup_{H_{0k}\cup H_{1k}}  2\log L_{\widehat{\bY}_{k}}( \bbeta_{k}, \sigma^2_{b,k}) - \sup_{H_{0,k}} 2\log L_{\widehat{\bY}_{k}}( \bbeta_{k}, \sigma^2_{b,k}), 
\end{equation}
where twice the pseudo log-likelihood is, up to an additive constant independent of the parameters,  $2\log L_{\widehat{\bY}_{k}}( \bbeta_{k}, \sigma^2_{b,k}) = -\log( \lvert \widehat{\bH}_{\sigma^2_{b,k}}\rvert  ) - (\widehat{\bY}_{k} -\widehat{\bX}_{k} \bbeta_{k})^{\top} \widehat{\bH}^{-1}_{\sigma^2_{b,k}}  (\widehat{\bY}_{k} - \widehat{\bX}_{k} \bbeta_{k} )$, where $\widehat{\bH}_{\sigma^2_{b,k}} = \M{I}_N+ \sigma^2_{b,k}\widehat{\bZ}_{k}\widehat{\bZ}_{k}^{\top}$. 

In steps (S1)-(S3) below, we will establish the connection between $pLRT_{N,k}$ and $\widehat{pLRT}_{N,k}$ using relation between $\widehat{\bY}_{k}$ and $\widehat{\bW}_{k}$ as in equation (\ref{eq: w_what_rel}). We will use the uniform convergence of the eigenfunctions in (S4) along with repeated application of the proof of Proposition $2.1$ of \cite{staicu2014likelihood} to prove the result.

\noindent\textbf{(S1)} For the sake of the proof and to match the notations with the proof of \cite{staicu2014likelihood}, we call $\sigma^2_{b,k}$ as $\lambda_k$. Please do not confuse this $\lambda_k$ with the one specified in the spectral decomposition of $\Xi(s,\sprime)$; section~\ref{subsec: basisfunc}. Note that the log-likelihood equation $L_{\bhatW_{k}}( \bbeta_{k}, \lambda_k)$ involves two unknown parameters $\bbeta_{k}$ and $\lambda_k$. Writing the form of the maximum likelihood estimator of $\bbeta_{k}$ as a function of $\lambda_k$,  $\bhbeta_{k,W}\fpr{\lambda_k} = \fpr{\widehat{\M{X}}_{k}^\top\widehat{\M{H}}^{-1}_{\lambda_k} \widehat{\M{X}}_{k}}^{-1}\widehat{\M{X}}_{k}^\top\widehat{\M{H}}^{-1}_{\lambda_k} \bhatW_{k}$, we get the profile likelihood for $\lambda_k$ as
\begin{equation}
    2\log L_{\bhatW_{k}}(\lambda_k) =  -\log \det\fpr{\widehat{\M{H}}_{\lambda_k}} -  \fpr{\bhatW_{k} - \widehat{\M{X}}_{k} \bhbeta_{k, W}(\lambda_k) }^{\top} \widehat{\M{H}}^{-1}_{\lambda_k}  \fpr{\bhatW_{k} - \widehat{\M{X}}_{k} \bhbeta_{k, W}(\lambda_k) }.
\end{equation}
Then, the proposed pseudo LRT can be partitioned as 
\begin{equation}
    \widehat{pLRT}_{N,k} =  \sup_{\lambda_k \geq 0} 2\log L_{\bhatW_{k}}(\lambda_k) - 2\log L_{\bhatW_{k}}(0)  + 2\log L_{\bhatW_{k}}(0) - 2 \log L_{\bhatW_{k}}^{0, N},
\end{equation}
where $L_{\bhatW_{k}}^{0, N}$ is the maximum value of the pseudo-likelihood $\log L_{\bhatW_{k}}( \bbeta_{k}, \sigma^2_{b,k})$ under $H_{0,k}$. In the same way as above, we can expand the pseudo-likelihood ratio test statistic using the unobserved response $\bhatY_k$, $pLRT_{N,k}$ as
\begin{equation}
    pLRT_{N,k} =  \sup_{\lambda_k \geq 0} 2\log L_{\widehat{\bY}_{k}}(\lambda_k) - 2\log L_{\widehat{\bY}_{k}}(0)  + 2\log L_{\widehat{\bY}_{k}}(0) - 2 \log L_{\widehat{\bY}_{k}}^{0, N},
\end{equation}
where $L_{\widehat{\bY}_{k}}^{0, N}$ is the maximum value of the pseudo-likelihood $\log L_{\widehat{\bY}_{k}}( \bbeta_{k}, \sigma^2_{b,k})$ under $H_{0,k}$ and 
\begin{equation}
    2\log L_{\widehat{\bY}_{k}}(\lambda_k) =  -\log \det\fpr{\widehat{\M{H}}_{\lambda_k}} -  \fpr{\widehat{\bY}_{k} - \widehat{\M{X}}_{k} \widehat{\V{\beta}}_{k, Y}(\lambda_k) }^{\top} \widehat{\M{H}}^{-1}_{\lambda_k}  \fpr{\widehat{\bY}_{k} - \widehat{\M{X}}_{k} \widehat{\V{\beta}}_{k, Y}(\lambda_k) },
\end{equation}
with $\bhbeta_{k,Y}\fpr{\lambda_k} = \fpr{\widehat{\M{X}}_{k}^\top\widehat{\M{H}}^{-1}_{\lambda_k} \widehat{\M{X}}_{k}}^{-1}\widehat{\M{X}}_{k}^\top\widehat{\M{H}}^{-1}_{\lambda_k} \bhatY_{k}$.

\noindent\textbf{(S2)} We will establish a relationship between $2\log L_{\bhatW_{k}}(\lambda_k)$ and $2\log L_{\bhatY_{k}}(\lambda_k)$, the profile-likelihood based on quasi-projection $\bhatW_k$ and unobserved $\bhatY_k$. Using (\ref{eq: w_what_rel}) we derive that
\begin{align*}
    \widehat{\M{X}}_{k}\bhbeta_{k,W}\fpr{\lambda_k} &= \widehat{\M{X}}_{k}\fpr{\widehat{\M{X}}_{k}^\top\widehat{\M{H}}^{-1}_{\lambda_k} \widehat{\M{X}}_{k}}^{-1}\widehat{\M{X}}_{k}^\top\widehat{\M{H}}^{-1}_{\lambda_k} \bhatW_{k} 
    = \widehat{\M{X}}_{k}\fpr{\widehat{\M{X}}_{k}^\top\widehat{\M{H}}^{-1}_{\lambda_k} \widehat{\M{X}}_{k}}^{-1}\widehat{\M{X}}_{k}^\top\widehat{\M{H}}^{-1}_{\lambda_k} \fpr{\bhatY_k + \widehat{\bdelta}_k} \\
    &= \widehat{\M{X}}_{k}\bhbeta_{k,Y}\fpr{\lambda_k} + \M{P}_{k, \lambda_k}\widehat{\bdelta}_k,
\end{align*}
with $\M{P}_{k, \lambda_k} =\widehat{\M{X}}_{k}\fpr{\widehat{\M{X}}_{k}^\top\widehat{\M{H}}^{-1}_{\lambda_k} \widehat{\M{X}}_{k}}^{-1}\widehat{\M{X}}_{k}^\top\widehat{\M{H}}^{-1}_{\lambda_k} $ being the projection matrix onto the column space of $\widehat{\M{X}}_{k}$. Using the above, the quadratic form in the profile-likelihood $2\log L_{\bhatW_{k}}(\lambda_k)$ can be re-expressed as
\begin{align}
     \nonumber &2\log L_{\bhatW_{k}}(\lambda_k) \\
     \nonumber &=  -\log \det\fpr{\widehat{\M{H}}_{\lambda_k}} - \fpr{\bhatW_{k} - \widehat{\M{X}}_{k} \bhbeta_{k,W}(\lambda_k) }^{\top} \widehat{\M{H}}^{-1}_{\lambda_k}  \fpr{\bhatW_{k}- \widehat{\M{X}}_{k} \bhbeta_{k,W}(\lambda_k) }  \\
    \nonumber &= -\log \det\fpr{\widehat{\M{H}}_{\lambda_k}}-\fpr{\bhatY_{k} + \widehat{\bdelta}_k- \widehat{\M{X}}_{k}\bhbeta_{k,Y}\fpr{\lambda_k} - \M{P}_{k, \lambda_k}\widehat{\bdelta}_k }^{\top} \widehat{\M{H}}^{-1}_{\lambda_k}  \fpr{\bhatY_{k} + \widehat{\bdelta}_k - \widehat{\M{X}}_{k}\bhbeta_{k,Y}\fpr{\lambda_k} - \M{P}_{k, \lambda_k}\widehat{\bdelta}_k } \\
    \nonumber &=  2\log L_{\bhatY_{k}}(\lambda_k) - 2 \fpr{\bhatY_{k} - \widehat{\M{X}}_{k}\bhbeta_{k,Y}\fpr{\lambda_k} }^{\top} \widehat{\M{H}}^{-1}_{\lambda_k}  \fpr{\M{I} - \M{P}_{k, \lambda_k}}\widehat{\bdelta}_k - \widehat{\bdelta}_k^\top\fpr{\M{I} - \M{P}_{k, \lambda_k}} \widehat{\M{H}}^{-1}_{\lambda_k}  \fpr{\M{I} - \M{P}_{k, \lambda_k}}\widehat{\bdelta}_k  \\
    \nonumber &= 2\log L_{\bhatY_{k}}(\lambda_k) - 2 \widehat{\bY}_k^\top\fpr{\M{I} - \M{P}_{k, \lambda_k}} \widehat{\M{H}}^{-1}_{\lambda_k}  \fpr{\M{I} - \M{P}_{k, \lambda_k}}\widehat{\bdelta}_k  - \widehat{\bdelta}_k^\top\fpr{\M{I} - \M{P}_{k, \lambda_k}} \widehat{\M{H}}^{-1}_{\lambda_k}  \fpr{\M{I} - \M{P}_{k, \lambda_k}}\widehat{\bdelta}_k \\
    \nonumber &= 2\log L_{\bhatY_{k}}(\lambda_k) - 2\widehat{\bY}_k^\top\widehat{\M{G}}_k\widehat{\M{G}}_k^\top\widehat{\bdelta}_k + 2\lambda_k\widehat{\bY}_k^\top\widehat{\M{G}}_k\widehat{\M{G}}_k^\top\widehat{\bZ}_k\fpr{\M{I}_Q + \lambda \widehat{\bZ}_k^\top\widehat{\M{G}}_k\widehat{\M{G}}_k^\top \widehat{\bZ}_k }^{-1}\widehat{\bZ}_k^\top\widehat{\M{G}}_k\widehat{\M{G}}_k^\top\widehat{\bdelta}_k \\
    & \qquad - \widehat{\bdelta}_k^\top\widehat{\M{G}}_k\widehat{\M{G}}_k^\top\widehat{\bdelta}_k + \lambda_k\widehat{\bdelta}_k^\top\widehat{\M{G}}_k\widehat{\M{G}}_k^\top\widehat{\bZ}_k\fpr{\M{I}_Q + \lambda \widehat{\bZ}_k^\top\widehat{\M{G}}_k\widehat{\M{G}}_k^\top \widehat{\bZ}_k }^{-1}\widehat{\bZ}_k^\top\widehat{\M{G}}_k\widehat{\M{G}}_k^\top\widehat{\bdelta}_k  \label{eqn : logL1rel},
\end{align}
where $\widehat{\M{G}}_k$ is such that $\widehat{\M{G}}_k\widehat{\M{G}}_k^\top = \M{I} - \M{P}_{\widehat{\bX}_k}$ and $\M{P}_{\widehat{\bX}_k}=\widehat{\bX}_k\fpr{\widehat{\bX}_k^\top\widehat{\bX}_k}^{-1}\widehat{\bX}_k^\top$ is the projection matrix onto column space of $\widehat{\bX}_k$ with $\widehat{\M{G}}_k^\top\widehat{\M{G}}_k= \M{I}_{N-2}$. The last step follows by repeated application of the Woodbury matrix inversion lemma and the explanation provided by \cite{staicu2014likelihood} in Equation (21) and (22). 
Then the first part of the $\widehat{pLRT}_{N,k}$ can be expanded as
\begin{align*}
    2\log L_{\bhatW_{k}}(\lambda_k) - 2\log L_{\bhatW_{k}}(0) 
    &= 2\log L_{\bhatY_{k}}(\lambda_k) - 2\log L_{\bhatY_{k}}(0)  + 2C_{1k}(\lambda_k) + C_{2k}(\lambda_k) \label{eqn : logL2rel},
\end{align*}
where,
$$C_{1k}(\lambda_k)  = \lambda_k\widehat{\bY}_k^\top\widehat{\M{G}}_k\widehat{\M{G}}_k^\top\widehat{\bZ}_k\fpr{\M{I}_Q + \lambda_k \widehat{\bZ}_k^\top\widehat{\M{G}}_k\widehat{\M{G}}_k^\top \widehat{\bZ}_k }^{-1}\widehat{\bZ}_k^\top\widehat{\M{G}}_k\widehat{\M{G}}_k^\top\widehat{\bdelta}_k,$$ and 
$$C_{2k}(\lambda_k) = \lambda_k\widehat{\bdelta}_k^\top\widehat{\M{G}}_k\widehat{\M{G}}_k^\top\widehat{\bZ}_k\fpr{\M{I}_Q + \lambda_k \widehat{\bZ}_k^\top\widehat{\M{G}}_k\widehat{\M{G}}_k^\top \widehat{\bZ}_k }^{-1}\widehat{\bZ}_k^\top\widehat{\M{G}}_k\widehat{\M{G}}_k^\top\widehat{\bdelta}_k.$$ 

Using the eigen decomposition of $\widehat{\bZ}_k^\top\widehat{\M{G}}_k\widehat{\M{G}}_k^\top \widehat{\bZ}_k=\widehat{\M{U}}_k\textrm{diag}\fpr{N^\varrho\widehat{\zeta}_{N, k1}, \dots, N^\varrho\widehat{\zeta}_{N, kQ} }\widehat{\M{U}}_k^\top $, we can succinctly write $C_{1k}$ and $C_{2k}$ as 
\begin{equation}\label{eqn: C1lC2k}
    C_{1k}(\lambda_k^\prime) = \sum_{q=1}^Q \frac{\lambda_k^\prime A_{N, kq}B_{N,kq}}{1+\lambda_k^\prime \widehat{\zeta}_{N, kq}}; \qquad C_{2k}(\lambda_k^\prime) = \sum_{q=1}^Q \frac{\lambda_k^\prime A_{N, kq}^2}{1+\lambda_k^\prime \widehat{\zeta}_{N, kq}},
\end{equation}
for some $\lambda_k^\prime \geq 0$ and $\M{A}_{k,N} = N^{-\varrho/2}\widehat{\M{U}}_k^\top\widehat{\bZ}_{k}^\top\widehat{\M{G}}_k\widehat{\M{G}}_k^\top \widehat{\bdelta}_{k}$, $\M{B}_{k,N} = N^{-\varrho/2}\widehat{\M{U}}_k^\top\widehat{\bZ}_{k}^\top\widehat{\M{G}}_k\widehat{\M{G}}_k^\top \widehat{\bY}_{k}$ and $\lambda_k^\prime = \lambda_k N^\varrho$. Moreover, following calculation in the proof of the Theorem 2 of \cite{staicu2014likelihood} (equation (20)-(24)) reveals that
\begin{equation*}
    2\sup_{\lambda_k \geq 0}\spr{\log L_{\bhatW_{k}}(\lambda_k) - \log L_{\bhatW_{k}}(0)} = \sup_{\lambda_k^\prime \geq 0} \spr{\hat{f}_N(\lambda_k^\prime) + 2C_{1k}(\lambda_k^\prime) + C_{2k}(\lambda_k^\prime)},
\end{equation*}
where 
\begin{equation*}
    \widehat{f}_N(\lambda_k^\prime) = \sum_{q=1}^Q \frac{\lambda_k^\prime \widehat{w}^2_{N,kq}}{1+\lambda_k^\prime\widehat{\zeta}_{N,kq}} - \sum_{l=1}^Q \log\fpr{1 + \lambda_k^\prime \widehat{\xi}_{N,kq}},
\end{equation*}
with $\widehat{\V{w}}_{N,k} = N^{-\varrho/2}\widehat{\bU}_k^\top\widehat{\bZ}_k^\top\widehat{\M{G}}_k\widehat{\M{G}}_k^\top\widehat{\bY}_k$ and $\widehat{\xi}_{N,kq}$ be the $q$th eigen value of $N^{-\varrho}\widehat{\bZ}_k^\top\widehat{\bZ}_k$, $q=1,\dots,Q$. Under the Assumptions \ref{assump: C1}-\ref{assump: C3}, following the lines of their paper,
\begin{equation*}
    \hat{f}_N(\lambda_k^\prime) + 2C_{1k}(\lambda_k^\prime) + C_{2k}(\lambda_k^\prime) \overset{d}{\longrightarrow} \sum_{q=1}^{Q} \dfrac{\lambda_k^\prime\zeta_{kq}\vartheta_{q}}{(1+\lambda_k^\prime \zeta_{kq})}  - \sum_{q=1}^{Q} \log (1+\lambda_k^\prime \xi_{ kq}),
\end{equation*}
where $\vartheta_{q} \overset{iid}{\sim} N(0,1), q=1,\dots,Q$ provided the quantities $C_{1k}(\lambda_k^\prime)$ and $C_{2k}(\lambda_k^\prime)$ converges in probability to zero (Slutsky's theorem). Assuming this is true (which we will prove in the last step), following the exact same lines of \cite{staicu2014likelihood}, by application of continuous mapping theorem and arguing that the sequence $\hat{f}_N(\lambda_k^\prime) + 2C_{1k}(\lambda_k^\prime) + C_{2k}(\lambda_k^\prime)$ is tight we can conclude
\begin{equation}\label{eqn : pLRTwPart1}
    2\sup_{\lambda_k \geq 0}\spr{\log L_{\bhatW_{k}}(\lambda_k) - \log L_{\bhatW_{k}}(0)} \overset{d}{\longrightarrow} \sup_{\lambda_k^\prime \geq 0}\spr{\sum_{q=1}^{Q} \dfrac{\lambda_k^\prime\zeta_{kq}\vartheta_{q}}{(1+\lambda_k^\prime \zeta_{kq})}  - \sum_{q=1}^{Q} \log (1+\lambda_k^\prime \xi_{ kq})}.
\end{equation}

\noindent\textbf{(S3)} Let's partition the design matrix as follows: $\bX_k =(\bX_{(1),k} \mid \bX_{(2),k})$, where $\bX_{(1),k}$ be the part corresponding to null hypothesis, only one column of $1$'s. Then maximum value of the twice of negative log-likelihood under the null model, $- 2 \log L_{\bhatW_{k}}^{0, N} = \bhatW_{k}^\top\fpr{\M{I}_N - \M{P}_{\widehat{\bX}_{(1),k}}}\bhatW_{k}$. Using ~(\ref{eq: w_what_rel}) again, we write,
\begin{align*}
    - 2 \log L_{\bhatW_{k}}^{0, N} &= \bhatW_{k}^\top\fpr{\M{I}_N - \M{P}_{\widehat{\bX}_{(1),k}}}\bhatW_{k} \\
    &= \bhatY_{k}^\top\fpr{\M{I}_N - \M{P}_{\widehat{\bX}_{(1),k}}}\bhatY_{k} + 2\bhatY_{k}^\top\fpr{\M{I}_N - \M{P}_{\widehat{\bX}_{(1),k}}}\widehat{\bdelta}_{k} + \widehat{\bdelta}_{k}^\top\fpr{\M{I}_N - \M{P}_{\widehat{\bX}_{(1),k}}}\widehat{\bdelta}_{k} \\
    &= - 2 \log L_{\bhatY_{k}}^{0, N} + 2\bhatY_{k}^\top\fpr{\M{I}_N - \M{P}_{\widehat{\bX}_{(1),k}}}\widehat{\bdelta}_{k} + \widehat{\bdelta}_{k}^\top\fpr{\M{I}_N - \M{P}_{\widehat{\bX}_{(1),k}}}\widehat{\bdelta}_{k},
\end{align*}
where $\widehat{\bX}_{(1),k} =\widehat{\M{\Sigma}}_{W, k}^{-1/2}\bX_{(1)} $ and $\M{P}_{\widehat{\bX}_{(1),k}} = \widehat{\bX}_{(1),k} (\widehat{\bX}_{(1),k}^\top \widehat{\bX}_{(1),k})^{-1}\widehat{\bX}_{(1),k}^\top$ be the projection onto the column space of $\widehat{\bX}_{(1)}$. Putting $\lambda_k = 0$ in equation (\ref{eqn : logL1rel}) we get
\begin{align*}
    2\log L_{\bhatW_{k}}(0) - 2 \log L_{\bhatW_{k}}^{0, N} &= 2\log L_{\widehat{\bY}_{k}}(0) - 2 \log L_{\widehat{\bY}_{k}}^{0, N}  + 2C_{3k} + C_{4k}, \label{eqn : logL3rel}
\end{align*}
where,
$$
C_{3k} = \bhatY_{k}^\top\fpr{\M{P}_{\widehat{\bX}_k} - \M{P}_{\widehat{\bX}_{(1),k}}}\widehat{\bdelta}_{k} \qquad \qquad C_{4k} = \widehat{\bdelta}_{k}^\top\fpr{\M{P}_{\widehat{\bX}_k} - \M{P}_{\widehat{\bX}_{(1),k}}}\widehat{\bdelta}_{k}. 
$$

By Theorem 2 of \cite{staicu2014likelihood} and Slutsky's theorem,
\begin{equation}\label{eqn : pLRTwPart2}
    2\log L_{\bhatW_{k}}(0) - 2 \log L_{\bhatW_{k}}^{0, N} \overset{d}{\longrightarrow} \chi^2_1;
\end{equation}
provided $C_{3k}$ and $C_{4k}$ converges in probability to zero. An independence between the two limiting random variables on the right hand side of those equation~(\ref{eqn : pLRTwPart1}) and~(\ref{eqn : pLRTwPart2}) can be established by the same logic presented there. Hence the proof of Theorem~\ref{theorem: pLRTnulldist}. It just remains to show that $C_{1k}(\lambda_k^\prime), C_{2k}(\lambda_k^\prime)$,
$C_{3k}$ and $C_{4k}$ converge in probability to zero. 

\noindent\textbf{(S4)} Since $\M{P}_{\widehat{\bX}_k} - \M{P}_{\widehat{\bX}_{(1),k}}$ is a projection matrix of rank $1$, there exists a $N \times 1$ matrix $\widehat{\bG}^{*}_{k}$ such that $\widehat{\bG}^{*}_{k}\widehat{\bG}^{*{^\top}}_{k} = \M{P}_{\widehat{\bX}_k} - \M{P}_{\widehat{\bX}_{(1),k}}$ and $\widehat{\bG}^{*{^\top}}_{k}\widehat{\bG}^{*}_{k} = 1$. This implies, $C_{3k} = \bhatY_{k}^\top\widehat{\bG}^{*}_{k}\widehat{\bG}^{*{^\top}}_{k}\widehat{\bdelta}_{k}$, $C_{4k} = \widehat{\bdelta}_{k}^\top\widehat{\bG}^{*}_{k}\widehat{\bG}^{*{^\top}}_{k}\widehat{\bdelta}_{k} $,  showing probability convergence of $C_{3k}$ and $C_{4k}$ is equivalent to showing $\bG^{*{^\top}}_{k}\widehat{\bdelta}_k = o_p(1)$. Thus it remains to show that,
\begin{equation}\label{eqn: showopfinalstep}
    N^{-\varrho/2}\widehat{\M{U}}_k^\top\widehat{\bZ}_{k}^\top\widehat{\M{G}}_k\widehat{\M{G}}_k^\top \widehat{\bdelta}_{k} = o_p(1) \qquad \qquad  \widehat{\bG}^{*{^\top}}_{k}\widehat{\bdelta}_k = o_p(1).
\end{equation}
Before proving this we will introduce additional notation. Let, $\widetilde{\bZ}_k$ and $\widetilde{\bG}_k$, $\widetilde{\bG}^*_k$ and $\widetilde{\bdelta}_{k}$ be defined similarly to $\widehat{\bZ}_k$, $\widehat{\bG}_k$, $\widehat{\bG}^*_k$ and $\widehat{\bdelta}_{k}$ but with the true covariance $\bSigma_{Y,k}$ pre-multiplied instead of $\widehat{\bSigma}_{W,k}$. By continuous mapping theorem and the Assumption~\ref{assump: C2}, it suffices to show that $
    N^{-\rho/2}\widetilde{\M{U}}_k^\top\widetilde{\bZ}_{k}^\top\widetilde{\M{G}}_k\widetilde{\M{G}}_k^\top \widetilde{\bdelta}_{k} = o_p(1)
$ and $\widetilde{\bG}^{*{^\top}}_{k}\widetilde{\bdelta}_k = o_p(1)$ (see the proof of \cite{staicu2014likelihood} for more details)

Under $H_{0k}$, $\delta_{ijk} = \int \mu_0(s) (\widehat{\phi}_k(s) - \phi_k(s))ds + \int \epsilon_{i}(s, t_{ij}) (\widehat{\phi}_k(s) - \phi_k(s))ds$. The first term of the summand is free of $(i,j)$. This means under the null, $
\bdelta_k = c^*\V{1}_N + \bdelta^*_k$, for some constant $c^*$, where $\delta^*_{ijk} = \int \epsilon_{i}(s, t_{ij}) (\widehat{\phi}_k(s) - \phi_k(s))ds$. 
Furthermore, $\widetilde{\bG}_k$ is the projection onto the null space of $\bSigma_{Y,k}^{-1/2}\bX$. From the construction of $\bX$, as $\V{1}_N \in \mathcal{C}(\bX)$, implies that $\widetilde{\M{G}}_k\widetilde{\M{G}}_k^\top \bSigma_{Y,k}^{-1/2}\V{1}_N = 0$. In the similar manner, we get $\widetilde{\M{G}}^{*}_k\widetilde{\bG}^{*{^\top}}_{k} \bSigma_{Y,k}^{-1/2}\V{1}_N = 0$. Since by Assumption~\ref{assump: C3} the matrix $N^{-\varrho}\widetilde{\M{U}}_k^\top\widetilde{\bZ}_{k}^\top\widetilde{\M{G}}_k\widetilde{\M{G}}_k^\top\widetilde{\bZ}_{k}\widetilde{\M{U}}_k$ stabilizes and the columns of $\widetilde{\M{G}}^{*}_k$ are orthonormal, proving~(\ref{eqn: showopfinalstep})  is equivalent to showing that for any matrix $N \times d$ matrix $\boldsymbol{\mathcal{W}} = [\V{\omega}_1, \dots, \V{\omega}_d]$ with $d$-fixed such that $\V{\omega}_l^\top\V{\omega}_{l^\prime} = O(1)$ as $n \to \infty$ for all $l, l^\prime \in \{1,2,\dots, d\}$, $\boldsymbol{\mathcal{W}}^\top\bdelta^*_k = o_p(1)$. To justify, observe that the $l$th element of $\boldsymbol{\mathcal{W}}^\top\bdelta^*_k$ is
\begin{align*}
    \lvert (\boldsymbol{\mathcal{W}}^\top\bdelta_k^*)_l \rvert &= \left\lvert \sum_{i=1}^n\sum_{j=1}^{m_i} \omega_{ij,l}\delta^*_{ijk} \right\rvert  = \left\lvert \sum_{i=1}^n\sum_{j=1}^{m_i} \omega_{ij,l} \int_{0}^1 \epsilon_{i}(s, t_{ij})(\widehat{\phi}_k(s) - \phi_k(s))ds  \right\rvert \\
    &\leq \sup_s \left\lvert \widehat{\phi}_k(s) - \phi_k(s) \right\rvert\sum_{i=1}^n\sum_{j=1}^{m_i} \left\lvert \omega_{ij,l}\right\rvert  \int_{0}^1 \left\lvert \epsilon_{i}(s, t_{ij})\right\rvert ds. 
\end{align*}
To show that $\sum_{i=1}^n\sum_{j=1}^{m_i} \left\lvert \omega_{ij,l}\right\rvert  \int_{0}^1 \left\lvert \epsilon_{i}(s, t_{ij})\right\rvert ds < \infty$ a.s as $n \to \infty$, we calculate the variance
\begin{align*}
    \textrm{Var}\fpr{\sum_{i=1}^n\sum_{j=1}^{m_i} \left\lvert \omega_{ij,l}\right\rvert  \int_{0}^1 \left\lvert \epsilon_{i}(s, t_{ij})\right\rvert ds } 
    &= \sum_{i=1}^n \textrm{Var}\fpr{\sum_{j=1}^{m_i} \left\lvert \omega_{ij,l}\right\rvert  \int_{0}^1 \left\lvert \epsilon_{i}(s, t_{ij})\right\rvert ds} \qquad  \qquad \qquad \qquad \\
    &\leq \sum_{i=1}^n \textrm{E}\fpr{\sum_{j=1}^{m_i} \left\lvert \omega_{ij,l}\right\rvert  \int_{0}^1 \left\lvert \epsilon_{i}(s, t_{ij})\right\rvert ds}^2 \\
    &\leq  \sum_{i=1}^n m_i\sum_{j=1}^{m_i}\omega_{ij,l}^2 \textrm{E}\fpr{\int_{0}^1 \left\lvert \epsilon_{i}(s, t_{ij})\right\rvert ds}^2 \\
    &\leq  \sum_{i=1}^n m_i\sum_{j=1}^{m_i}\omega_{ij,l}^2 \textrm{E}\lVert \epsilon_i \rVert^2 \\
    &\leq \fpr{\sup_i m_i} \textrm{E}\lVert \epsilon \rVert^2 \sum_{i=1}^n\sum_{j=1}^{m_i} \omega_{ij,l}^2 \\
    &< \infty \quad a.s \quad \textrm{as } n \to \infty.
\end{align*}
where the first line is due to independence of the error across $i$, third line follows by Hölder's inequality. The proof is now complete by the uniform convergence of the eigenfunctions $\widehat{\phi}_k(s)$ and Assumption~\ref{assump: B3}. 
\end{proof}

\section{Justification for Assumption~\ref{assump: C2}} \label{sec: supp_IntJustSigma}
In this section, we will provide a justification for the validity of Assumption~\ref{assump: C2}. Suppose the covariance function of $\epsilon_{ik}(t)$, $\gamma_k(t,\tprime)$ in model~(\ref{eqn: derivemodel}) is estimated through a local linear smoother using the quasi projections $W_{k,ij}$ as follows:. 


In the first step, we estimate the mean $\eta_k(t)$ as
 \begin{equation} \label{eqn: etaktkernel}
          \widehat{\eta}_k(t, \bW_k) = \widehat{a}_{\eta 0} = \argmin_{a_{\eta0}, a_{\eta1}}\;  \sum_{i=1}^n v_{\eta i} \sum_{j=1}^{m_i}   K_{\heta}\left(T_{ij}-t\right)\left[W_{k,ij} - a_{\eta0} - a_{\eta1} (T_{ij}-t) \right]^2,
 \end{equation}
and in the second step, a smoothed version of $\gamma_k(t,\tprime)$ is estimated using the raw covariances $\widetilde{C}_{k,ij\jprime} = (\Wkij - \widehat{\eta}_k(t_{ij}))(W_{k,i\jprime} - \widehat{\eta}_k(t_{i\jprime}))$, $j, \jprime = 1,\dots, m_i, j \neq \jprime, i=1,\dots,n$ as
 \begin{align}
     \nonumber \widehat{\gamma}_k(t,\tprime, \bW_k) = \widehat{a}_{\gamma 0} &= \argmin_{a_{\gamma0}, a_{\gamma1}, a_{\gamma2}}  \;\sum_{i=1}^n v_{\gamma i} \sum_{j \neq \jprime}^{m_i}   \left[\widetilde{C}_{k,ij\jprime} - a_{\gamma0} - a_{\gamma1} (T_{ij}-t) -  a_{\gamma2} (T_{i\jprime}-\tprime) \right]^2 \\
     & \qquad \qquad \qquad \qquad \times K_{\hGam}\left(T_{ij}-t \right)K_{\hGam}\left(T_{i\jprime}-\tprime \right). \label{eqn: gammaktkernel}
 \end{align}
where $v_{\eta i}$ and $v_{\gamma i}$ are the weights applied to each observations in the smoother, For different choices of weights see \cite{zhang2016sparse}. By Proposition $3.3$ of \cite{staicu2014likelihood}, Assumption~\ref{assump: C2} will be satisfied in our case if we can show that under certain regularity conditions $\widehat{\gamma}_k(t,\tprime, \bW_k)$ is a uniformly consistent estimator of $\gamma_k(t,\tprime)$ at a rate $n^{-\alpha}$ for some $\alpha$ that is not too small. Uniform consistency of the estimator of the mean and covariance function is shown under different sparsity level of repeated measurements $T_{ij}$ by \cite{yao2005functional, li2010uniform} and \cite{zhang2016sparse} to name a few. The idea of the proof is similar for both mean and covariance. So, we will only justify the steps to show the uniform convergence of $\widehat{\eta}_k(t, \bW_k)-\eta_k(t)$. Note that
\begin{equation*}
    \widehat{\eta}_k(t, \bW_k) = \frac{\mathfrak{R}_{0,k}\mathfrak{V}_{2}-\mathfrak{R}_{1,k}\mathfrak{V}_{1}}{\mathfrak{V}_{2}\mathfrak{V}_{0}-\mathfrak{V}_{1}^2} .
\end{equation*}
where for $p = 0,1,2$,
\begin{align*}
    \mathfrak{R}_{p,k} &= \sum_{i=1}^n v_i \sum_{j=1}^{M_i} K_{\heta}\left(T_{ij}-t\right) \fpr{\frac{T_{ij}-t}{\heta}}^p W_{k, ij} \\
        \mathfrak{V}_{p} &= \sum_{i=1}^n v_i \sum_{j=1}^{M_i} K_{\heta}\left(T_{ij}-t\right) \fpr{\frac{T_{ij}-t}{\heta}}^p.
\end{align*}
Then, we can write
\begin{align*}
    &\widehat{\eta}_k(t, \bW_k) - \eta_k(t) \\
    &= \fpr{\mfV_0\mfV_2 - \mfV_1^2}^{-1}\tpr{\fpr{\mfR_{0,k} - \eta_k(t)\mfV_0 - \heta\eta_k^\prime(t)\mfV_1}\mfV_2 - \fpr{\mfR_{1,k} - \eta_k(t)\mfV_1 - \heta\eta_k^\prime(t)\mfV_2}\mfV_1}.
\end{align*}
Following the steps of either \cite{li2010uniform} or \cite{zhang2016sparse}, one can derive that the quantities $\mfV_p, p=0,1,2$ are uniformly bounded and the denominator of the above expression $\mfV_0\mfV_2 - \mfV_1^2$ uniformly bounded away from zero. However, the uniform convergence is driven by the first quantity in the numerator i.e. $\mfR_{0,k} - \eta_k(t)\mfV_0 - \heta\eta_k^\prime(t)\mfV_1$. However, since $\mfR_{0,k}$ involves the quasi projections $W_{k,ij}$ which are not independently distributed across $i$, many key steps in the proof, like Bernstein type of concentration inequality, fail miserably. However, if we define,
\begin{align*}
    \mftR_{p,k} &= \sum_{i=1}^n v_i \sum_{j=1}^{M_i} K_{\heta}\left(T_{ij}-t\right) \fpr{\frac{T_{ij}-t}{\heta}}^p Y_{k, ij}.
\end{align*}
which is a function of the unobserved projected response $Y_{k,ij}$, the almost sure uniform convergence rates can be derived for the quantity $\mftR_{0,k} - \eta_k(t)\mfV_0 - \heta\eta_k^\prime(t)\mfV_1$ without any theoretical pitfall, because $Y_{k,ij}$'s are independent across $i$. Moreover, by Assumption~\ref{assump: B2} and application of Hölder's inequality the difference
\begin{align*}
    \sup_t \abs{\mftR_{0,k}(t) - \mfR_{0,k}(t)} 
    &\leq \sum_{i=1}^n v_i \sum_{j=1}^{m_i} K_{\heta}\left(T_{ij}-t\right) \abs{ Y_{k, ij} -  W_{k, ij}} \\
    &\leq \sum_{i=1}^n v_i \sum_{j=1}^{m_i} K_{\heta}\left(T_{ij}-t\right) \int_{0}^1 \abs{Y_{ij}(s)}\abs{\widehat{\phi}_k(s) - \phi_k(s)} ds \\
    &\leq \sup_{s}\;\abs{\widehat{\phi}_k(s) - \phi_k(s)} M_K\sum_{i=1}^n v_i \sum_{j=1}^{m_i} \int_{0}^1 \abs{Y_{ij}(s)}ds \\
    &\leq  \sup_{s}\;\abs{\widehat{\phi}_k(s) - \phi_k(s)} M_K\fpr{\sum_{i=1}^n v_i \sum_{j=1}^{m_i} \fpr{\int_{0}^1 \abs{Y_{ij}(s)}ds}^\tau}^{1/\tau} \\
    &\leq \sup_{s}\;\abs{\widehat{\phi}_k(s) - \phi_k(s)} M_K\fpr{2^{\tau-1}\sum_{i=1}^n v_i \sum_{j=1}^{m_i} \spr{\norm{\mu}^\tau + \norm{\epsilon_i}^\tau} }^{1/\tau} \\
    &\leq \sup_{s}\;\abs{\widehat{\phi}_k(s) - \phi_k(s)} M_K\fpr{2^{\tau-1}\spr{\norm{\mu}^\tau + \sup_n (n \max_i v_i m_i) \frac{1}{n}\sum_{i=1}^n  \norm{\epsilon_i}^\tau} }^{1/\tau} \\
    &= \mathcal{O}(\sqrt{\log n/n} + \alpha_n) \quad \textrm{a.s}.
\end{align*}
is uniformly convergent with an order that is driven by the uniform rates of the eigen function $\phi_k(s)$. Similarly we can show that $\sup_t \abs{\mftR_{1,k}(t) - \mfR_{1,k}(t)} = \mathcal{O}(\sqrt{\log n/n} + \alpha_n)$ a.s. Let $\widehat{\eta}_k(t, \bY_k)$ be the estimator $\eta_k(t)$ obtained by replacing $W_{k,ij}$ by $Y_{k,ij}$ in (\ref{eqn: etaktkernel}). This means if under certain conditions, $\sup_t \abs{\widehat{\eta}_k(t, \bY_k) - \eta_k(t)} = \mathcal{O}(n^{-\alpha})$, then $\sup_t \abs{\widehat{\eta}_k(t, \bW_k) - \eta_k(t)} = \mathcal{O}(n^{-\alpha} + \sqrt{\log n/n} + \alpha_n)$ a.s, where $\alpha_n$ is the rate for the estimation of the bivariate mean function. Following Theorem (2b) of \cite{chen2012modeling}, when the design of $t$ is sparse, an optimal rate for $\widehat{\mu}(s,t)$ is $\alpha_n \approx n^{-2/5}$. Thus, an uniform convergence rate of the order $n^{-\min\{\alpha, 2/5\}}$ for the estimator $\widehat{\eta}_k(t, \bW_k)$ is achievable. Similar justification can be provided the estimator $\widehat{\gamma}_k(t,\tprime,\bW_k)$ involving the quasi projections $W_{k,ij}$, and hence for the eigen functions of $\gamma_k(t,\tprime)$ justifying the validity of Assumption~\ref{assump: C2}.

\section{Additional simulation results} \label{sec: supp_additionalsimresults}

{\color{black}\subsection{Validity of the p-values} \label{sec: supp_validity_pvalues}
 Figure~\ref{fig:qqplot} displays the plot of uniform quantiles versus ordered p-values obtained from the individual pseudo-likelihood ratio tests under the null hypothesis, for different sample size and sparsity levels, based on 5,000 replications. In each subplot, the two colored lines correspond to the two p-values from pseudo LRT conducted by projecting the data onto two eigenfunctions, based on a PVE of $90\%$. The linear pattern of each of these plots indicates that the asymptotic null distribution of $\widehat{pLRT}_{N,k}$ in Theorem~\ref{theorem: pLRTnulldist} also has good finite sample properties. Figure~\ref{fig:qqplot_additional} presents the quantile-quantile plot of two additional p-values obtained by selecting a higher PVE of $99\%$, estimating about $3$ to $4$ eigenfunctions, thereby showing the validity of the null distribution under finite samples for all $k=1,\dots, K$.
\begin{figure}
    \centering
    \subfloat{{\includegraphics[scale=0.25]{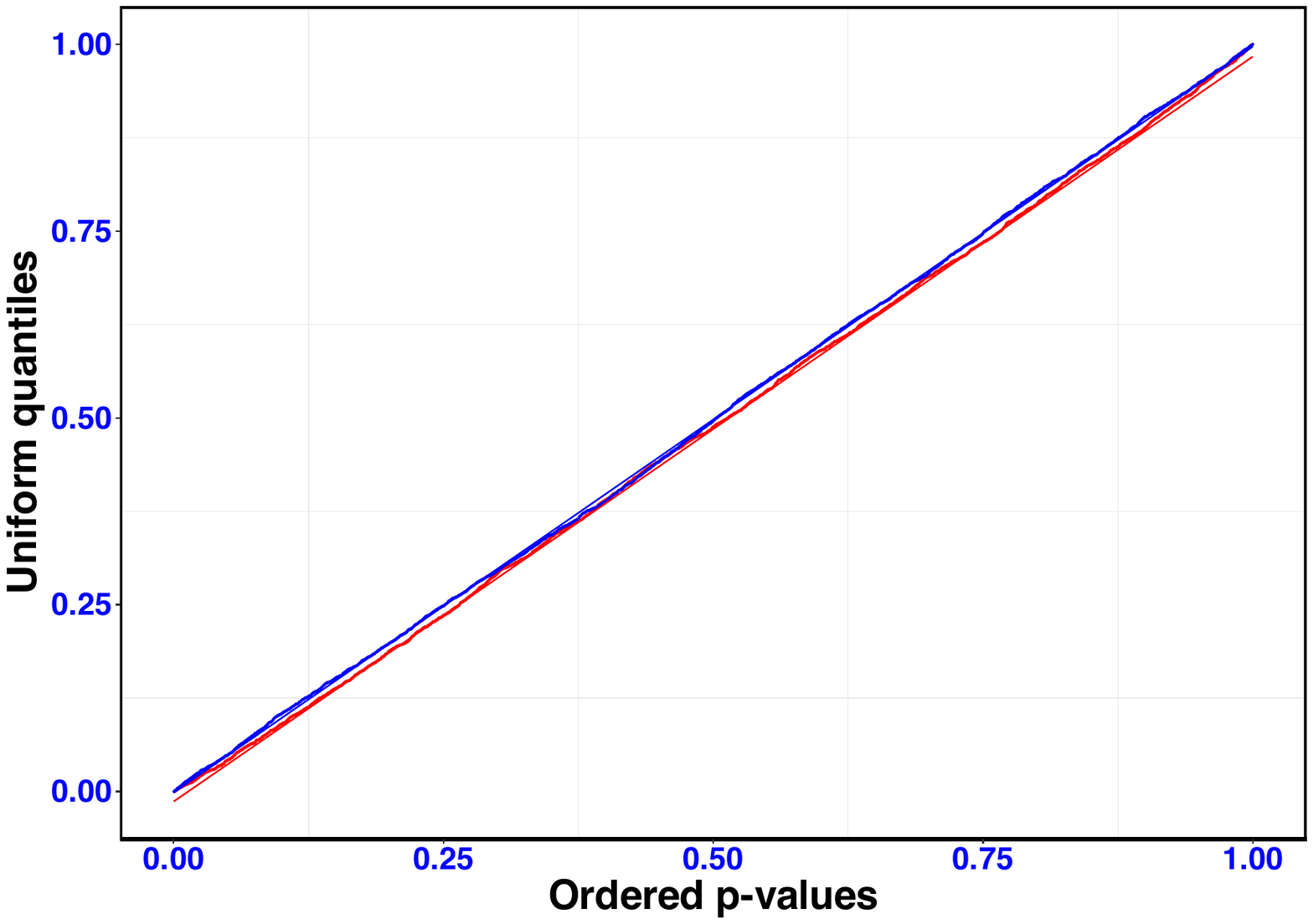} }} \qquad
    \subfloat{{
    \includegraphics[scale=0.25]{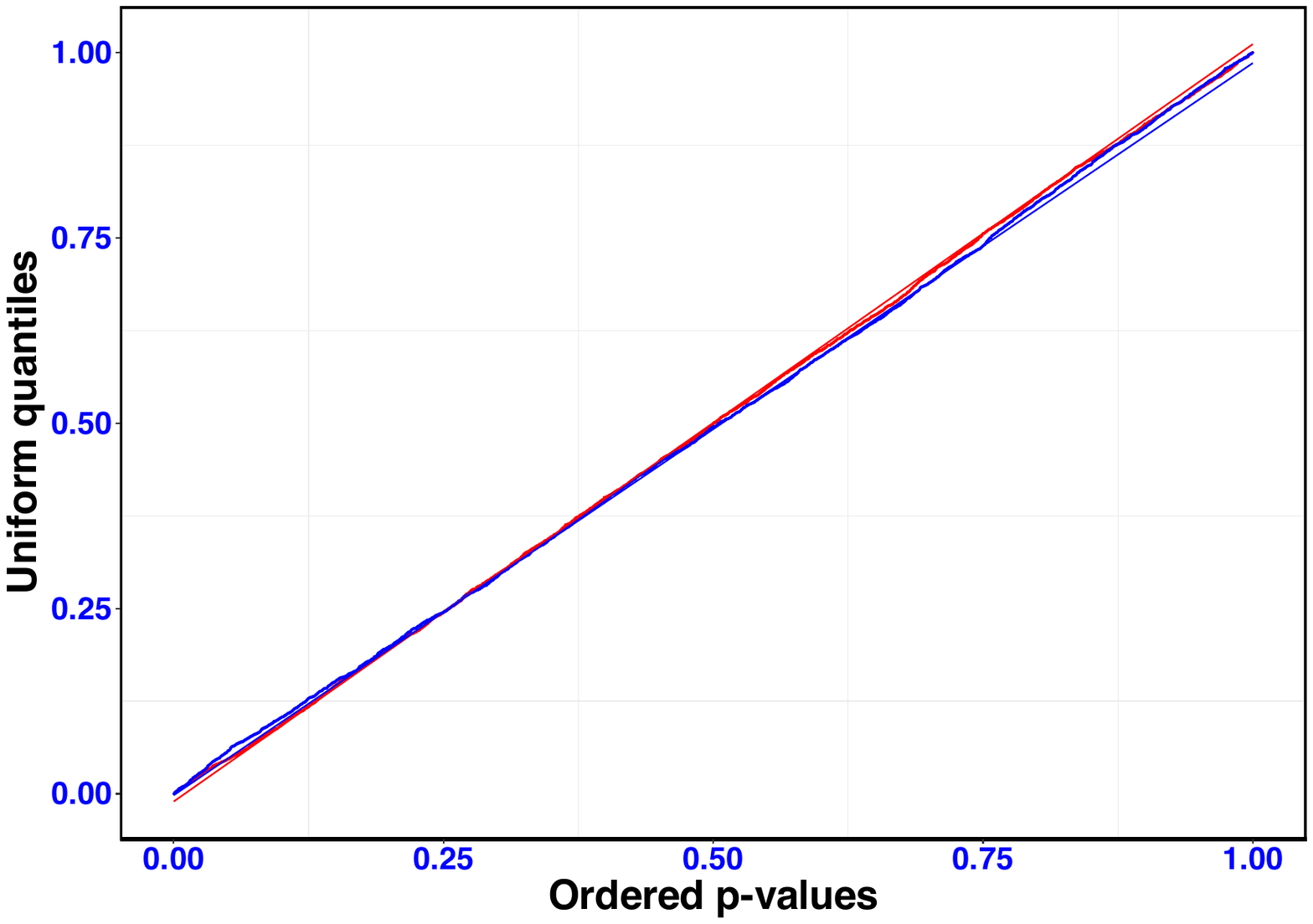} }}\\
\subfloat{{\includegraphics[scale=0.25]{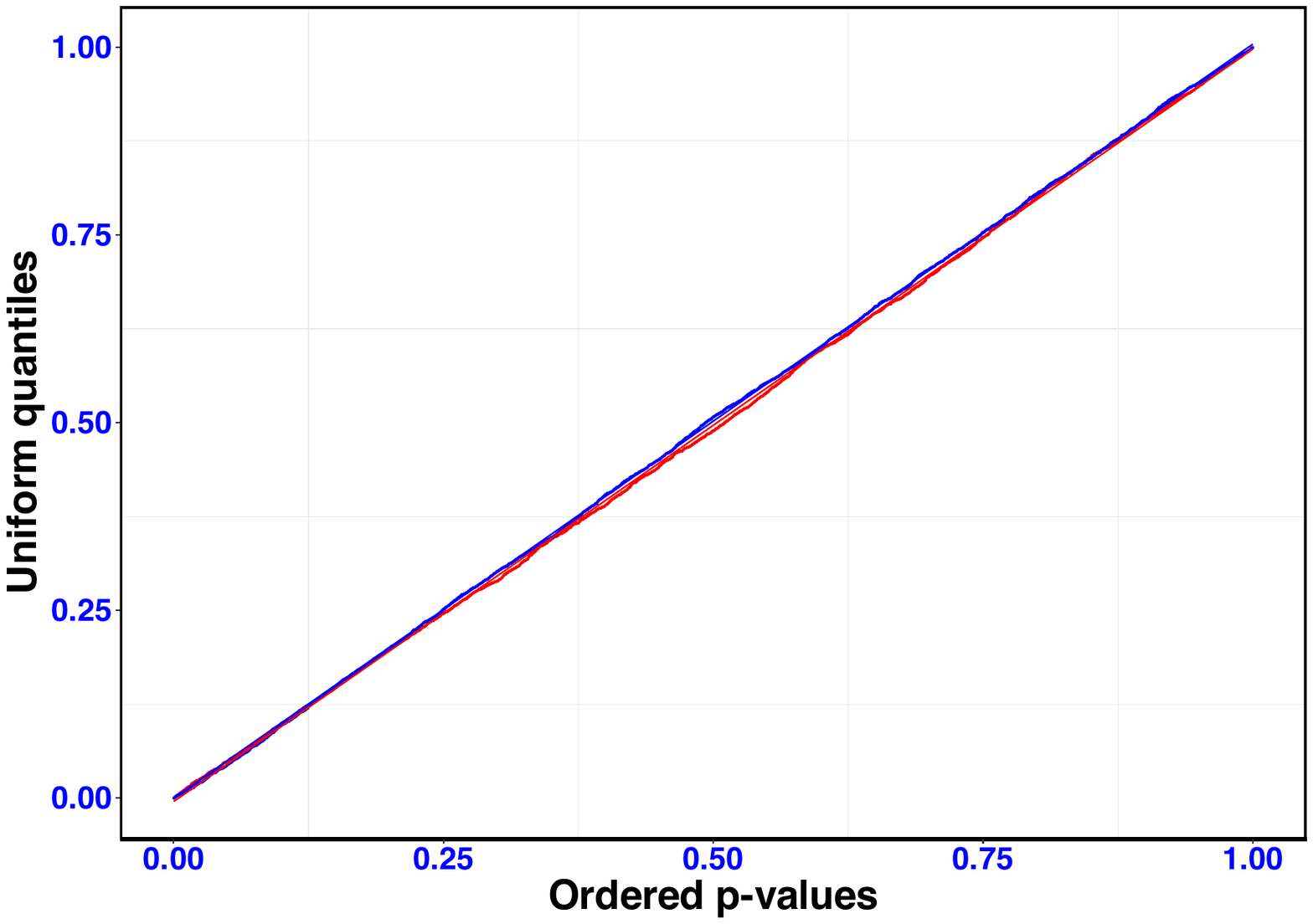} }} \qquad
    \subfloat{{
    \includegraphics[scale=0.25]{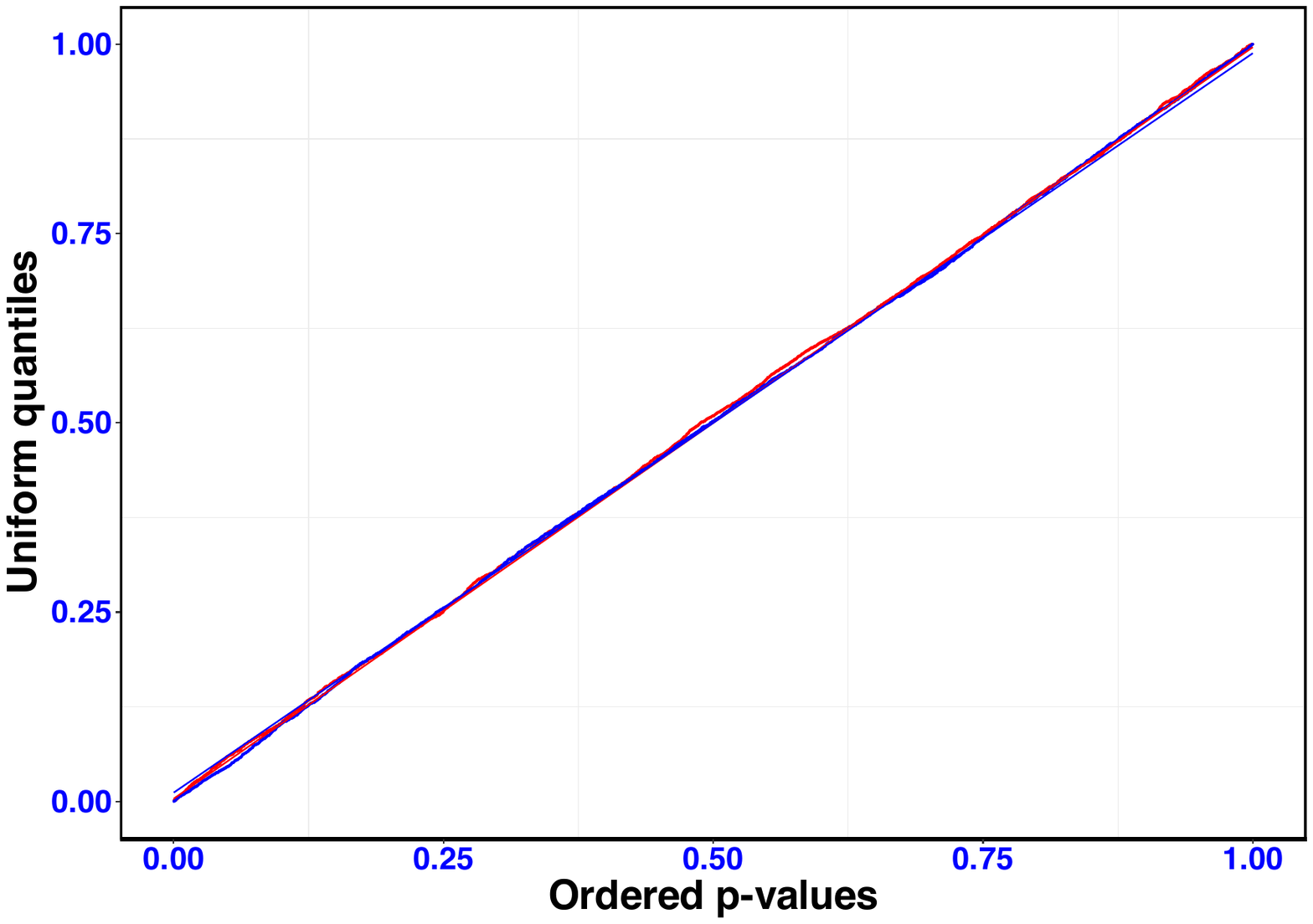} }}
    \caption{\textcolor{black}{The plot of the ordered p-values obtained from the individual pseudo-likelihood ratio-tests versus the quantiles of Uniform$[0,1]$ distribution. The top panel corresponds to {\em moderate} sparsity and the bottom panel to {\em low} sparsity level. The left panel is for $n=100$, and the right panel is for $n=200$. A PVE of $ 90\%$ selects $K=2$ to $3$ eigenfunctions. In each graph, the two colors correspond to the first two p-values obtained by performing the pseudo-likelihood ratio test ($\widehat{pLRT}_{N,k} : k=1, \dots, 2$). The red colors refer to the pseudo LRT for $k=1$ (first eigenfunction), and the blue color for $k=2$ (second eigenfunction).}}
\label{fig:qqplot}.
\end{figure}

\begin{figure}
    \centering
    \subfloat{{\includegraphics[scale=0.25]{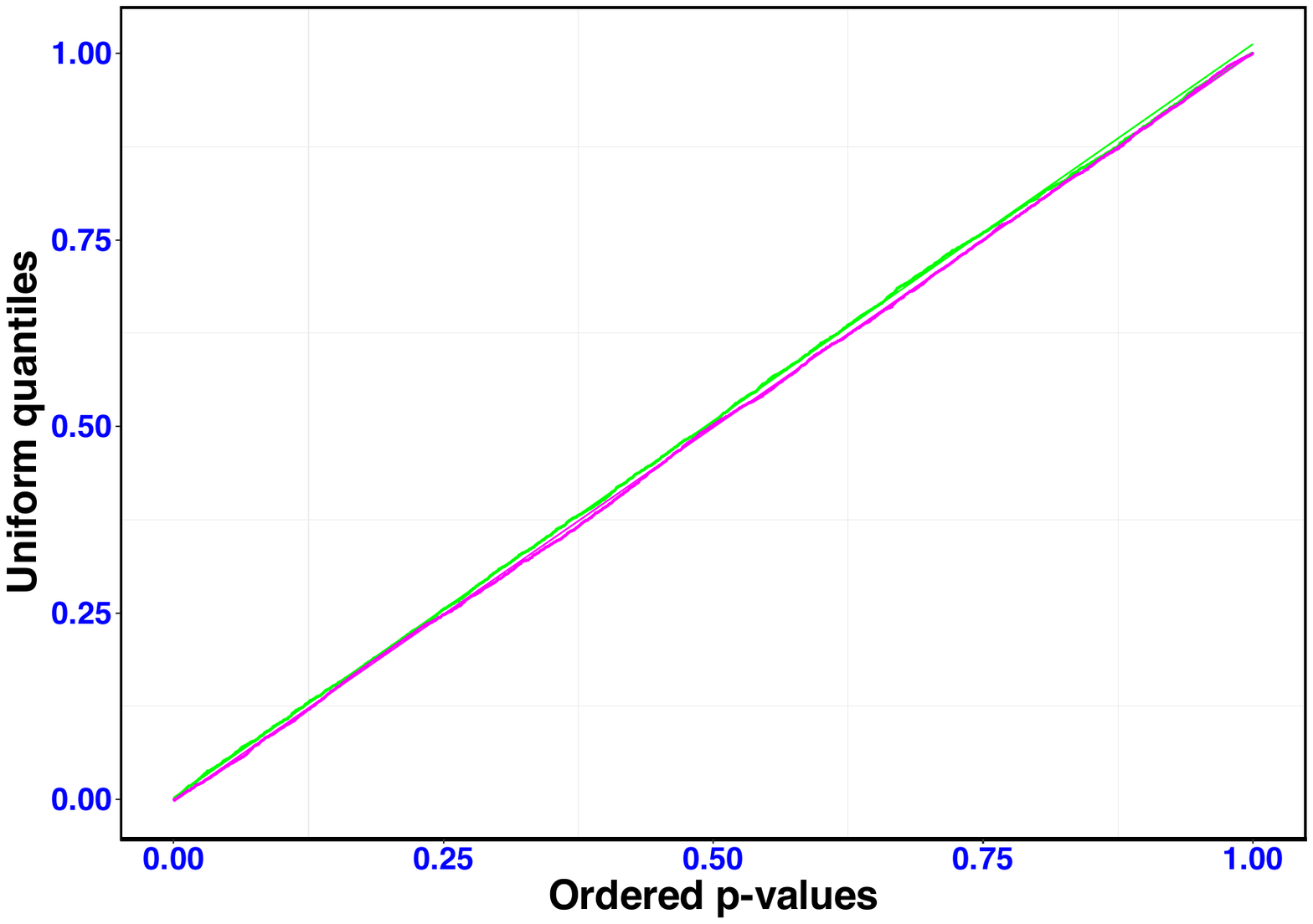} }} \qquad
    \subfloat{{
    \includegraphics[scale=0.25]{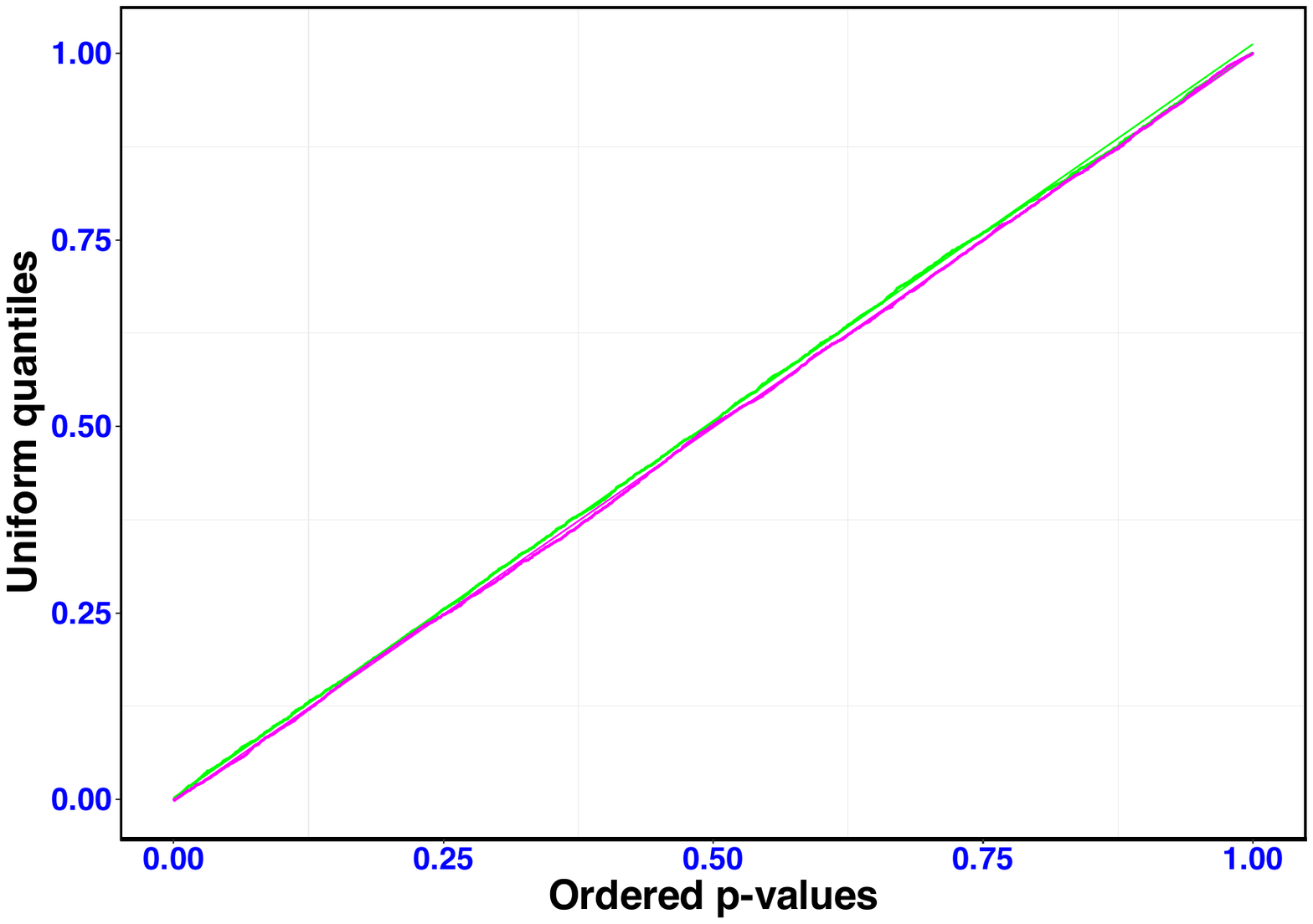} }}\\
\subfloat{{\includegraphics[scale=0.25]{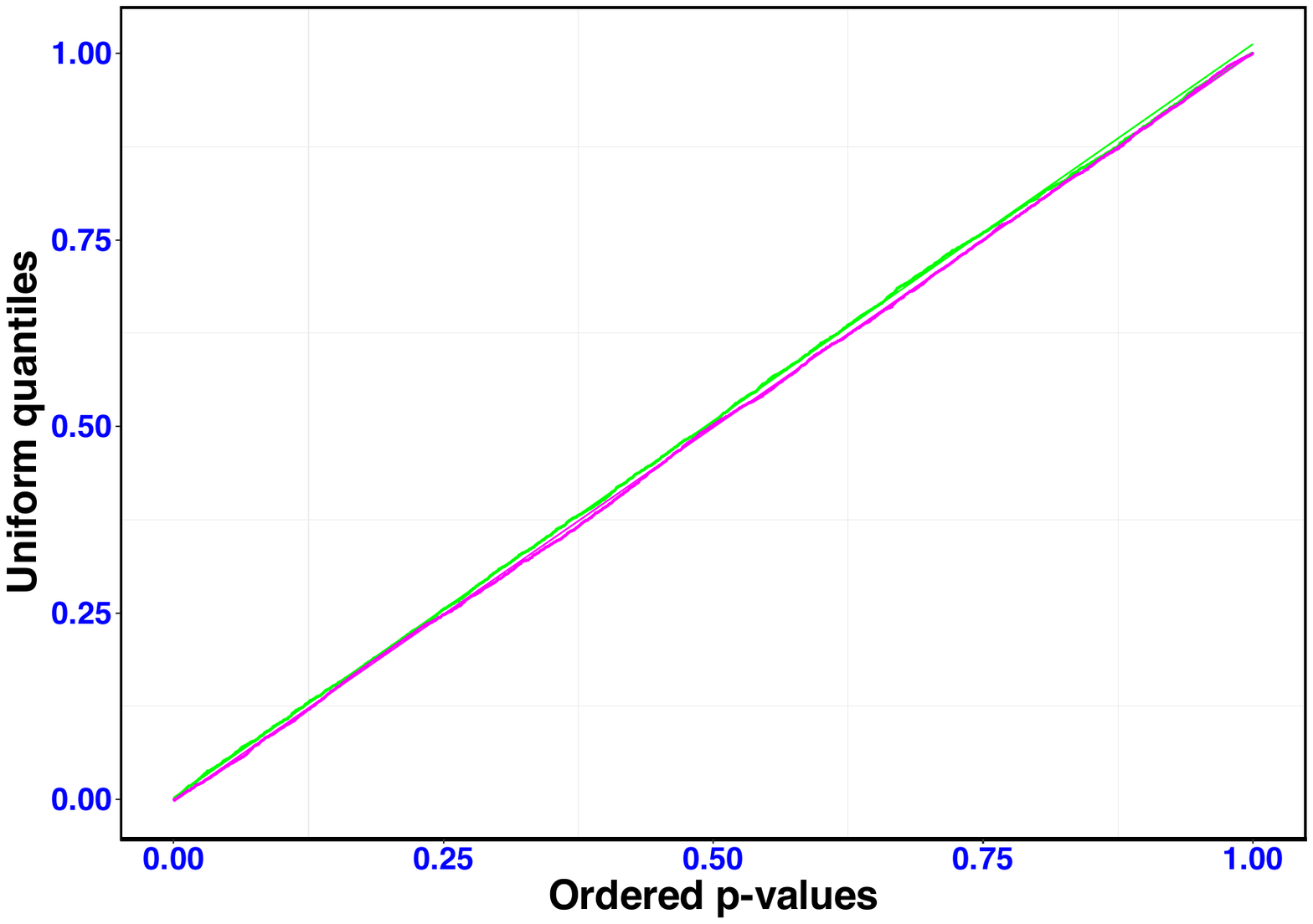} }} \qquad
    \subfloat{{
    \includegraphics[scale=0.25]{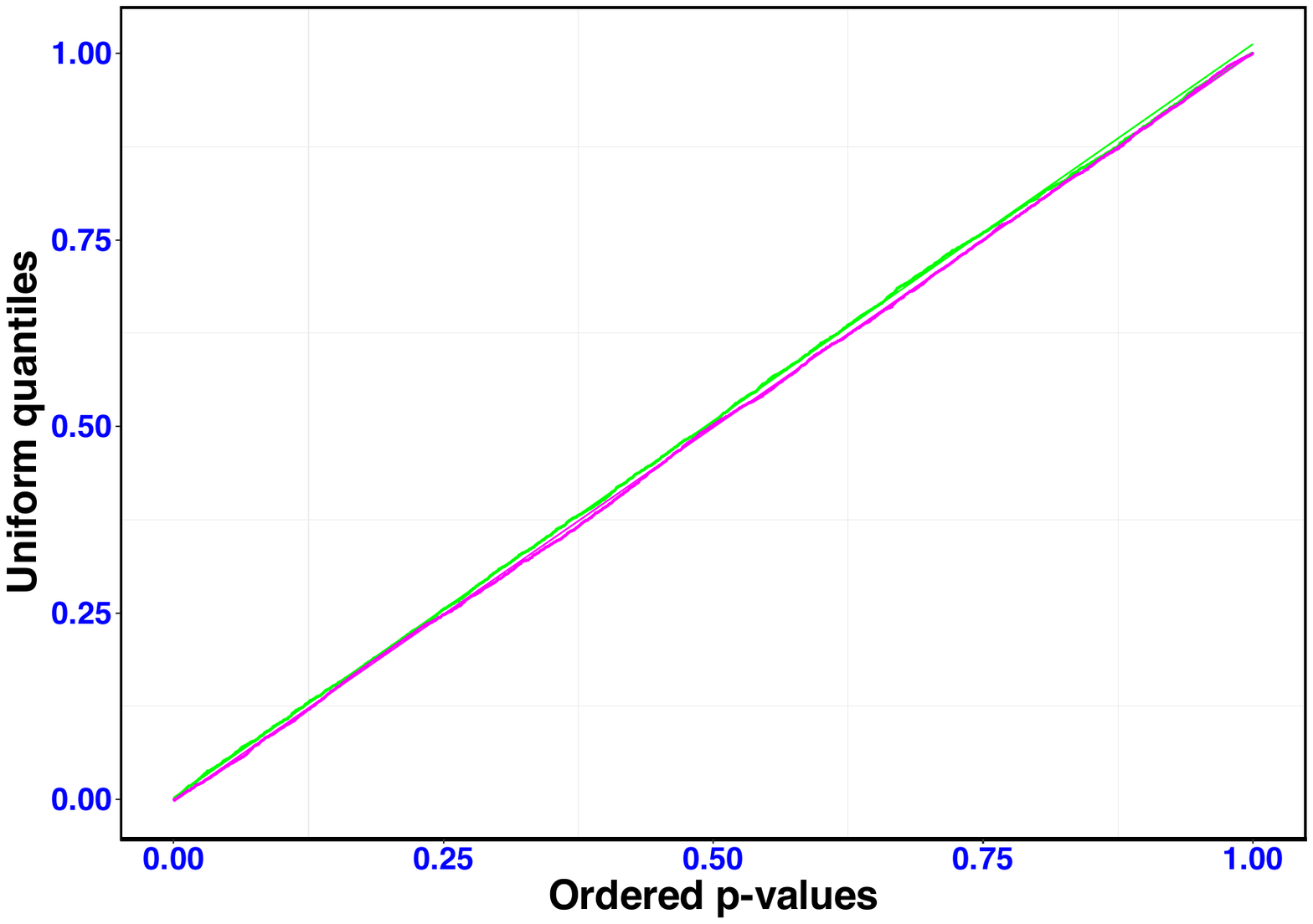} }}
    \caption{\textcolor{black}{The plot of the ordered p-values obtained from the pseudo-likelihood ratio-test across the quantiles of Uniform(0,1) distribution. The upper panel corresponds to {\em moderate} sparsity and the lower panel to {\em low} sparsity level. The left panel is for $n=100$, and the right panel is for $n=200$. In each graph, the two colors correspond to the two \textbf{additional} p-values of the pseudo LRT obtained by setting a higher PVE of $99\%$. The p-values corresponding to the third eigenfunction are colored green, and magenta for the fourth functions.}}
\label{fig:qqplot_additional}.
\end{figure}
}

{\color{black}\subsection{Sensitivity of PROFIT to number of knots $Q$} \label{sec: supp_senstivity_Q}
The placement of knots is at equispaced quantiles of the set of the pooled time points; as it is standard in the semiparametric literature. Table~\ref{tab: supp_sens_Q} shows the performance of PROFIT for different numbers of knots $Q$ used to represent $\eta_k(t)$. As expected, the results indicate that changing the number of knots does not affect either the test's size or its power. Consistent with the literature, as long as a reasonably large number of knots is used, increasing it further does not have any noticeable impact on PROFIT's performance.
\begin{table}[]
\centering
    \caption{Effect of number of knots ($Q$) on the power of the test at $\alpha=0.05$. The value of PVE is set at $90\%$. The columns with $\delta=0$ correspond to the case when the null hypothesis is true.}
    \label{tab: supp_sens_Q}
\begin{tabular}{ccccccccccccc}
\hline \hline 
\multicolumn{12}{c}{$m_i \sim \{8,\dots, 12\}$}                                                      \\ \hline
\multicolumn{3}{c}{\multirow{2}{*}{\begin{tabular}[c]{@{}c@{}} $Q$ \end{tabular}}} & & \multicolumn{4}{c}{$n = 100$} & & \multicolumn{4}{c}{$n = 200$} \\ 
\cline{5-13} 
\multicolumn{3}{c}{}           & & $\delta=0$ & $0.4$ & $0.8$ & $1.2$  & & $\delta=0$ & $0.4$ & $0.8$ & $1.2$ \\ \hline
\multicolumn{3}{c}{$10$}  & &  0.05 &	0.17	& 0.52	& 0.88 &	& 0.05 &	0.30	& 0.82 & 0.99  \\
\multicolumn{3}{c}{$20$}  & & 0.05 &	0.17	& 0.52	& 0.88 &	& 0.05 &	0.31	& 0.82 & 0.99   \\
\multicolumn{3}{c}{$30$} & & 0.05 &	0.17	& 0.52	& 0.88 &	& 0.05 &	0.31	& 0.82 & 0.99 	 \\ \hline
\multicolumn{12}{c}{$m_i \sim \{15,\dots, 20\}$}                                                      \\ \hline
\multicolumn{3}{c}{\multirow{2}{*}{\begin{tabular}[c]{@{}c@{}} $Q$ \end{tabular}}} & & \multicolumn{4}{c}{$n = 100$} & & \multicolumn{4}{c}{$n = 200$} \\ 
\cline{5-13} 
\multicolumn{3}{c}{}           & & $\delta=0$ & $0.05$ & $0.1$ & $0.2$  & & $\delta=0$ & $0.05$ & $0.1$ & $0.2$ \\ \hline
\multicolumn{3}{c}{$10$}  & &  0.05 &	0.27	& 0.82	& 1.00 &	& 0.04 &	0.52	& 0.99 & 1.00  \\
\multicolumn{3}{c}{$20$}  & & 0.05 &	0.27	& 0.82	& 1.00 &	& 0.04 &	0.52	& 0.99 & 1.00   \\
\multicolumn{3}{c}{$30$} & & 0.05 &	0.27	& 0.82	& 1.00 &	& 0.04 &	0.52	& 0.99 & 1.00 	 \\ \hline
\end{tabular}
\end{table}
}

\subsection{Additional results related to empirical size of PROFIT, ZC-BT, and ZC-MT} \label{sec: supp_additional_res_ZCMT_ZCBT}

Table \ref{tab: proposed only all} presents the Type-1 error rates of the PROFIT corresponding to nominal levels $\alpha = 0.01, 0.05, 0.10,$ and $0.15$ for different sample sizes $n$ and number of profiles per subject $m_{i}$ to illustrate the effect of the estimation of the eigen function $\widehat{\phi}_k(s)$ and the covariance function of the projected response, $\gamma_k(t,\tprime)$. 
Table~\ref{tab: size for all competitive methods} presents the empirical sizes of the \verb|ZC-MT| and \verb|ZC-BT| methods.

\begin{table}[H]
	\centering
	\caption{The empirical Type-1 error rates of the proposed method based on 5000 simulations. Standard errors are presented in parentheses.}  
	\scalebox{0.8}{
		\begin{tabular}{ccc cccc}
			\hline	\hline
			\multicolumn{7}{c}{$m_1 \sim \{8, \ldots, 12\}$} \\	\hline
			$n$ & $\phi_k(s)$-choice & $\gamma_{k}(t,t')$-choice & $\alpha = 0.01$ & 	$\alpha = 0.05$ & 	$\alpha = 0.10$ & 	$\alpha = 0.15$ \\ 
			\hline	\hline
			100 & True & True  & 0.010  (0.001) & 0.047  (0.003) & 0.093  (0.004) & 0.141  (0.005) \\ 
			& True & Est. & 0.013  (0.002) & 0.058 (0.003) & 0.105  (0.004) & 0.155  (0.005)  \\ 
			& Est. & Est.&  0.017  (0.002) & 0.063  (0.003) & 0.112  (0.004) & 0.156  (0.005) \\ \hline
			
			200 & True & True & 0.010  (0.001) & 0.048  (0.003) & 0.099  (0.004) & 0.147  (0.005)  \\ 
			& True & Est. &  0.012  (0.002) & 0.051  (0.003) & 0.102  (0.004) & 0.154 (0.005) \\ 
			& Est. & Est.& 0.012  (0.002) & 0.052  (0.003) & 0.103  (0.004) & 0.151  (0.005) \\ \hline
			
			300 & True & True  & 0.008  (0.001) & 0.049  (0.003) & 0.098  (0.004) & 0.145  (0.005)\\ 
			& True & Est. & 0.008  (0.001) & 0.051  (0.003) & 0.099  (0.004) & 0.147  (0.005) \\ 
			& Est. & Est.&0.008  (0.001) & 0.050  (0.003) & 0.097  (0.004) & 0.146  (0.005) \\ \hline
			
			400 & True & True  & 0.012  (0.002) & 0.051  (0.003) & 0.100  (0.004) & 0.147  (0.005)  \\ 
			& True & Est. & 0.013  (0.002) & 0.054  (0.003) & 0.098  (0.004) & 0.144  (0.005) \\ 
			& Est. & Est. & 0.013  (0.002) & 0.055  (0.003) & 0.098  (0.004) & 0.142  (0.005) \\ \hline\hline
			
			\multicolumn{7}{c}{$m_1 \sim \{15, \ldots, 20\}$} \\	\hline
			$n$ & $\phi_k(s)$-choice & $\gamma_{k}(t,t')$-choice & $\alpha = 0.01$ & 	$\alpha = 0.05$ & 	$\alpha = 0.10$ & 	$\alpha = 0.15$ \\ 				\hline\hline
			100 & True & True  & 0.011  (0.001) & 0.055  (0.003) & 0.103  (0.004) & 0.156  (0.005) \\ 
			& True & Est. &0.011  (0.002) & 0.058  (0.003) & 0.110  (0.004) & 0.160  (0.005) \\ 
			& Est. & Est.& 0.018  (0.002) & 0.066  (0.004) & 0.122  (0.005) & 0.171  (0.005) \\ \hline
			
			200 & True & True  & 0.010  (0.001) & 0.048  (0.003) & 0.100  (0.004) & 0.148  (0.005)\\ 
			& True & Est. & 0.011  (0.001) & 0.051  (0.003) & 0.104  (0.004) & 0.156 (0.005) \\ 
			& Est. & Est. & 0.012  (0.002) & 0.053  (0.003) & 0.106  (0.004) & 0.156  (0.005)\\ \hline
			
			300 & True & True & 0.010  (0.001) & 0.050  (0.003) & 0.103  (0.004) & 0.154  (0.005) \\ 
			& True & Est.  &0.011  (0.001) & 0.050  (0.003) & 0.107  (0.004) & 0.155  (0.005)  \\ 
			& Est. & Est.&0.011  (0.001) & 0.052  (0.003) & 0.109  (0.004) & 0.155  (0.005) \\ \hline
			
			400 & True & True &0.012  (0.002) & 0.054  (0.003) & 0.100  (0.004) & 0.150  (0.005) \\ 
			& True & Est.  &0.012  (0.002) & 0.053  (0.003) & 0.102  (0.004) & 0.151  (0.005)\\ 
			& Est. & Est. & 0.012  (0.002) & 0.056  (0.003) & 0.102  (0.004) & 0.153  (0.005)  \\
			\hline\hline
		\end{tabular}
	} \label{tab: proposed only all}
\end{table}

\begin{table}[H]
\centering
\caption{The empirical Type-1 error rates of the ZC-MT and ZC-BT methods based on 5000 simulations. Standard errors are presented in parentheses.}
\scalebox{0.85}{
\begin{tabular}{cc | cccc}
\hline \hline
\multicolumn{6}{c}{ZC-MT} \\
\hline
&& $\alpha = 0.01$ & 	$\alpha = 0.05$ & 	$\alpha = 0.10$ & 	$\alpha = 0.15$  \\
\hline
$m_{i} \sim \{8, \ldots, 12 \}$	&	$n = 100$   & 0.017  (0.002) & 0.066  (0.004) & 0.120  (0.005) & 0.169  (0.005) \\ 
&	$n = 200$    & 0.017  (0.002) & 0.063  (0.003) & 0.117  (0.005) & 0.168  (0.005) \\ 
&	$n = 300$   & 0.014  (0.002) & 0.065  (0.003) & 0.114  (0.004) & 0.164  (0.005) \\
&	$n = 400$   & 0.012  (0.002) & 0.062  (0.003) & 0.111  (0.004) & 0.157  (0.005) \\ 
\hline
$m_{i} \sim \{15, \ldots, 20 \}$	&	$n = 100$    & 0.009  (0.001) & 0.049  (0.003) & 0.091  (0.004) & 0.134  (0.005) \\ 
&	$n = 200$    & 0.007  (0.001) & 0.038  (0.003) & 0.075  (0.004) & 0.117  (0.005) \\ 
&	$n = 300$    & 0.008  (0.001) & 0.039  (0.003) & 0.081  (0.004) & 0.122  (0.005) \\ 
&	$n = 400$  & 0.007 (0.001) & 0.038  (0.003) & 0.078  (0.004) & 0.121  (0.005) \\ 
\hline \hline
\multicolumn{6}{c}{ZC-BT} \\
\hline
				&& $\alpha = 0.01$ & 	$\alpha = 0.05$ & 	$\alpha = 0.10$ & 	$\alpha = 0.15$  \\
				\hline
				$m_{i} \sim \{8, \ldots, 12 \}$	&	$n = 100$   & 0.005  (0.001) & 0.017  (0.002) & 0.033  (0.003) & 0.048  (0.003) \\ 
				&	$n = 200$     & 0.005  (0.001) & 0.017  (0.002) & 0.032  (0.002) & 0.047  (0.003) \\ 
				&	$n = 300$   & 0.004  (0.001) & 0.019  (0.002) & 0.035  (0.003) & 0.051  (0.003) \\ 
				&	$n = 400$   & 0.003  (0.001) & 0.017  (0.002) & 0.032  (0.002) & 0.048  (0.003) \\ 
				\hline
				$m_{i} \sim \{15, \ldots, 20 \}$	&	$n = 100$    & 0.007  (0.001) & 0.034  (0.003) & 0.062  (0.003) & 0.091  (0.004) \\ 
				&	$n = 200$    & 0.006  (0.001) & 0.028  (0.002) & 0.055  (0.003) & 0.084  (0.004) \\ 
				&	$n = 300$    & 0.007  (0.001) & 0.029  (0.002) & 0.055  (0.003) & 0.086  (0.004) \\ 
				&	$n = 400$ & 0.005 (0.001) & 0.027 (0.002) & 0.055  (0.003) & 0.087  (0.004) \\  
				\hline\hline

\end{tabular}} \label{tab: size for all competitive methods}
\end{table}

{\color{black}
\subsection{Comparison of the performance of PROFIT with data-driven basis and preset basis} \label{sec: supp_compare_presetbasis}
As one Reviewer suggested for the existing choice of our simulation design in Section~\ref{sec: SimStudy} of the manuscript, one could use Fourier basis to project the response trajectories and perform the testing procedure and indeed achieve great performance - in terms of both Type-1 error and power for $K$ small but even for $K$ as large as $30$. The mean function considered in Section \ref{sec: SimStudy} is $\mu(s,t) = \cos(\pi s/2) + 5\delta(t/4-s)^3$, which is a smooth function of $s$ and $t$. \\
However, we want to emphasize that such performance is conditional on using a basis informed by the knowledge of the true mean function, or its structure in terms of how it varies over $\mathcal{S}$. If instead the mean function $\mu(s,t)$ is not smooth in $s$, using Fourier basis would result in poor performance, irrespective of how large $K$ we select. To demonstrate this, we conduct additional simulation studies where we consider a different smoothness structure for the mean function, keeping all the other parameters of simulation design fixed. Specifically, consider a mean function that is a piecewise smooth function of $s$, i.e. $\mu(s,t) = t\{-0.2 + 0.5s + 0.1(s-0.5)_{+}\} + 0.5 + 0.25s + 0.1s^2 + 0.05(s-0.5)^2_{+}$. We distinguish these two mean structures in terms of their overall smoothness, and denote the two cases as {\em smooth} and {\em piecewise-smooth} mean.}

\begin{figure}[!h]
    \centering
    \subfloat[{\em Smooth} mean]{\includegraphics[scale= 0.35]{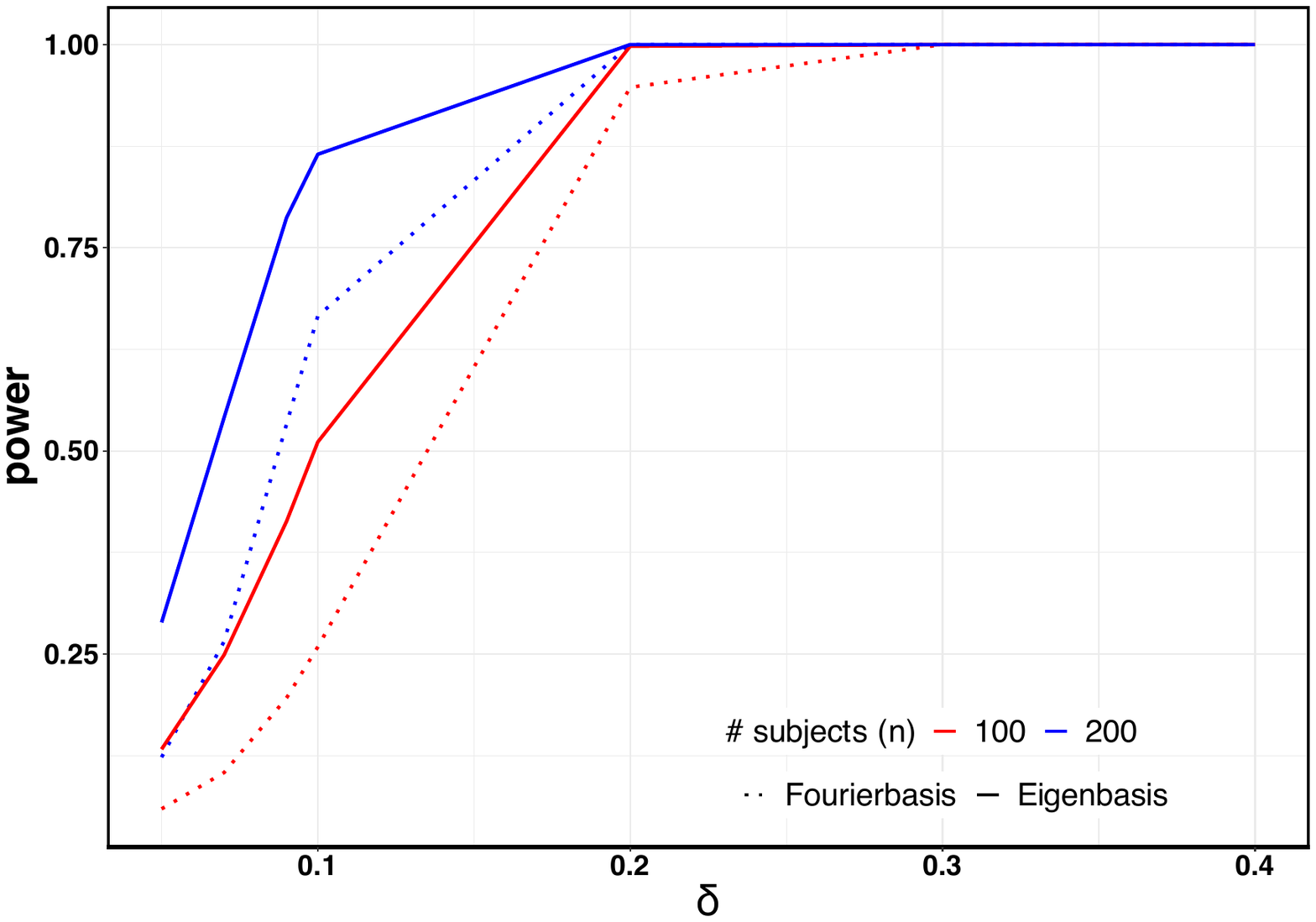}} \quad 
        \subfloat[{\em Piecewise smooth} mean]{\includegraphics[scale= 0.35]{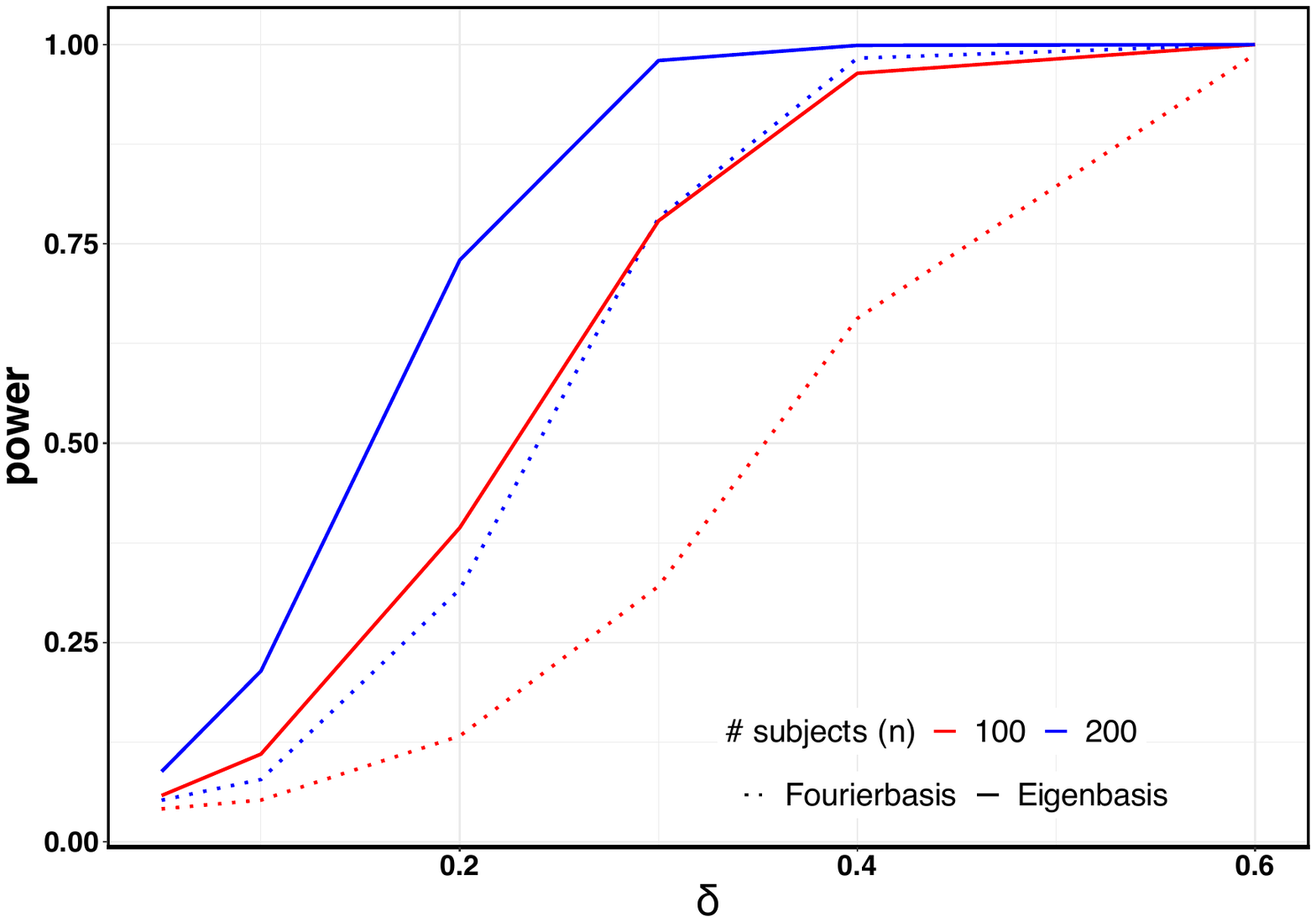}}
    \caption{A comparison of the power function of PROFIT versus the projection-based test using 30 preset Fourier bases, for two different choices of the mean function, (a) where the mean function is a smooth function of $s$ and $t$, (b) where the mean function is piecewise smooth function of $s$, but a smooth function of $t$. The significance level is chosen to be $0.05$. The numbers are calculated based on $1000$ simulations. PVE of projection-based test is set as $99\%$, leading to four eigenfunctions. The number of observations per subject is fixed at {\em moderate} sparsity level, i.e., $m_i \sim \{8, \dots, 12\}$.}
    \label{fig: supp_fourier_vs_eigen}
\end{figure}

{\color{black}Inspired by the Reviewer's suggestion, we consider 30 Fourier basis to project the data and use the pseudo-likelihood ratio test on the projected data to infer about the null hypothesis, using a Bonferroni's correction with $K = 30$. Table \ref{tab: supp_size_fourierbasis} presents the empirical Type-1 error of the PROFIT using 30 Fourier basis, for both {\em smooth} and {\em piecewise-smooth} mean cases. The numbers suggests that PROFIT with preset basis is also a valid testing procedure. The empirical size properties of PROFIT with data-driven bases for {\em piecewise-smooth} mean case is also presented in Table~\ref{tab: supp_size_eigenbasis_nonsmooth}, for completeness.

Figure \ref{fig: supp_fourier_vs_eigen} presents a comparison of the power performance of PROFIT using 30 Fourier bases and the data-driven eigenbases estimated from a PVE of $99\%$. The upper panel of the figure presents the power comparison for the {\em smooth} mean case, showcasing that the power of the test using Fourier bases iss competitive to data-driven basis. However, when the mean function is not smooth throughout the entire interval of $[0,1]$, the test's power drops significantly when using 30 Fourier bases, as shown in the bottom panel of Figure~\ref{fig: supp_fourier_vs_eigen}.}

\begin{table}
	\centering
	\caption{The empirical Type-1 error rates of projection-based test using 30 Fourier basis. Standard errors are presented in parentheses.}  
		\begin{tabular}{cc cccc}
  \hline\\
  \multicolumn{6}{c}{$m_i \sim \{8, \dots, 12\}$} \\
			\hline	\hline
			\multicolumn{6}{c}{{\em Smooth} mean} \\	\hline
			& & $\alpha = 0.01$ & 	$\alpha = 0.05$ & 	$\alpha = 0.10$ & 	$\alpha = 0.15$ \\ 
			\hline	\hline
	$n = $	100	 &&  0.017  (0.002) & 0.040  (0.003) & 0.093  (0.004) & 0.123  (0.005) \\  
	$n = $	150	 &&  0.017  (0.002) & 0.038  (0.003) & 0.089  (0.004) & 0.117  (0.005) \\ 
	$n = $	200	 && 0.017  (0.002) & 0.040  (0.003) & 0.088  (0.004) & 0.119  (0.005) \\  
	$n = $	300	 && 0.016  (0.002) & 0.041  (0.003) & 0.094  (0.004) & 0.122  (0.005) \\  
			\hline	\hline
			\multicolumn{6}{c}{{\em Piecewise-smooth} mean} \\	\hline
			& & $\alpha = 0.01$ & 	$\alpha = 0.05$ & 	$\alpha = 0.10$ & 	$\alpha = 0.15$ \\ 
			\hline	\hline
	$n = $	100	 &&  0.017  (0.002) & 0.040  (0.003) & 0.093  (0.004) & 0.123  (0.005) \\  
	$n = $	150	 &&  0.017  (0.002) & 0.038  (0.003) & 0.089  (0.004) & 0.117  (0.005) \\ 
	$n = $	200	 && 0.017  (0.002) & 0.040  (0.003) & 0.088  (0.004) & 0.119  (0.005) \\  
	$n = $	300	 && 0.016  (0.002) & 0.041  (0.003) & 0.094  (0.004) & 0.122  (0.005)  \\  
			\hline	\hline \\ \\
		\end{tabular} \\
  		\begin{tabular}{cc cccc}
    \hline\\
                \multicolumn{6}{c}{$m_i \sim \{15, \dots, 20\}$} \\
			\hline	\hline
			\multicolumn{6}{c}{{\em Smooth} mean} \\	\hline
			& & $\alpha = 0.01$ & 	$\alpha = 0.05$ & 	$\alpha = 0.10$ & 	$\alpha = 0.15$ \\ 
			\hline	\hline
	$n = $	100	 &&  0.017  (0.002) & 0.041  (0.003) & 0.092  (0.004) & 0.121  (0.005) \\  
	$n = $	150	 &&  0.018  (0.002) & 0.039  (0.003) & 0.096  (0.004) & 0.125  (0.005) \\ 
	$n = $	200	 && 0.014  (0.002) & 0.039  (0.003) & 0.098  (0.004) & 0.125  (0.005) \\  
	$n = $	300	 && 0.015  (0.002) & 0.041  (0.003) & 0.094  (0.004) & 0.119  (0.005) \\  
			\hline	\hline
			\multicolumn{6}{c}{{\em Piecewise-smooth} mean} \\	\hline
			& & $\alpha = 0.01$ & 	$\alpha = 0.05$ & 	$\alpha = 0.10$ & 	$\alpha = 0.15$ \\ 
			\hline	\hline
	$n = $	100	 &&  0.017  (0.002) & 0.041  (0.003) & 0.092  (0.004) & 0.121  (0.005) \\  
	$n = $	150	 &&  0.018  (0.002) & 0.039  (0.003) & 0.096  (0.004) & 0.125  (0.005) \\ 
	$n = $	200	 && 0.014  (0.002) & 0.039  (0.003) & 0.098  (0.004) & 0.125  (0.005) \\  
	$n = $	300	 && 0.015  (0.002) & 0.041  (0.003) & 0.094  (0.004) & 0.120  (0.005)  \\  
			\hline	\hline
		\end{tabular}
	\label{tab: supp_size_fourierbasis}
\end{table}

\begin{table}
	\centering
	\caption{The empirical Type-1 error rates of PROFIT across different sparsity levels of the observations and number of subjects when the mean function is {piecewise-smooth} function of $s$. Standard errors are presented in parentheses. The numbers are obtained from 5000 replications.}  
		\begin{tabular}{cc cccc}
			\hline	\hline
			\multicolumn{6}{c}{$m_i \sim \{8,\dots, 12\}$} \\	\hline
			& & $\alpha = 0.01$ & 	$\alpha = 0.05$ & 	$\alpha = 0.10$ & 	$\alpha = 0.15$ \\ 
			\hline	\hline
	$n = $	100	 &&  0.013  (0.002) & 0.051  (0.003) & 0.095  (0.004) & 0.145  (0.005) \\  
	$n = $	150	 &&  0.008  (0.001) & 0.049  (0.003) & 0.094  (0.004) & 0.145  (0.005) \\ 
	$n = $	200	 && 0.010  (0.001) & 0.047  (0.003) & 0.089  (0.004) & 0.134  (0.005) \\  
	$n = $	300	 && 0.009  (0.001) & 0.048  (0.003) & 0.095  (0.004) & 0.138  (0.005) \\  
			\hline	\hline
			\multicolumn{6}{c}{$m_i \sim \{15,\dots, 20\}$} \\	\hline
			& & $\alpha = 0.01$ & 	$\alpha = 0.05$ & 	$\alpha = 0.10$ & 	$\alpha = 0.15$ \\ 
			\hline	\hline
	$n = $	100	 &&  0.008  (0.001) & 0.049  (0.003) & 0.090  (0.004) & 0.134  (0.005) \\  
	$n = $	150	 &&  0.009  (0.001) & 0.043  (0.003) & 0.089  (0.004) & 0.138  (0.005) \\ 
	$n = $	200	 && 0.009  (0.001) & 0.045  (0.003) & 0.088  (0.004) & 0.132  (0.005) \\  
	$n = $	300	 && 0.009  (0.001) & 0.045  (0.003) & 0.092  (0.004) & 0.136  (0.005)  \\  
			\hline	\hline
		\end{tabular}
	\label{tab: supp_size_eigenbasis_nonsmooth}
\end{table}

\end{document}